\documentclass[prl,aps,amsfonts,showpacs,superscriptaddress,preprintnumbers,nofootinbib,10pt,twocolumn]{revtex4-2}
\usepackage{array}[=2016-10-06]

\renewcommand{\thesection}{\arabic{section}}
\renewcommand{\thesubsection}{\thesection.\arabic{subsection}}

\newcommand{\secpar}[1]{\medskip\textbf{#1.---}\ignorespaces} 

\makeatletter
\let\@afterindenttrue\@afterindentfalse
\makeatother

\usepackage{tensor}
\usepackage{tikz}
\usetikzlibrary{shapes,arrows,positioning}
\usepackage{graphicx}
\usepackage[dvipsnames]{xcolor}
\usepackage{rotating}
\usepackage{bm,amsmath,amssymb}
\allowdisplaybreaks[4]
\usepackage[utf8]{inputenc}
\usepackage{footmisc}
\usepackage{soul}
\usepackage[colorlinks=true, urlcolor=RoyalBlue, linkcolor=RubineRed, citecolor=RoyalBlue, hyperindex=true, linktocpage=true,linktoc=all]{hyperref}
\usepackage{orcidlink} 
\usepackage{physics}
\usepackage{braket}
\usepackage{balance} 
\usepackage{dsfont} 

\usepackage{booktabs}
\usepackage{tabularx}
\usepackage{array}
\usepackage{multirow}

\usepackage{bibentry}

\newcommand{\mn}{{\mu\nu}}

\newcommand{\klambda}{{\mathbf{k},\lambda}}
\newcommand{\klambdaprime}{{\mathbf{k}',\lambda'}}
\newcommand{\klambdadoubleprime}{{\mathbf{k}'',\lambda''}}
\newcommand{\klambdazero}{{\mathbf{k}_0,\lambda_0}}
\newcommand{\kzero}{{\mathbf{k_0}}}
\newcommand{\kprime}{{\mathbf{k}'}}
\newcommand{\kdoubleprime}{{\mathbf{k}''}}
\newcommand{\lambdazero}{{\lambda_0}}
\newcommand{\lambdaprime}{{\lambda'}}
\newcommand{\lambdadoubleprime}{{\lambda''}}
\newcommand{\deltathree}{\delta^{(3)}}
\newcommand{\hc}{\text{h.c.}}
\newcommand{\Hilbert}{\mathcal{H}}

\renewcommand{\order}[1]{{{(#1)}}} 
\newcommand{\onematrix}{\mathds{1}} 

\makeatletter

\newcommand{\supplementtableofcontents}{%
  \begingroup
  \setcounter{tocdepth}{2}
  \@starttoc{stoc}%
  \endgroup
}

\newcommand{\startsupplementcontents}{%
  \let\standardaddcontentsline\addcontentsline
  \renewcommand{\addcontentsline}[3]{%
    \def\currentextension{##1}%
    \def\tocextension{toc}%
    \ifx\currentextension\tocextension
      \standardaddcontentsline{stoc}{##2}{##3}%
    \else
      \standardaddcontentsline{##1}{##2}{##3}%
    \fi
  }%
}

\makeatother

\begin{document}


\title{The small effect of graviton-induced decoherence}

\author{Guri K.~Buza\;\orcidlink{0000-0003-0297-1591}}
\email[]{guri.buza@fmf.uni-lj.si}
 \affiliation{Faculty of Mathematics and Physics, University of Ljubljana, Jadranska 19, SI-1000 Ljubljana, Slovenia}

\author{Marko Toroš\;\orcidlink{0000-0001-8089-8997}}
\email[]{marko.toros@fmf.uni-lj.si}
\affiliation{Faculty of Mathematics and Physics, University of Ljubljana, Jadranska 19, SI-1000 Ljubljana, Slovenia}

\begin{abstract}
\noindent
We derive and investigate the reduced dynamics of a nonrelativistic quantum matter system interacting with quantized gravitational waves. Because gravitational radiation couples tidally to the mass quadrupole moment, we emphasize that the leading interaction is quadratic in the mechanical displacement. Rather than treating it as a complication to be removed by linearization, we retain this full quadratic dependence, which gives rise to two-phonon rather single-phonon processes typical in dipolar interactions. We clarify its relation to the effective linear coupling that often appears in optomechanical literature, but which conceals the delicate quantum signatures of the full quadratic coupling. Using the Born-Markov and secular approximations, we microscopically derive Lindblad master equations for the matter dynamics induced by graviton baths in multi-mode vacuum, coherent, number, thermal, and squeezed states. We treat the ultraviolet-divergent Lamb-shift Hamiltonian by an explicit renormalization procedure, with the mechanical frequency identified as the physical frequency. Remarkably, owing to the nature of the interaction, the matter Hilbert space decomposes into even- and odd-parity sectors for all bath states considered. For a coherent graviton bath, the decoherence is identical to that of the vacuum, while the bath one-point function produces a coherent-shift Hamiltonian that exactly reproduces the tidal interaction with a classical gravitational wave. For a vacuum graviton bath, we recover the coherence protection between the two lowest mechanical number states, while number and thermal graviton baths remove this protection and enhance decoherence, although the effect remains small for realistic parameters. By contrast, a squeezed graviton bath enlarges the coupled coherence structure, and gives rise to a novel, coherence-protected sector spanned by dressed dark states.
\end{abstract}

\maketitle

\section{Introduction}

The detection of gravitational waves was made possible by characterizing their interaction with optomechanical systems \cite{ligo_virgo_2016_observation_of_gravitational_waves_from_a_binary}. Further understanding of the interaction between gravitational radiation and nonrelativistic matter systems in the quantum regime provides a new avenue for exploring fundamental physics. In the framework used for present-day detectors, the relevant optical and mechanical degrees of freedom may be treated quantum mechanically, and their sensitivity is ultimately limited by quantum measurement noise, while the incident gravitational wave is ordinarily retained as a classical field~\cite{danilishin_khalili_2012_quantum_measurement_theory_in_gravitationalwave,arvanitaki_geraci_2013_detecting_highfrequency_gravitational_waves_with_opticallylevitated}. At the same time, the quantization of weak gravitational fields has a long history, beginning with the early analyses of Bronstein \cite{bronstein_2012_republication_of_quantum_theory_of_weak}, Fierz-Pauli \cite{fierz_pauli_1939_on_relativistic_wave_equations_for_particles} and Gupta~\cite{gupta_1952_quantization_of_einsteins_gravitational_field}, among others. Taken together, these developments motivate the question of how the reduced dynamics of a quantum detector depends on the quantum state of the incoming gravitational radiation.

The quantization of the gravitational field in the weak-field regime, and its coupling to matter, have been studied from several complementary perspectives. Within the theory of open quantum systems, the gravitational degrees of freedom have been traced out to derive master equations for quantum matter and to characterize gravitationally-induced decoherence and dissipation~\cite{blencowe_2013_effective_field_theory_approach_to_gravitationally,anastopoulos_hu_2013_a_master_equation_for_gravitational,oniga_wang_2016_quantum_gravitational_decoherence_of_lighta,toros_mazumdar_bose_2024_loss_of_coherence_and_coherencea,nandi_sahu_majhi_ea_2026_stateselective_signatures_of_quantum_and_classical,hsiang_cho_hu_2024_graviton_physics_quantum}. 
Complementary studies have investigated graviton-induced noise and the response of detectors to particular quantum states of the gravitational field~\cite{parikh_wilczek_zahariade_2021_signatures_of_the_quantization_of_gravity,das_parikh_wilczek_ea_2025_squeezed_states_in_gravity,kanno_soda_tokuda_2021_noise_and_decoherence_induced_by_gravitons}. A further line of work has explored whether individual gravitons, or more generally signatures of quantized linearized gravity, could be accessed using matter-based quantum sensors, including atomic, trapped-mechanical, and phononic systems~\cite{boughn_rothman_2006_aspects_of_graviton_detection_graviton,carney_domcke_rodd_2024_graviton_detection_and_the_quantization,tobar_berg_2025_the_gravitophononic_effect_a_quantum,tobar_manikandan_beitel_ea_2024_detecting_single_gravitons_with_quantum,sen_gangopadhyay_bhattacharyya_2024_quantum_gravity_signatures_in_gravitational}.

In this work, we develop a unified open-quantum-system treatment of a
nonrelativistic quantum harmonic oscillator interacting with quantized
gravitational waves in multi-mode vacuum, coherent, number, thermal, and
squeezed states. Starting from the universal weak-field coupling we derive the tidal interaction and retain its full
quadratic dependence on the mechanical displacement. Using the Born--Markov and secular
approximations, we trace out the graviton field and derive the corresponding
Lindblad master equations and decoherence rates. 
In addition, the derivation gives rise to an ultraviolet divergent Lamb-shift Hamiltonian, which we resolve through an explicit renormalization procedure. We construct a counterterm Hamiltonian and impose a renormalization condition that identifies the mechanical frequency appearing in the matter Hamiltonian as the physical frequency, thereby obtaining a finite renormalized matter Hamiltonian.

The resulting dynamics separates naturally into a universal structure fixed
by the quadratic interaction and a state-dependent structure determined by the
graviton correlations. As a consequence of the quadratic interaction, the matter Hilbert space decomposes into dynamically invariant
even- and odd-parity sectors for every bath state considered. For a vacuum graviton bath, we recover the recently derived two-phonon decay dynamics, including the coherence protection between the two lowest mechanical number states~\cite{toros_mazumdar_bose_2024_loss_of_coherence_and_coherencea}. For a coherent
graviton bath, the nonvanishing bath one-point function produces a
coherent-shift Hamiltonian that reproduces the classical tidal interaction,
while the connected two-point function yields the same dissipator as the
vacuum. Number and thermal graviton baths introduce two-phonon excitation,
remove the vacuum coherence protection, and couple the lowest coherence to
an infinite hierarchy of higher coherences. Although number and thermal graviton
baths enhance decoherence, the effect remains small for realistic
parameters. 
Finally, the anomalous correlations of a
squeezed graviton bath enlarge the coherence ladder into a coupled coherence
block and support a novel coherence-protected sector formed by
parity-resolved dressed dark states.

The remainder of this work is organized as follows.
In Section~\ref{sec_nonrelativistic_matter_gravitons}, we formulate the matter and graviton degrees of freedom and derive their tidal interaction. We further elaborate on the nature of this interaction in
Section~\ref{sec_nature_of_the_interaction}, where we discuss the intrinsically
quadratic nature of this coupling and its relation to effective linearized
descriptions. 
In Section~\ref{sec_reduced_dynamics_matter_system}, we derive the general
reduced dynamics using standard methods in open quantum systems under the Born-Markov and secular approximations.
With this framework at hand, in Section~\ref{sec_state_of_gravitons}, we specialize the master equation to the different graviton bath states and investigate their dynamics.
We conclude in
Section~\ref{sec_conclusions_outlook}. Detailed derivations and intermediate
calculations are provided in Sections~\ref{app_matter_hamiltonian}--\ref{sec_final_master_equations} of the \hyperref[sec_supplementary]{Supplemental Material}~\cite{supp}.

\section{Nonrelativistic matter and gravitons}\label{sec_nonrelativistic_matter_gravitons}
Because gravitational radiation couples to the mass quadrupole moment, a single isolated particle cannot interact radiatively with the gravitational field. Its motion corresponds only to free fall of the center of mass, i.e.~a geodesic. A minimal setup possessing a dynamical quadrupole moment is a pair of coupled particles whose relative separation provides the necessary coupling~\cite{eddington_1922_the_propagation_of_gravitational_waves,maggiore_2007_gravitational_waves_vol,misner_thorne_wheeler_1973_gravitation}. Indeed, a passing gravitational wave produces tidal forces that alternately stretch and squeeze the relative separation of the two particles, thereby introducing a time-dependent quadrupole moment. Denoting the relative coordinate by $x$ and its momentum by $p$, the matter Hamiltonian reads \cite[\ref{sec_going_to_center_of_mass}]{supp},
\begin{equation}\label{eq_matter_Hamiltonian}
    H^\order{m} = \frac{p^2}{2m} + \frac{1}{2}m \omega^2_m x^2,
\end{equation}
where $m$ is the reduced mass of the coupled two-particle system and $\omega_m$ denotes the mechanical angular frequency. We have omitted the center-of-mass term, whose free dynamics, importantly, decouples from the interaction~\cite[\ref{sec_decoupling_of_center_of_mass_quadrupole}]{supp}. Such a description applies, for example, to two coupled macroscopic particles, such as nanoparticles, whose interaction can be approximately linearized around equilibrium~\cite{rieser_ciampini_rudolph_ea_2022_tunable_lightinduced_dipoledipole,deplano_pontin_ranfagni_ea_2024_coulomb_coupling_between,cao_guo_tan_2025_entangling_two_levitated_particles_in_free,cosco_pedernales_plenio_2021_enhanced_force_sensitivity_and_entanglement}, effectively yielding a description of the relative coordinate trapped in a harmonic potential \cite[\ref{sec_matter_hamiltonian_generalized}]{supp}. Equivalently, the relevant mechanical degree of freedom can be thought of as implemented by a single nanoparticle confined in a harmonic trap~\cite{ashkin_dziedzic_1971_optical_levitation_by_radiation_pressure,millen_monteiro_pettit_ea_2020_optomechanics_with_levitated_particles,delic_reisenbauer_dare_ea_2020_cooling_of_a_levitated_nanoparticle}, or by a cavity mirror whose displacement is modeled as a mechanical harmonic oscillator~\cite{bose_jacobs_knight_1997_preparation_of_nonclassical_states_in_cavities,abbott_abbott_adhikari_ea_2009_observation_of_a_kilogram-scale_oscillator,aspelmeyer_kippenberg_marquardt_2014_cavity_optomechanics}. However, the coordinate $x$ then represents the separation between the mechanical object and the reference worldline singled out by the trap minimum. In this sense, the coupling remains to a relative coordinate, as required by the tidal nature of the gravitational interaction.
Upon canonical quantization, the relative mode is promoted to an operator describing quantized excitations of the mechanical mode~\cite[\ref{sec_quantization_modes}]{supp}. We introduce the bosonic creation and annihilation operators $b^\dagger$ and $b$, such that the Hamiltonian takes the form
\begin{equation}\label{eq_matter_Hamiltonian_modes}
    H^\order{m} = \hbar \omega_m b^\dagger b,
\end{equation}
where we have omitted writing the zero-point contribution. The matter Hamiltonian is thus diagonal in the oscillator number basis and simply counts the number of mechanical excitations of the relative mode. In terms of modes, the relative coordinate is written as 
\begin{equation}\label{eq_x_in_modes}
    x = x_\text{ZPF} (b+b^\dagger), \quad x_\text{ZPF} = \sqrt{{\hbar}/{2m\omega_m}},
\end{equation}
where $x_\text{ZPF}$ denotes the zero-point fluctuation (ZPF) amplitude. In particular, the latter plays a role in determining the characteristic coupling strength to fields that interact with the quantum harmonic oscillator~\cite{aspelmeyer_kippenberg_marquardt_2014_cavity_optomechanics, chan_alegre_safavinaeini_ea_2011_laser_cooling_of_a_nanomechanical,teufel_donner_li_ea_2011_sideband_cooling_micromechanical_motion_to_the_quantum,safavinaeini_chan_hill_ea_2012_measurement_of_the_quantum_zeropoint}.

We now turn our attention to the gravitational degrees of freedom~\cite[\ref{sec_from_einstein_hilbert_to_fierz_pauli}]{supp}. 
Starting from the vacuum Einstein--Hilbert action~\cite{einstein1916HamiltonPrincipleGeneral,
hilbert1915FoundationsPhysics}, we expand around flat Minkowski spacetime, $g_{\mu\nu}=\eta_{\mu\nu}+h_{\mu\nu}$, and work to quadratic order in the weak-field perturbation~\cite{carroll2014spacetimeandgeometry,wald1984generalrelativity,weinberg_1972_gravitation_and_cosmology_principles_and_applications,brizuela_martingarcia_marugan_2009_xpert_computer_algebra_for_metric}, thereby recovering the massless Fierz-Pauli action~\cite{fierz_pauli_1939_on_relativistic_wave_equations_for_particles,ortin_2015_gravity_and_strings}. 
Rather than covariantly quantizing the full tensor perturbation and imposing subsidiary conditions~\cite{gupta_1952_quantization_of_einsteins_gravitational_field, bose_mazumdar_schut_ea_2022_mechanism_for_the_quantum_natured}, we directly and noncovariantly quantize the reduced sector with the gauge redundancy already eliminated~\cite{bronstein_2012_republication_of_quantum_theory_of_weak, toros_mazumdar_bose_2024_loss_of_coherence_and_coherencea}.   In the transverse-traceless (TT) gauge, only the two physical degrees of freedom remain~\cite{eddington_1922_the_propagation_of_gravitational_waves,maggiore_2007_gravitational_waves_vol,feynman1995feynmanlecturesongravitation}, and the resulting quadratic gravitational action reads~\cite[\ref{sec_equations_of_motion_TT_gauge}]{supp},
\begin{equation}\label{eq_gravity_action}
    S^{(g)} = - \frac{c^3}{64 \pi G} \int d^4 x \,\partial_\rho h_\mn \,\partial^\rho h^\mn.
\end{equation}
The classical plane-wave solutions are then expanded in modes whose amplitudes are promoted, upon canonical quantization, to annihilation and creation operators satisfying the usual bosonic commutation relations~\cite[\ref{sec_quantization_modes_gravity}]{supp}. The perturbation may therefore be written as
\begin{equation}\label{eq_hij_in_modes}
    h_{ij}(t,\mathbf x) = \sum_{\lambda} \int d^3 k \, \mathcal{G}^{\mathbf k,\lambda}_{ij}  g_\klambda e^{-i(\omega_k t - \mathbf k \cdot \mathbf x)} + \text{h.c.},
\end{equation}
where $g_\klambda$ has the effect of annihilating a graviton of momentum $\mathbf k$ and polarization $\lambda$. The sum over polarizations $\lambda = \{+,\times\}$ accounts for the two independent TT degrees of freedom of the gravitational field, commonly referred to as the plus and cross polarizations~\cite{press_thorne_1972_gravitationalwave_astronomy}. 
The prefactor is given by $\mathcal{G}^{\mathbf k,\lambda}_{ij} \equiv \sqrt{{G \hbar}/({\pi^2 c^2 \omega_k})} \, e_{ij}^\lambda(\mathbf n)$~\cite{oniga_wang_2016_quantum_gravitational_decoherence_of_lighta}, and correctly reproduces the classical result~\cite{toros_mazumdar_bose_2024_loss_of_coherence_and_coherencea}. The dispersion relation $\omega_k = c k$ with $k \equiv |\mathbf k|$ reflects the massless nature of gravitons and $e_{ij}^\lambda(\mathbf n)$ denotes the basis polarization tensor associated with the plus and cross modes propagating in the $\mathbf n = \mathbf k /k$ direction~\cite[\ref{sec_polarization_tensor}]{supp}.
The resulting Hamiltonian (omitting the zero-point contribution~\cite{bronstein_2012_republication_of_quantum_theory_of_weak}) is that of a collection of independent harmonic modes
\begin{equation}\label{eq_graviton_hamiltonian}
    H^\order{g} = \sum_{\lambda} \int d^3 k \, \hbar \omega_k \, g_\klambda^\dagger g_\klambda.
\end{equation}
Therefore, the gravitational Hamiltonian is diagonal in the graviton number basis~\cite{oniga_wang_2016_quantum_gravitational_decoherence_of_lighta}.

The coupling between the mechanical detector mode and the gravitational field can be characterized through the universal weak-field coupling interaction~\cite{basile_buoninfante_filippo_ea_2025_lectures_in_quantum_gravity},
\begin{equation}\label{eq_interaction_action}
    S_I
    =
    \frac{1}{2c}
    \int d^4x\,
    h_{\mu\nu}T^{\mu\nu}.
\end{equation}
The energy-momentum tensor $T^{\mu\nu}$ is understood to be evaluated at zeroth order in the metric perturbation. It is also worth emphasizing that the metric perturbation appearing here is treated as a free incoming radiative field, rather than as the retarded field sourced by the matter system under consideration~\cite{weinberg_1972_gravitation_and_cosmology_principles_and_applications}.
The effective point particle of mass $m$, has the energy-momentum tensor~\cite{weinberg_1972_gravitation_and_cosmology_principles_and_applications,bose_mazumdar_schut_ea_2022_mechanism_for_the_quantum_natured,maggiore_2007_gravitational_waves_vol},
\begin{equation}\label{eq_energy_momentum_point_particle_covariant}
    T^{\mu\nu}(x)
    =
    c^2 \int d\tau\,
    \frac{
        p^\mu(\tau)p^\nu(\tau)
    }{
        \sqrt{-p_\lambda p^\lambda}
    }
    \delta^{(4)}(x-z(\tau)).
\end{equation}
The quantity $z^\mu(\tau)$ denotes the worldline of the particle, $\tau$ is its proper time, with $u^\mu = dz^\mu/d\tau$ as its four-velocity;  $({-p_\lambda p^\lambda})^{1/2} = m c$ is the invariant rest mass of the effective point particle, with $p^\mu=m u^\mu$. After the $\tau$-integration, we take the spatial part of $z^\mu(\tau)$ to be the relative displacement $\mathbf{x}(t)$~\cite[\ref{sec_energy_momentum_tensor_specialization}]{supp}. 

Substituting the point-particle energy-momentum tensor given by Eq.~\eqref{eq_energy_momentum_point_particle_covariant} into the minimal-coupling action in Eq.~\eqref{eq_interaction_action}, in the TT gauge, and taking the nonrelativistic limit gives the corresponding velocity coupling, familiar from classical literature~\cite{maggiore_2007_gravitational_waves_vol}. We then pass to Fermi local normal coordinates centered on the center-of-mass worldline, which provide the natural frame of the detector~\cite{rakhmanov_2014_ferminormal_optical_and_wavesynchronous_coordinates}; we take the reference worldline to be $\mathbf x = 0$. In this frame, a total-derivative rearrangement, performed consistently to linear order in the perturbation, yields the tidal interaction \cite{garciachung_carney_mertens_ea_2022_what_do_gravitational,tobar_berg_2025_the_gravitophononic_effect_a_quantum,boughn_rothman_2006_aspects_of_graviton_detection_graviton}. Specialized to a mechanical mode along $x$, the the resulting tidal interaction Hamiltonian reads,
\begin{equation}\label{eq_interaction_Hamiltonian_derived}
    H_I = - \frac{m}{4} \, \ddot{h}_{xx}(t,\mathbf 0) \, x^2.
\end{equation}
This form of the interaction arises due to the fact the oscillator couples to the local tidal component of the gravitational perturbation through its quadrupole. Further details, including alternative derivations, are provided in the
Supplemental Material~\cite[\ref{sec_interaction_Hamiltonian_TT_to_FNC}--\ref{sec_interaction_Hamiltonian_general_relativistic_point_particle}]{supp}. Inserting the free mode expansions given by Eqs.~\eqref{eq_x_in_modes} and \eqref{eq_hij_in_modes} and passing to the interaction picture ($I$), gives 
\begin{subequations}\label{eq_interaction_Hamiltonian_interaction_picture}
\begin{equation}
    H_{II}(t) = H_{II}^{(m)}(t) \otimes H_{II}^{(g)}(t),
\end{equation}
where we have defined
\begin{equation}
\begin{aligned}
    H_{II}^{(m)}(t) &\equiv  \left( b \, e^{-i\omega_m t}+\hc \right)^2, \\
    H_{II}^{(g)}(t) &\equiv  \sum_\lambda \int d^3 k \,G_{\mathbf k}^\lambda \left(g_\klambda e^{- i \omega_k t}+\hc \right).
\end{aligned}
\end{equation}
\end{subequations}
Here, the first subscript $I$ labels the interaction Hamiltonian, while the second denotes the interaction picture.
The physical constant, frequency dependence, and the polarization tensor have been absorbed into the interaction coupling
\begin{equation}
    G_{\mathbf k}^\lambda \equiv \frac{m}{4}\, x_\text{ZPF}^2 \, \mathcal{G}_{xx}^{\mathbf k, \lambda} \,\omega_k^2 = \sqrt{\frac{G \hbar^3 \omega_k^3}{64\pi^2 c^2 \omega_m^2}}\,e_{xx}^\lambda(\mathbf n).
\end{equation}
In summary, the total Hamiltonian of the coupled system, $\Hilbert = \Hilbert^\order{m} \otimes \Hilbert^\order{g}$, is therefore $ H = H^\order{m} \otimes 
    \onematrix^\order{g} + \onematrix^\order{m} \otimes H^\order{g} + H_I$.

\section{Nature of the interaction}\label{sec_nature_of_the_interaction}
It is worth pausing for a moment to emphasize the nature of the tidal interaction in
Eq.~\eqref{eq_interaction_Hamiltonian_derived}.  A noteworthy comparison is with the electromagnetic interaction familiar from quantum optics.  There, the fundamental coupling is between the electromagnetic field and the
matter current, and in the long-wavelength limit this reduces to the
electric-dipole interaction~\cite{breuer_petruccione_2002_the_theory_of_open_quantuma}, which is linear in the charged particle's position operator~\cite{gerry_knight_2008_introductory_quantum_optics,duric_2004_advanced_astrophysics,walls_milburn_2008_quantum_optics}. For gravitational waves, instead, the corresponding universal coupling is to the energy-momentum tensor. 
Here lies the crucial difference: in the local inertial frame of a freely falling detector, unlike the electromagnetic case, the uniform part of the gravitational field can be transformed away, and the leading-order interaction is tidal. This distinction is sometimes less explicit in effective detector descriptions where the gravitational wave is written as a linear drive on a mechanical displacement, in close analogy with an electromagnetic dipole coupling \cite{tobar_manikandan_beitel_ea_2024_detecting_single_gravitons_with_quantum,tobar_berg_2025_the_gravitophononic_effect_a_quantum,danilishin_khalili_2012_quantum_measurement_theory_in_gravitationalwave,abrahao_coradeschi_micolfrassino_ea_2023_the_quantum_optics_of_gravitational,ma_blair_zhao_ea_2014_extraction_of_energy_from_gravitational,hsiang_cho_hu_2024_graviton_physics_quantum}. In the formulation used here, the tidal interaction makes explicit that the leading coupling is quadratic in the relative coordinate~\cite{boughn_rothman_2006_aspects_of_graviton_detection_graviton,toros_mazumdar_bose_2024_loss_of_coherence_and_coherencea,nandi_sahu_majhi_ea_2026_stateselective_signatures_of_quantum_and_classical} (cf.~\cite{sen_gangopadhyay_bhattacharyya_2024_quantum_gravity_signatures_in_gravitational,trenggana_zen_ariwahjoedi_2025_quantum_gravity_signatures_of_gravitons}).

It is perhaps worth noting that the linear couplings are not inconsistent with the quadratic tidal interaction. They arise by expanding the physical separation $x = L+ y$, around a nonzero classical equilibrium value $L$ with the quantum coordinate $y$ describing fluctuations around that separation, in the regime $|L|\gg\sqrt{\langle y^2\rangle}$. Equivalently, at the mode level, we could write $b = B + \delta b$, where $B$ is the complex classical amplitude and $\delta b$ describes quantum fluctuations around that mean. 
Since $x^2=L^2+2Ly+y^2$, this approximation retains the length-enhanced linear term but somewhat obscures the intrinsically quadratic coupling to the fluctuations, and hence, in open-quantum-systems language~\cite{schaller_2014_open_quantum_systems_far_from_equilibrium}, the associated two-phonon processes remain concealed. By contrast, the present description retains the full quadratic dependence, and thereby exposes the distinctive two-phonon processes which remains largely unexplored in graviton-matter interactions. In quantum-field-theoretic terms~\cite{ryder_1996_quantum_field_theory}, the minimal vertex is one-to-two, with one graviton interacting with two phonons \cite{biswas_bose_mazumdar_ea_2023_gravitational_optomechanics_photonmatter}, which distinguishes it from the photon-phonon one-to-one vertex~\cite{aspelmeyer_kippenberg_marquardt_2014_cavity_optomechanics}.

\section{Reduced dynamics of the matter system}\label{sec_reduced_dynamics_matter_system}
The total density operator in the interaction picture, $\rho_I(t)$, obeys the Liouville--von Neumann equation,
\begin{equation}\label{eq_liou_von}
    \dot{\rho_I}(t) = -\frac{i}{\hbar} \, [H_{II}(t), \rho_I(t)].
\end{equation}
Here the interaction Hamiltonian in the interaction picture is given in Eq.~\eqref{eq_interaction_Hamiltonian_interaction_picture}. Tracing out the graviton bath will give the matter state $\rho_I^\order{m}(t) \equiv \tr_g \rho_I(t)$, whose reduced dynamics we now derive using standard methods in the theory of open quantum systems~\cite{breuer_petruccione_2002_the_theory_of_open_quantuma,manzano_2020_a_short_introduction_to_the_lindblad, lidar_2020_lecture_notes_on_the_theorya,schaller_2014_open_quantum_systems_far_from_equilibrium}.

We begin by formally integrating the Liouville--von Neumann equation \eqref{eq_liou_von} from $0$ to $t$ and iterating once, by substituting the resulting expression for $\rho_I(t)$ back into the right-hand side. We further assume that the total initial density operator is factorized into a matter and a graviton bath part $\rho_I(0)=\rho_I^\order{m}(0)\otimes\rho_I^\order{g}(0)$ \cite{colla_neubrand_breuer_2022_initial_correlations_in_open_quantum}. Indeed, it is easy to imagine that the mechanical detector is prepared independently of the incoming gravitons, so that any correlations arise only from the subsequent interaction. This yields, upon tracing over the graviton bath, the following evolution for the reduced density matrix,
\begin{multline}\label{eq_born_markov_integrated}
    \dot\rho_I^\order{m}(t)
    =
    -\frac{i}{\hbar}\,
    \tr_g\bigl[H_{II}(t),\rho_I^\order{m}(0)\otimes\rho_I^\order{g}(0)\bigr] \\
    -\frac{1}{\hbar^2}\int_0^t d\tau\,
    \tr_g
        \bigl[H_{II}(t),\,[H_{II}(\tau),\,\rho_I(\tau)]\bigr].
\end{multline}
Under the factorization assumption, inserting Eq.~\eqref{eq_interaction_Hamiltonian_interaction_picture} in the first-order term, one finds it to be proportional to the bath one-point function, 
\begin{equation}\label{eq_onepoint_correlation_function}
    C^\order{g}(t) \equiv \text{tr}_g \bigl( H_{II}^\order{g}(t) \rho_I^\order{g}(0) \bigr),
\end{equation}
which we take to be vanishing without loss of generality.\footnote{\label{footnote_one_point_function}When the bath one-point function does not vanish, its effect can be isolated by decomposing the bath operator into its mean value and fluctuations, $H_{II}^{(g)}(t) = C^{(g)}(t)\,\onematrix^\order{g}+\Delta H_{II}^{(g)}(t)$. The mean-field part gives a coherent matter Hamiltonian contribution, generally time-dependent, even in the Schrödinger picture, $H_\text{CS}(t) \equiv H_I^{(m)}\,C^{(g)}(t)$, which contributes in addition to the free matter Hamiltonian. The remaining, honest interaction, $\tilde{H}_{I} \equiv H_I^{(m)}\otimes\Delta H_I^{(g)}$, has vanishing bath one-point function, so the usual Born--Markov derivation applies to this residual fluctuation interaction. Consequently, the bath two-point function entering the dissipative kernel is effectively the connected correlator, namely the ordinary two-point function with the product of one-point functions subtracted.}
We further invoke the weak-coupling limit. Using again the Liouville-von Neumann equation \eqref{eq_liou_von}, now integrated from $\tau$ to $t$, we express $\rho_I(\tau)$ in terms of $\rho_I(t)$, insert this into the remaining second-order term in Eq.~\eqref{eq_born_markov_integrated}, and neglect higher-order contributions in the coupling. Doing so effectively amounts to replacing $\rho_I(\tau)$ with~$\rho_I(t)$ under the integral, yielding,
\begin{multline}\label{eq_born_markov_integrated_2}
    \dot\rho_I^\order{m}(t)
    =
    -\frac{1}{\hbar^2}\int_0^t d\tau\,
    \tr_g
        \bigl[H_{II}(t),\\\,[H_{II}(\tau),\,\rho_I(t)]\bigr] + \mathcal{O}\left[(G_{\mathbf k}^\lambda)^3\right].
\end{multline}
This further motivates the fact that the state of the graviton bath remains essentially unchanged by the coupling to the matter system which we approximate as stationary, i.e.~$\rho_I(t)=\rho_I^\order{m}(t)\otimes\rho_I^\order{g}(0)$ (the Born approximation). Although the interaction generates matter--graviton correlations, their back-action on the bath state enters only at higher order in the weak coupling; to second order, the bath may therefore be kept fixed while these correlations still mediate decoherence and dissipation in the reduced matter dynamics.

\begin{table*}[t]
\centering
\small
\renewcommand{\arraystretch}{2.0}
\begin{tabular}{
>{\raggedright\arraybackslash}m{0.5\linewidth}
>{\raggedright\arraybackslash}m{0.23\linewidth}
>{\raggedright\arraybackslash}m{0.23\linewidth}
}
\hline\hline
Master equations
&
Decoherence rates
&
Jump operators
\\
\bottomrule
\(
\dot{\rho}^{(m)}_\mathcal{V}(t)
=
-\frac{i}{\hbar}
[
H^{(m)},\rho^{(m)}_\mathcal{V}(t)
]
+
\gamma_\mathcal{V}
\mathcal{D}[L_\mathcal{V}]
\rho^{(m)}_\mathcal{V}(t)
\)
&
\(
\gamma_\mathcal{V}
=
\frac{32}{15}t_P^2\omega_m^3
\)
&
\(
L_\mathcal{V}=b^2
\)
\\
\bottomrule

\(
\dot{\rho}^{(m)}_\mathcal{C}(t)
=
-\frac{i}{\hbar}
[
H^{(m)}+H_\text{CS}(t),
\rho^{(m)}_\mathcal{C}(t)
]
+
\gamma_\mathcal{C}
\mathcal{D}[L_\mathcal{C}]
\rho^{(m)}_\mathcal{C}(t)
\)
&
\(
\gamma_\mathcal{C}
=
\gamma_\mathcal{V}
\)
&
\(
L_\mathcal{C}=b^2
\)
\\[-0.1em]
\midrule
\(
\dot{\rho}^{(m)}_\mathcal{N}(t)
=
-\frac{i}{\hbar}
[
H^{(m)},\rho^{(m)}_\mathcal{N}(t)
]
+
\displaystyle
\sum_{k=\downarrow,\uparrow}
\gamma_{\mathcal{N},k}
\mathcal{D}[L_{\mathcal{N},k}]
\rho^{(m)}_\mathcal{N}(t)
\)
&
\(
\begin{aligned}
\gamma_{\mathcal{N},\downarrow}
&=
\gamma_\mathcal{V}
\left[
1+n(2\omega_m)
\right]
\\
\gamma_{\mathcal{N},\uparrow}
&=
\gamma_\mathcal{V} \,
n(2\omega_m)
\end{aligned}
\)
&
\(
\begin{aligned}
L_{\mathcal{N},\downarrow}
&=
b^2
\\
L_{\mathcal{N},\uparrow}
&=
b^{\dagger 2}
\end{aligned}
\)
\\[0.9em]
\midrule
\( \dot{\rho}^{(m)}_\mathcal{T}(t) = -\frac{i}{\hbar} [ H^{(m)},\rho^{(m)}_\mathcal{T}(t) ] + \displaystyle \sum_{k=\downarrow,\uparrow} \gamma_{\mathcal{T},k} \mathcal{D}[L_{\mathcal{T},k}] \rho^{(m)}_\mathcal{T}(t) \) & \( \begin{aligned} \gamma_{\mathcal{T},\downarrow} &= \gamma_\mathcal{V} \left[ 1+n_\beta(2\omega_m) \right] \\ \gamma_{\mathcal{T},\uparrow} &= \gamma_\mathcal{V} \, n_\beta(2\omega_m) \end{aligned} \) & \( \begin{aligned} L_{\mathcal{T},\downarrow} &= b^2 \\ L_{\mathcal{T},\uparrow} &= b^{\dagger 2} \end{aligned} \) 
\\[1em]
\midrule

\(
\dot{\rho}^{(m)}_\mathcal{S}(t)
=
-\frac{i}{\hbar}
[
H^{(m)},\rho^{(m)}_\mathcal{S}(t)
]
+
\gamma_\mathcal{S}
\mathcal{D}[L_\mathcal{S}]
\rho^{(m)}_\mathcal{S}(t)
\)
&
\(
\gamma_\mathcal{S}
=
\gamma_\mathcal{V}
\)
&
\(
\begin{aligned}
L_\mathcal{S}
=&{}
\cosh{r(2\omega_m)}\,b^2
\\
-&
e^{i\phi(2\omega_m)}\sinh r(2\omega_m)\,
b^{\dagger 2}
\end{aligned}
\)
\\[1em]

\hline\hline
\end{tabular}
\caption{
Master equations governing the reduced dynamics of the matter system for multi-mode vacuum ($\mathcal{V}$), coherent ($\mathcal{C}$), number ($\mathcal{N}$), thermal ($\mathcal{T}$), and squeezed ($\mathcal{S}$) graviton bath states \cite[\ref{eq_expectation_values_dissipators}]{supp}, written in the
general Lindblad form given in Eq.~\eqref{eq_master_equation_general_diagonal_lindblad}, after the renormalization procedure \cite[\ref{sec_lamb_shift_Hamiltonian_renormalized}]{supp}. The matter Hamiltonian $H^{(m)}$ is given in Eq.~\eqref{eq_matter_Hamiltonian_modes}. The Planck time is $t_P=(G\hbar/c^5)^{1/2} \approx 5.39 \times 10^{-44} \text{ s}$, and $\omega_m$ denotes the mechanical angular frequency. For the coherent bath, $H_\text{CS}(t)$ is the time-dependent coherent Hamiltonian drive given in Eq.~ \eqref{eq_coherent_shift_definition_main} (cf. Eq.~\eqref{eq_coherent_shift_Hamiltonian_multimode} Eq.~\eqref{eq_coherent_shift_Hamiltonian_classical}). For the number state, $n(2\omega_m)$ denotes the dimensionless Fock occupation number of the resonant graviton modes. For the thermal graviton bath state, $n_\beta(2\omega_m) = 1/(e^{2\beta \hbar \omega_m/k_B}-1)$ denotes the Bose-Einstein thermal mean occupation number, with inverse temperature $\beta = 1/(k_B T)$ and Boltzmann's constant $k_B$, evaluated at the same resonant frequency. For the squeezed state, $r(2\omega_m)$ and $\phi(2\omega_m)$ denote the dimensionless squeezing amplitude and the squeezing phase, respectively, at the same resonant frequency.
}
\label{tab_master-equation-summary}
\end{table*}

We then change variables from absolute times to the time difference, $\tau \to t- \tau$, and extend the upper integration limit $t\to\infty$, assuming sufficiently fast decay of the bath correlation functions (the Markov approximation).  Finally, having inserted the interaction Hamiltonian~\eqref{eq_interaction_Hamiltonian_interaction_picture}
into the second-order term, we obtain the master equation
\begin{multline}\label{eq_master_equation_D}
    \dot\rho_I^\order{m}(t)
    =
    -\frac{1}{\hbar^2}
    \int_0^\infty d\tau \\
    \Bigl(
        D^\order{g}(t,t- \tau)\, 
        \Big[ H_{II}^\order{m}(t), H_{II}^\order{m}(t-\tau) \, \rho_I^\order{m}(t) \Big]
        \\ - D^{\order{g}}(t, t-\tau)^*\,
        \left[ H_{II}^\order{m}(t),\rho_I^\order{m}(t) \,H_{II}^\order{m}(t-\tau) \right]
    \Bigr).
\end{multline}
Here, we have defined the kernel representing the graviton bath two-point correlation function,
\begin{equation}\label{eq_two_point_bath_conn_or_not_conn}
    D^\order{g}(t,t-\tau)
    \equiv
    \tr_g \, \Bigl[ H_{II}^\order{g}(t)\,H_{II}^\order{g}(t-\tau)\,\rho_I^\order{g}(0) \Bigr].
\end{equation}
Thus the specific form of the resulting master equation depends on the state of the graviton bath through its correlation functions~\cite{wangsness_bloch_1953_the_dynamical_theory_of_nuclear, redfield_1957_on_the_theory_of_relaxation}; see \cite[\ref{eq_expectation_values_dissipators}]{supp}.

Nevertheless, after applying the corresponding secular approximation \cite[\ref{sec_secular_approximation}]{supp}, and transforming back to the Schr\"odinger picture, in all our cases the reduced dynamics will obey the general Lindblad master equation (Table~\ref{tab_master-equation-summary})~\cite{lindblad_1976_on_the_generators_of_quantuma,gorini_kossakowski_sudarshan_1976_completely_positive_dynamical_semigroups_of_n-levela},
\begin{multline}\label{eq_master_equation_general_diagonal_lindblad}
    \dot{\rho}^\order{m}(t) =-\frac{i}{\hbar} \Big[H^\order{m} +H_\text{CS}(t)  \\ +H_\text{LS}+H_\text{CT}  , \rho^\order{m}(t)\Big]
+ \sum_k \gamma_k \,\mathcal{D}[L_k]\,\rho^\order{m}(t).
\end{multline}
Let us discuss each of its individual terms. The matter Hamiltonian $H^{(m)}$ is given in Eq.~\eqref{eq_matter_Hamiltonian}. The term $H_\text{CS}(t)$ denotes the time-dependent coherent-shift Hamiltonian which appears when the one-point expectation value of the graviton bath is nonvanishing (in which case $D$ is to be calculated as a connected correlator; see Footnote \ref{footnote_one_point_function}),
\begin{equation}\label{eq_coherent_shift_definition_main}
    H_\text{CS}(t)=H_I^{(m)} \,C^{(g)}(t).
\end{equation}
Here, $H_I^{(m)}$ is given in Eq. \eqref{eq_interaction_Hamiltonian_interaction_picture}, in the Schrödinger picture, and the bath one-point correlation function $C^{(g)}(t)$ is given in Eq.~\eqref{eq_onepoint_correlation_function}. $H_\text{CS}(t)$ can be viewed as the gravitational analogue of the coherent drive generated by a nonzero mean input field in quantum-optical open systems~\cite{gardiner_collett_1985_input_and_output_in_damped,carmichael_1999_statistical_methods_in_quantum_optics, gardiner_zoller_2010_quantum_noise_a_handbook_of_markovian}. On the other hand, the term $H_\text{LS}$ denotes the Lamb-shift Hamiltonian coming from off-shell contributions, which in the present model is ultraviolet (UV) divergent, as is often the case for Lamb-shift terms obtained from principal-value integrals~\cite{breuer_petruccione_2002_the_theory_of_open_quantuma}. A common treatment is to introduce a phenomenological UV cutoff, as in Bethe's original treatment and related open-system approaches~ \cite{breuer_petruccione_2002_the_theory_of_open_quantuma,malcolm_sharmila_wang_ea_2024_detecting_quantum_vacuum_fluctuations_of_the_electromagnetic,vats_john_busch_2002_theory_of_fluorescence_in_photonic,liu_song_wang_ea_2010_observation_of_lamb_shift_and_modified}. Here, instead, we introduce a
counterterm Hamiltonian $H_\text{CT}$, which absorbs the UV-divergent
terms in $H_\text{LS}$, up to the relevant order in the perturbation~\cite{correa_glatthard_2025_potential_renormalisation_lamb_shift_and_meanforce,ryder_1996_quantum_field_theory,shankar_2017_quantum_field_theory_and_condensed}. We perform the renormalization procedure such
that the frequency $\omega_m$ appearing in \(H^\order{m}\) is the
observed physical frequency. In particular, we regulate the Lamb-shift Hamiltonian with an intermediate UV cutoff $\Lambda$ and introduce a cutoff-dependent counterterm Hamiltonian $H_\text{CT}(\Lambda)$, chosen such that
\begin{equation}\label{eq_renormalization_condition_main}
    \lim_{\Lambda\to\infty}
    \left(
        H^\order{m}
        +H_\text{LS}(\Lambda)
        +H_\text{CT}(\Lambda)
    \right)
    =
    H^\order{m}.
\end{equation}
This defines our renormalization condition. Full details of the procedure are
provided in the Supplemental Material~\cite[\ref{sec_lamb_shift_Hamiltonian_renormalized}]{supp}. The coefficients $\gamma_k$ are the rates associated with the dissipative channels labeled by $k$. The corresponding dissipators,
\begin{equation}\label{eq_dissipator_definition}
    \mathcal{D}[L_k]\,\rho^\order{m}(t) \equiv L_k \, \rho^\order{m}(t) \,L_k^\dagger - \frac{1}{2} \big\{ L_k^\dagger \,L_k ,\rho^\order{m}(t)\big\},
\end{equation}
written with a diagonalized Kossakowski matrix, have jump operators $L_k$ that are quadratic in the matter phonon operators (Table~\ref{tab_master-equation-summary}).

\section{State of gravitons}\label{sec_state_of_gravitons}

Of course, in principle, we are unaware of the quantum state of the gravitational field relevant to a given experiment. The choice of graviton bath is therefore not just a technical assumption, but a way of parameterizing both the physical character of the gravitational source and our ignorance of its occupation, phase, direction, and correlations. We model this uncertainty by  considering the graviton bath to be in a vacuum, coherent~\cite{glauber_1963_coherent_and_incoherent_states_of_the_radiation}, number~\cite{fock_1932_konfigurationsraum_und_zweite,neumann_1939_on_infinite_direct_products}, thermal~\cite{neumann_1927_thermodynamik_quantenmechanischer_gesamtheiten}, and squeezed state~\cite{stoler_1970_equivalence_classes_of_minimum_uncertainty, walls_1983_squeezed_states_of_light}, and identify their unique dynamical signatures (cf.~\cite{parikh_wilczek_zahariade_2021_signatures_of_the_quantization_of_gravity,das_parikh_wilczek_ea_2025_squeezed_states_in_gravity,hertzberg_litterer_2023_bound_on_quantum_fluctuations_in_gravitational,moreira_celeri_2024_decoherence_of_a_composite_particle}). Precise definitions of the graviton bath states are provided in the Supplemental Material~\cite[\ref{eq_expectation_values_dissipators}]{supp}.

A natural starting point is to take the graviton bath to be in the vacuum state. This isolates the irreducible quantum contribution of the gravitational field and is analogous to the electromagnetic vacuum in quantum optics, which already gives rise to spontaneous emission, vacuum noise, and radiative damping~\cite{breuer_petruccione_2002_the_theory_of_open_quantuma}. Indeed, vacuum fluctuations alone are sufficient to generate dissipation and decoherence~\cite{anastopoulos_hu_2013_a_master_equation_for_gravitational}. As studied in Ref.~\cite{toros_mazumdar_bose_2024_loss_of_coherence_and_coherencea}, however, the same quadratic structure that gives rise to dissipation also profoundly implies a nontrivial coherence-protection mechanism. In the matter number-state basis, the vacuum jump operator is $L_\mathcal{V}=b^2$ (Table~\ref{tab_master-equation-summary}), so the bath induces two-phonon decays. In particular, the lowest states, are dark, satisfying,
\begin{equation}
\begin{aligned}
    L_\mathcal{V} \ket{0}=0, \qquad 
    L_\mathcal{V} \ket{1}=0.
\end{aligned}
\end{equation}
Thus the full qubit matter state
\begin{equation}\label{eq_initial_state_superposition_general}
    \rho_{\vartheta,\varphi}^{(m)}
\equiv
|\psi_{\vartheta,\varphi}^{(m)}\rangle\langle\psi_{\vartheta,\varphi}^{(m)}|,
\end{equation}
with $|\psi_{\vartheta,\varphi}^{(m)}\rangle
=
\cos({\vartheta}/{2})\,|0\rangle
+
e^{i\varphi}\sin({\vartheta}/{2})\,|1\rangle$, is coherence-protected for arbitrary $\vartheta$ and $\varphi$.

Beyond this vacuum coherence protection, the quadratic nature of the interaction also implies a more general parity selection rule: the matter Hilbert space splits into dynamically invariant even~($+$) and odd~($-$) parity ladders, 
\begin{equation}
    \Hilbert^{(m)} =\Hilbert^{(m)}_+ \oplus \Hilbert^{(m)}_-. 
\end{equation}
Remarkably, this parity decomposition is not special to the vacuum bath, but persists for all graviton bath states considered here. 

\begin{figure*}[!t]
    \centering
    \includegraphics[width=0.38\linewidth]{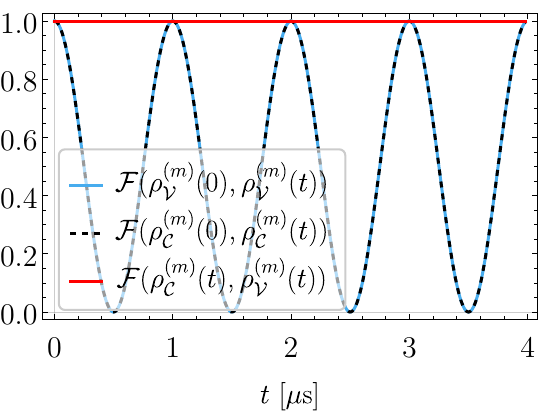}
    \hspace{1.5cm}
    \includegraphics[width=0.38\linewidth]{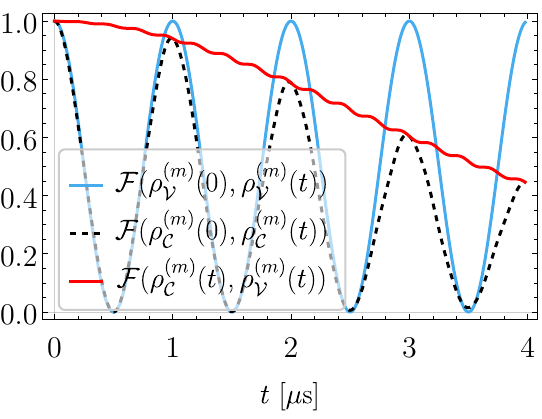}
    \caption{
    Survival of the time-evolved initial coherence-protected matter state $\rho^{(m)}_{\mathcal{V,\mathcal{C}}}(0)=\rho^{(m)}_{\vartheta,\varphi}$ given in Eq.~\eqref{eq_initial_state_superposition_general}, with $\vartheta=\pi/2$ and $\varphi = 0$, in the case of vacuum ($\mathcal{V}$) and coherent ($\mathcal{C}$) graviton bath states, quantified by fidelity $\mathcal{F}$. The dynamics are obtained by solving the Lindblad master equation in Eq.~\eqref{eq_master_equation_general_diagonal_lindblad} (see Table~\ref{tab_master-equation-summary}). We take $\omega_m=2\pi \times 1\text{ MHz}$, with $\omega_{k_0}=2\omega_m$, $\theta=0$, and plus polarization $e_{xx}^{+}=1$, corresponding to gravitons propagating along the $z$-direction. 
    \textbf{Left:} Coherent graviton state matched to a realistic gravitational-wave strain $h_+=10^{-21}$, giving $|\alpha| = 1.99\times10^{13}$. In this case, the coherent and vacuum evolutions are indistinguishable. 
    \textbf{Right:} Exaggerated coherent amplitude, corresponding to $|\alpha|=1.59\times10^{33}$, or equivalently an effective gravitational strain $h_+ =0.08$, chosen such that the fidelity decreases to less than half its initial value over the relevant timescale. In this strong-field case the coherent-shift Hamiltonian contributes a substantial deviation.}
    \label{fig_fidelity_comparison}
\end{figure*}

The coherent graviton bath provides an important benchmark because it does not amplify the intrinsic graviton-induced decoherence. As is clear from the master equation in this case (Table~\ref{tab_master-equation-summary}), the dissipative part is exactly the same as in the vacuum case: the coherent displacement only contributes an additional time-dependent coherent drive $H_\text{CS}(t)$ \cite[\ref{sec_coherent_bath_consequences}]{supp}. The latter is nevertheless of particular physical importance. Upon identifying the corresponding classical (CL) gravitational wave with the one-point function of the metric perturbation in the coherent state $\alpha$, 
\begin{equation}
    h_{ij}^\text{CL}(t,\mathbf x) \equiv \langle h_{ij}(t,\mathbf x)\rangle_\alpha,
\end{equation}
the coherent-shift Hamiltonian is precisely the tidal Hamiltonian \eqref{eq_interaction_Hamiltonian_derived} for a classical gravitational wave,
\begin{equation}\label{eq_coherent_shift_Hamiltonian_classical}
    H_\text{CS}(t)=- \frac{m}{4}\,\ddot h_{xx}^{\rm CL}(t,\mathbf 0)\,x^2 .
\end{equation}
To see this more clearly, let us restrict ourselves to the deterministic Hamiltonian part of the reduced dynamics. Then the Hamilton’s equations generated by $H^{(m)}+H_{\mathrm{CS}}(t)$, given in Eqs.~\eqref{eq_matter_Hamiltonian} and \eqref{eq_coherent_shift_Hamiltonian_classical}, respectively, give
\begin{equation}
\ddot x+\omega_m^2x
=
\frac{1}{2}\ddot h_{xx}^{\mathrm{CL}}(t,\mathbf 0)\,x.
\end{equation}
This is exactly the geodesic-deviation equation supplemented by the mechanical restoring force. In the freely falling limit, $\omega_m\rightarrow0$, it reduces to the standard equation describing the relative response of freely falling test masses to a classical gravitational wave~\cite{parikh_wilczek_zahariade_2021_signatures_of_the_quantization_of_gravity}. In contrast to stochastic formulations of quantum geodesic deviation, the graviton fluctuations in the present open-quantum-system treatment are encoded in the Lindblad dissipator rather than in an explicit stochastic force.

Furthermore, the identification allows us to trade the coherent-state eigenvalue for the observable gravitational-wave strain. To make this relation explicit, let us restrict for simplicity to a monochromatic plus-polarized wave propagating in the $z$-direction. Then,
\begin{equation}
    h_{xx}^\text{CL}(t,\mathbf 0) = h_+ \cos(\omega_{k_0}t-\theta),
\end{equation}
where $h_+$ is the gravitational-wave strain amplitude. With the corresponding coherent state sharply peaked around the plus-polarized mode $(\mathbf k_0,\lambda_0=+)$, we thus obtain \cite{supp}, 
\begin{equation}
    h_+ = 2\mathcal{G}_{xx}^{\mathbf{k}_0,+}|\alpha|
= \frac{16\omega_m}{\hbar\omega_{k_0}^2}\,G_{k_0}^+|\alpha| .
\end{equation}
Although the coherent bath does not enhance the decoherence rate, the additional drive moves the state $\rho_{\vartheta,\varphi}^{(m)}$, given in Eq.~\eqref{eq_initial_state_superposition_general}, out of the decoherence-free subspace. That is, the coherent-shift Hamiltonian drives the system into states that are no longer protected from the vacuum dissipator. For realistic strains, however, such as $h_+ = 10^{-21}$~\cite{ligo_virgo_2016_observation_of_gravitational_waves_from_a_binary}, this effect is completely negligible on the relevant timescales \cite{delic_reisenbauer_dare_ea_2020_cooling_of_a_levitated_nanoparticle}, as quantified by the fidelity~\cite{jozsa_1994_fidelity_for_mixed_quantum_states}, shown in Fig.~\ref{fig_fidelity_comparison}.

\begin{figure*}[!t]
    \centering
    \includegraphics[width=0.38\linewidth]{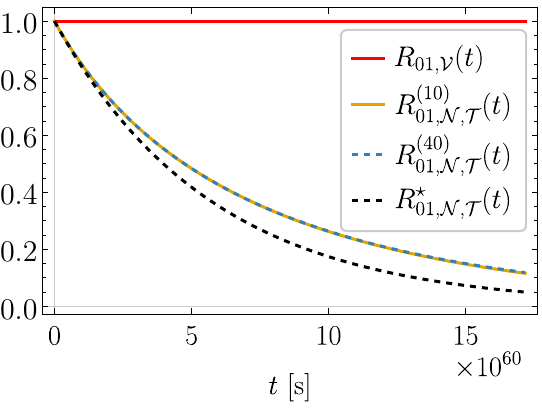}
    \hspace{1.5cm}
    \includegraphics[width=0.38\linewidth]{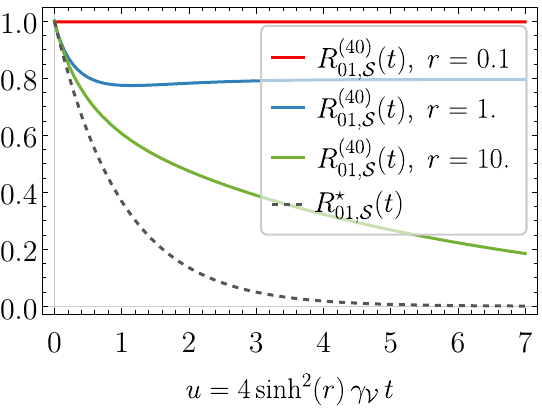}
    \caption{
    Survival of the bare encoded matter coherence $\rho_{01}$ in number and thermal ($\mathcal{N},\mathcal{T}$) and squeezed ($\mathcal{S}$) graviton baths, quantified by the coherence ratio $R_{01}$ defined in Eq.~\eqref{eq_coherence_ratio_p_q} (cf. Eqs.~\eqref{eq_coherence_ratio_N_star}-\eqref{eq_coherence_ratio_S_truncated}). The dynamics are obtained by solving the respective Lindblad master equation in Eq.~\eqref{eq_master_equation_general_diagonal_lindblad} (see Table~\ref{tab_master-equation-summary}), in the matter number-state basis. We take $\omega_m=2\pi\times1\,\text{MHz}$, so that the resonant gravitons
have $\omega_k=2\omega_m$. For the number and thermal baths, we use
$n=2.8\times10^4$, corresponding in the thermal case to
$T=2.725\,\text{K}$.
    \textbf{Left:} The truncated-ladder results for $M=10$ and $M=40$ are indistinguishable, demonstrating numerical convergence. The short-time exponential reproduces the initial slope and departs from the full dynamics once higher coherences become populated. For comparison, it is shown that vacuum ($\mathcal{V}$) coherence remains exactly protected.
    \textbf{Right:} The truncated-block results for $M=40$, in the interaction picture, for different values of the squeezing parameter $r$. Although the curves initially follow the short-time exponential, higher-coherence feedback produces qualitatively different dynamics at later times. In particular, the bare coherence undershoots its asymptotic value and then increases toward the contribution associated with the dressed dark sector.
    }
    \label{fig_coherence_ratio_comparison}
\end{figure*}

The number and thermal graviton baths produce a richer dissipative structure. They introduce not only the two-phonon decay channel, but also the two-phonon excitation channel  (Table~\ref{tab_master-equation-summary}). Since these two cases have the same formal structure we can discuss them simultaneously via the single occupation number $n$. To diagnose the loss of coherence, in the matter number-state basis, we define $\rho_{pq}(t) \equiv \langle p|\rho^{(m)}(t)|q\rangle $, where the coherences are given by its off-diagonal elements. The governing dynamics then implies an infinite coherence ladder. In particular, the encoded coherence $\rho_{01}$ is not closed under time evolution, but it is coupled to higher coherences, $\rho_{23},\rho_{45}$, which in turn are coupled to still higher coherences, and so on \cite[\ref{sec_number_thermal_consequences}]{supp}. This parity-ladder structure is already present in the vacuum case, but now the excitation channel makes the ladder dynamically relevant. Since the free Hamiltonian only rotates the phase of $\rho_{pq}$, we observe that the magnitude of the matrix element nicely isolates the dissipative loss of coherence and provides a natural measure for the latter. We therefore define the coherence ratio
\begin{equation}\label{eq_coherence_ratio_p_q}
    R_{pq}(t) \equiv \frac{|\rho_{pq}(t)|}{|\rho_{pq}(0)|}.
\end{equation}
The normalization, in turn, removes the dependence on the initial coherence magnitude, so that the coherence ratio measures the coherence magnitude at time $t$, relative to the initial coherence magnitude.
For the particular coherence $\rho_{01}$, let us take the same initial condition as in the vacuum case, implying that all the higher coherences vanish, $\rho_{23}(0)=\cdots=0$. Then, expanding the finite-occupation dynamics around $t = 0$, we find that the first departure from the vacuum-protected result is given by the short-time exponential (denoted by a superscript $\star$)~\cite[\ref{sec_number_thermal_consequences}]{supp},
\begin{equation}\label{eq_coherence_ratio_N_star}
    R_{01,\mathcal{N},\mathcal{T}}^{\star}(t)
=
e^{-4 \gamma_\mathcal{V} n \, t}.
\end{equation}
This expression smoothly reproduces the vacuum case. Indeed, since there is no excitation channel, i.e.~$n=0$, we obtain $R_{01,\mathcal{V}}(t)=1$. Although the vacuum result is exact, the short-time exponential only captures the initial slope of the coherence loss for number and thermal baths. The initially vanishing coherences become populated during the evolution and feed back into $\rho_{01}$. Therefore the full dynamics requires solving the infinite coherence ladder. In practice, we can solve this hierarchy numerically by truncating the ladder to some finite $M$. That is, we retain the coherences $\rho_{2j,2j+1}$ for $j = 0,\ldots,M-1$, and denote the resulting finite-ladder coherence ratio by,
\begin{equation}\label{eq_coherence_ratio_N_truncated}
    R_{01,\mathcal N,\mathcal T}^{(M)}(t)
\equiv
\frac{|\rho_{01,\mathcal N,\mathcal T}^{(M)}(t)|}{|\rho_{01,\mathcal N,\mathcal T}(0)|}.
\end{equation}
Number and thermal baths therefore have two related effects. First, they remove the exact vacuum protection of the lowest coherence $\rho_{01}$. Second, they enhance the dissipative dynamics of the higher coherences through the rates $\gamma_\downarrow$ and $\gamma_\uparrow$. Thus the occupation number $n$ acts as an enhancement of the vacuum rate $\gamma_\mathcal{V}$. Nevertheless, for the benchmark parameters considered here, the effect remains extremely small. For an oscillator frequency $\omega_m = 2\pi \times 1 \text{ MHz}$, the relevant occupation number is the occupation of resonant gravitons. To obtain a sense of the relevant orders of magnitude, let us assign the graviton bath the CMB temperature, $T = 2.725 \text{ K}$, gives $n \approx 2.8 \times 10^4$. Even with this enhancement, the corresponding decoherence timescale is of order $10^{60} \text{ s}$, as shown in Fig.~\ref{fig_coherence_ratio_comparison} (left). By contrast, obtaining an order-one deviation on a $\text{$\mu$s}$ timescale requires $n \approx 10^{71}$ resonant gravitons, corresponding, if interpreted thermally, to an effective bath temperature $T\approx 10^{67}\text{ K}$.

The squeezed graviton bath contains the same two-phonon decay and excitation channels, but, importantly, they combine into a single jump operator that is a coherent superposition of two-phonon annihilation and creation processes (Table~\ref{tab_master-equation-summary}). These correlations have two qualitatively distinct consequences. First, they enlarge the coherence structure beyond the nearest-neighbor ladder encountered for number and thermal baths. Second, as we elaborate below, although they similarly remove the vacuum protection of the bare coherence, they introduce a new form of dissipative protection associated with a dressed dark sector.

We begin with the evolution of the bare coherence $\rho_{01}$. For the same initially encoded qubit, all higher coherences vanish at $t=0$, and the departure from the vacuum-protected result is similarly given by a short-time exponential (denoted by a superscript $\star$) \cite[\ref{sec_squeezed_bath_consequences}]{supp},
\begin{equation}\label{eq_coherence_ratio_S_star}
R_{01,\mathcal S}^{\star}(t)
=
e^{
-4\gamma_{\mathcal V}\sinh^2r\,t}.
\end{equation}
As in the number and thermal cases, this expression captures only the initial slope. In the squeezed case, however, it can also be qualitatively misleading, since a simple monotonic decay does not necessarily persist at longer times. The full dynamics is richer in the way coherences are coupled. In addition to being coupled to $\rho_{23},\rho_{45}$, the anomalous correlations generate off-ladder elements such as $\rho_{05}$ and $\rho_{41}$. These coherences are populated during the evolution and feed back into $\rho_{01}$, which, as shown below, can decay initially and later increase. The evolution of $\rho_{01}$ in the squeezed case belongs to an infinite coherence block rather than a single ladder. In practice, we solve this hierarchy numerically by truncating the connected coherence sector at some finite $M$. We retain the coherences \(\rho_{2a,2b+1}\), with $a,b=0,\ldots,M-1$, and discard couplings to matrix elements outside this truncated sector. Similarly to before, we denote the resulting finite-sector coherence ratio by
\begin{equation}\label{eq_coherence_ratio_S_truncated}
    R_{01,\mathcal S}^{(M)}(t)
    \equiv
    \frac{
    \left|\rho_{01,\mathcal S}^{(M)}(t)\right|
    }{
    \left|\rho_{01,\mathcal S}(0)\right|
    }.
\end{equation}
To isolate the dissipative effects, we work in the interaction picture and solve the truncated coherence block with $M = 40$, taking $\phi = 0$, and $r = 0.1,1,10$, representing weak, intermediate, and strong squeezing, respectively. As shown in Fig.~\ref{fig_coherence_ratio_comparison} (right), at short times, the curves follow the exponential approximation, as required by their common initial slope. At later times, however, the approximation fails qualitatively. In particular, the coherence may initially decay below its asymptotic value and subsequently increase and saturate. It is perhaps important to note that this recoherence is a result of redistribution of coherence among the coupled matrix elements.

This nonmonotonic behavior, consisting of an initial decay followed by recoherence and eventual saturation, may be viewed as pointing to an underlying protected structure of the squeezed dissipator. Indeed, the squeezed jump operator admits two linearly independent dark states, one in each parity sector, satisfying \cite[\ref{sec_squeezed_bath_consequences}]{supp}
\begin{equation}
\begin{aligned}
    L_{\mathcal S}\ket{D_+}=0, \qquad
    L_{\mathcal S}\ket{D_-}=0.
\end{aligned}
\end{equation} 
Thus the full dressed qubit matter state,
\begin{equation}\label{eq_dressed_dark_qubit}
    \rho^{(m)}_{D;\vartheta,\varphi}
=
\ket{\psi^{(m)}_{D;\vartheta,\varphi}}
\bra{\psi^{(m)}_{D;\vartheta,\varphi}},
\end{equation}
with $\ket{\psi^{(m)}_{D;\vartheta,\varphi}}
=
\cos(\vartheta/2)\ket{D_+}
+
e^{i\varphi}\sin(\vartheta/2)\ket{D_-}$, is coherence-protected. The two dark states are dressed versions of the vacuum states $\ket{0}$ and $\ket{1}$, to which they continuously reduce as $r \to 0$. Their leading number-state components are
\begin{align}
    \ket{D_+}
&=
\mathcal N_+
\left(
\ket{0}
+
\frac{e^{i\phi}\tanh r}{\sqrt6}\ket{4}
+\cdots
\right), \\
\ket{D_-}
&=
\mathcal N_-
\left(
\ket{1}
+
\sqrt{\frac{3}{10}}\,e^{i\phi}\tanh r\ket{5}
+\cdots
\right),
\end{align}
where $\mathcal{N}_\pm$ are the normalization factors \cite[\ref{sec_squeezed_bath_consequences}]{supp}. The coherence between these two dressed states is exactly preserved by the squeezed dissipator, namely $R_{D_+D_-,\mathcal{S}}(t) = 1$. It is useful to compare this dressed protection with the vacuum protection of the bare qubit. Although  both are exactly preserved by their respective dissipators, the $01$ matrix element of the dressed dark qubit, given in Eq.~\eqref{eq_dressed_dark_qubit} is reduced relative to that of the bare qubit, given in Eq.~\eqref{eq_initial_state_superposition_general}, according to~\cite[\ref{sec_squeezed_bath_consequences}]{supp}
\begin{equation}
    {|\langle0|\rho^{(m)}_{D;\vartheta,\varphi}|1\rangle|
}/{
|\langle0|\rho^{(m)}_{\vartheta,\varphi}|1\rangle|
}
=
\mathcal N_+\mathcal N_-.
\end{equation}
This quantity approaches unity as $r\to 0$, and decreases with squeezing. Squeezing therefore transfers the vacuum protection from the bare qubit to a dressed qubit, while leaving only a reduced protected contribution in the bare basis.

It is important to emphasize that this protection is exact for the squeezed dissipator alone. The free oscillator Hamiltonian does not leave the dressed dark sector invariant, since it rotates the different number-state components of the dressed states at different rates. That is, the protected dressed coherence is not automatically available in the lab frame; it requires a phase reference to the squeezed gravitational correlations. Establishing such a phase reference would therefore seem to present an interesting challenge.

\section{Conclusion and outlook}\label{sec_conclusions_outlook}

In this work, we studied graviton-induced decoherence of a nonrelativistic quantum matter system within the framework of linearized quantum gravity and open quantum systems. The matter degree of freedom was modeled as a quantum harmonic oscillator, while the transverse-traceless gravitational field was represented as a multi-mode graviton bath. Starting from the universal weak-field coupling to the energy-momentum tensor, we derived the tidal interaction between the oscillator and the gravitational radiation. Because this interaction couples to the mass quadrupole moment, it is quadratic in the mechanical displacement and generates two-phonon rather than dipole-like single-phonon processes. We retain this quadratic dependence, which remains largely unexplored in the literature, as a consequence of which, the matter Hilbert space decomposes into dynamically invariant even- and odd-parity sectors.

Using the Born--Markov and secular approximations, we provided a microscopic derivation of the Lindblad master equations for graviton baths in multi-mode vacuum, coherent, number, thermal, and squeezed states. The derivation gives rise to an UV-divergent Lamb-shift contribution which was absorbed into a counterterm through an explicit renormalization procedure that identifies the mechanical frequency with its observed value. For a coherent graviton bath, the connected two-point function yields the same dissipator as the vacuum, while the nonvanishing bath one-point function generates a coherent-shift Hamiltonian that reproduces the tidal interaction with a classical gravitational wave. We recovered the vacuum protection of the lowest mechanical number-state sector. Number and thermal graviton baths introduce two-phonon excitation, remove this protection, and couple the lowest coherence to an infinite hierarchy of higher coherences. Although number and thermal graviton baths enhance the decoherence rate, the effect remains small for realistic mechanical frequencies and gravitational environments. For a squeezed graviton bath, the anomalous correlations enlarge the coherence ladder into a coupled coherence block. At the same time, we find that the squeezed dissipator supports a new coherence-protected sector formed by two dressed dark states, one in each parity sector, whose mutual coherence is exactly preserved by the dissipative dynamics.

Several extensions remain open.
The rich structure of the resulting master equations warrants further investigation, particularly the coupled coherence ladders and blocks, their long-time dynamics, and the role of parity in organizing the dissipative evolution. More generally, connecting the effective harmonic-oscillator description used here to a fully relativistic microscopic theory of matter would further clarify the origin and range of validity of the uncovered effects. 
Extending the present treatment to detector models with more realistic geometries and mechanical mode structures may clarify which of these state-dependent signatures are most robust and experimentally relevant. The dressed dark sector of the squeezed bath also raises the interesting question of how phase-sensitive gravitational correlations could be referenced or accessed experimentally, since the free oscillator Hamiltonian does not by itself preserve this sector in the laboratory frame. 

The obtained results hence show that different quantum states of gravitational radiation leave qualitatively distinct signatures in the reduced dynamics of quantum matter. The challenge is to devise an experiment that would become sensitive directly, or perhaps in an indirect way, to these effects.


\medskip
\begin{acknowledgments}
\secpar{Acknowledgments}
G.K.B.~acknowledges support from the research programme N1-0392 and P1-0416 of the Slovenian Research and Innovation Agency (ARIS). M.T.~acknowledges funding from the Slovenian Research and Innovation
Agency (ARIS) under contracts N1-0392, P1-0416, SN-ZRD/22-27/0510
(RSUL Toro\v{s}). 
\end{acknowledgments}

\bibliography{bib}

\onecolumngrid
\clearpage
\section*{Supplemental Material}\label{sec_supplementary}

\begingroup
\hypersetup{linkcolor=black}
\supplementtableofcontents
\endgroup
\startsupplementcontents


\setcounter{section}{0}
\renewcommand{\thesection}{\Alph{section}}
\renewcommand{\thesubsection}{\arabic{subsection}}

\setcounter{equation}{0}
\renewcommand{\theequation}{S\arabic{equation}}

\section{Nonrelativistic matter and gravitons: Matter Hamiltonian}\label{app_matter_hamiltonian}

\subsection{Going to the center-of-mass coordinates}\label{sec_going_to_center_of_mass}

The Hamiltonian for two coupled harmonic oscillators is given by
\begin{equation}\label{eq_hamiltonian_two_coupled}
H = \frac{p_1^2}{2 m_1} + \frac{p_2^2}{2 m_2}
+ \frac{k}{2}(x_1 - x_2)^2,
\end{equation}
where $k$ can be thought of as an effective spring constant. Now we can define the relative and center-of-mass coordinates
\begin{subequations}\label{eq_coordinates_com_rel}
\begin{align}
x = x_1 - x_2, \quad
p_{\mathrm{cm}} = p_1 + p_2, \quad
x_{\mathrm{cm}} = \frac{m_1 x_1 + m_2 x_2}{m_1 + m_2}, \quad
p &= \frac{m_2 p_1 - m_1 p_2}{m_1 + m_2}.
\end{align}
Solving for the original variables 
\begin{align}
x_1 = \frac{m_2 \,x}{m_1 + m_2}  + x_{\mathrm{cm}}, \quad
x_2 = -\frac{m_1\,x}{m_1 + m_2} + x_{\mathrm{cm}}, \quad
p_1 = p + \frac{m_1\,p_{\mathrm{cm}}}{m_1 + m_2} , \quad
p_2 = -p + \frac{m_2\, p_{\mathrm{cm}}}{m_1 + m_2},
\end{align}
\end{subequations}
and substituting into the Hamiltonian gives
\begin{equation}
H = \frac{1}{2}\frac{m_1+m_2}{m_1 m_2}p^2
+ \frac{1}{2}\frac{1}{m_1+m_2}p_{\mathrm{cm}}^2
+ \frac{1}{2} k x^2.
\end{equation}

Now we define total and reduced masses
\begin{align}\label{eq_mass_total_and_reduced}
M = m_1 + m_2, \qquad
m = \frac{m_1 m_2}{m_1 + m_2}.
\end{align}
in which case the Hamiltonian reads
\begin{equation}
H = \frac{p^2}{2m}
+ \frac{p_{\mathrm{cm}}^2}{2M}
+ \frac{k}{2} x^2 .
\end{equation}

Defining the harmonic angular frequency referred to in the main text,
\begin{equation}\label{eq_angular_harmonic_frequency}
\omega_m = \sqrt{\frac{k}{m}},
\end{equation}
for the final expression we get
\begin{equation}
H =
\frac{p^2}{2m}
+ \frac{1}{2}m \omega_m^2 x^2 +\frac{p_{\mathrm{cm}}^2}{2M}.
\end{equation}
Omitting the center-of-mass term, this is exactly \eqref{eq_matter_Hamiltonian} (cf.~\eqref{eq_eom_for_x_xcm_uncoupled_eqs} and~\eqref{eq_effectively_coupled_Hamiltonian}).

\subsection{Quantization modes}\label{sec_quantization_modes}

We want to quantize
\begin{equation}
    H^\order{m} = \frac{p^2}{2m} + \frac{1}{2} m \omega_m^2 x^2,
\end{equation}
We introduce the ladder operators and the dimensionless quantities for $x$
\begin{equation}
    X = b^\dagger + b, \qquad x = x_\text{ZPF}\, X, \qquad x_\text{ZPF} = \sqrt{\frac{\hbar}{2 m \omega_m}},
\end{equation}
and for $p$
\begin{equation}
    P = i (b^\dagger - b), \qquad p = \delta_p \, P, \qquad \delta_p = \frac{\hbar}{2} \frac{1}{x_\text{ZPF}} = \sqrt{\frac{\hbar m \omega_m}{2}},
\end{equation}
Then for the Hamiltonian we obtain
\begin{equation}
    H^\order{m} = \frac{\hbar \omega_m}{4} (X^2 + P^2) = \hbar \omega_m b^\dagger b.
\end{equation}
where we have omitted the zero-point energy.

The commutation relations are
\begin{equation}
    [x,p] = i \hbar \,\onematrix, \qquad [x,x]=[p,p]=0.
\end{equation}
And it follows that
\begin{equation}
    [b,b^\dagger] = \onematrix, \qquad [b,b]=[b^\dagger,b^\dagger] = 0.
\end{equation}

\subsection{Decoupling of the center-of-mass coordinate: only the quadrupole radiates}\label{sec_decoupling_of_center_of_mass_quadrupole}

The mass multipole moment of order $n$ of the source is given by
\begin{equation}
    I^{i_1 \cdots i_n} = \int d^3x \, T^{00} \, x^{\langle i_1}\cdots x^{i_n\rangle},
\end{equation}
where the angular brackets $\langle \;\rangle$ denote the symmetric trace-free projection. For $n = 0, 1, 2$ this gives the monopole, dipole and quadrupole moments, respectively,
\begin{equation}
\begin{aligned}
    M &= \int d^3x \, T^{00}, \\
    D^i &= \int d^3 x \, T^{00} \, x^i, \\
    Q^{ij} &= \int d^3 x \, T^{00} \, \left(x^i x^j - \frac{1}{3} \delta^{ij} \, x_k x^k \right).
\end{aligned}
\end{equation}
In the slow-motion limit, we have $T^{00} = \rho$, where $\rho$ is the mass density of the source. Specializing for motion along the $x$-direction, for a system of two point particles with masses $m_1$ and $m_2$ at positions $x_1$ and $x_2$, the mass density is given by
\begin{equation}
    \rho = m_1 \, \delta(x- {x}_1) + m_2 \, \delta({x}- {x}_2).
\end{equation}
The corresponding monopole, dipole, and quadrupole moments (denoting $ D \equiv D^x $ and $ Q \equiv Q^{xx}$) are then
\begin{equation}
\begin{aligned}
    M &= m_1 + m_2, \\
    D &= m_1 x_1 + m_2 x_2, \\
    Q &= m_1 x_1^2 + m_2 x_2^2.
\end{aligned}
\end{equation}
This makes it apparent that the monopole moment $M$ is simply the total mass of the system, the dipole moment $D$ represents the center-of-mass position, and the quadrupole moment $Q$ characterizes the time-dependent shape or spatial distribution of the mass. In the center-of-mass coordinates we obtain
\begin{equation}\label{eq_quadrupole_D_and_x}
\begin{aligned}
    Q = \frac{D^2}{M} + m x^2,
\end{aligned}
\end{equation}
where we have used $x_\text{cm} = D/M$. But, importantly, in our case, the equation of motion for the center of mass reads $\ddot{x}_\text{cm} =0$ (cf.~Eq.~\eqref{eq_eom_for_x_xcm_uncoupled_eqs}). This clearly shows why the center-of-mass motion, unlike the relative motion, decouples from gravitational radiation: because the gravitational-wave field is proportional to the second time derivative of the quadrupole \cite{maggiore_2007_gravitational_waves_vol},
\begin{equation}
    h_{ij}(t,\mathbf x) = \frac{1}{|\mathbf x|} \frac{2 G}{c^4} \ddot{Q}_{ij}\left(t-{|\mathbf x|}/{c}\right)
\end{equation}
and the center-of-mass dipole satisfies $\ddot{D} = 0$, the first term in Eq.~\eqref{eq_quadrupole_D_and_x} does not contribute to radiation.


\subsection{Matter Hamiltonian for two trapped nanoparticles coupled to each other}\label{sec_matter_hamiltonian_generalized}

For completeness, we now consider the more general situation in which
each nanoparticle is confined by its own harmonic trapping potential.
This is the experimentally relevant setting, since levitated
nanoparticles must be trapped in order to be controlled and measured~\cite{deplano_pontin_ranfagni_ea_2024_coulomb_coupling_between}. The Hamiltonian for two nanoparticles trapped with frequencies
$\Omega_1$ and $\Omega_2$, and mutually coupled with
an effective spring constant $k$, can be written as
\begin{equation}\label{eq_hamiltonian_two_coupled_trapped_first}
    H = \frac{p_1^2}{2m_1} + \frac{1}{2} m_1 \,\Omega_1^2 \, x_1^2 + \frac{p_2^2}{2 m_2} + \frac{1}{2} m_2 \, \Omega_2^2 \, x_2^2
    + \frac{k}{2}(x_1-x_2)^2.
\end{equation}
It is easy to see that for $\Omega_1 = \Omega_2 = 0$, the above reduces to Eq.~\eqref{eq_hamiltonian_two_coupled}, which denotes the Hamiltonian for two coupled harmonic oscillators. Using the relative and center-of-mass coordinates \eqref{eq_coordinates_com_rel}, the definitions for the total and reduced mass \eqref{eq_mass_total_and_reduced}, and the definition for the harmonic angular frequency \eqref{eq_angular_harmonic_frequency}, we rewrite the Hamiltonian as
\begin{equation}\label{eq_hamiltonian_two_coupled_trapped}
    H = \frac{p^2}{2m} + \frac{1}{2}m\,(\omega_m^2 + \Omega_\text{rel}^2) \, x^2
    +
    \frac{p_\text{cm}^2}{2M}+ \frac{1}{2} M \,\Omega_\text{cm}^2 \, x_\text{cm}^2 + m \,x \, x_\text{cm} (\Omega_1^2 - \Omega_2^2),
\end{equation}
where we have defined
\begin{equation}
    \Omega^2_\text{rel} \equiv \frac{m_2 \Omega_1^2 + m_1 \Omega_2^2}{M}
    ,\quad
    \Omega^2_\text{cm} \equiv \frac{m_1 \Omega_1^2 + m_2 \Omega_2^2}{M}.
\end{equation}
The coupled equations of motion for the relative and center-of-mass coordinate are obtained to be
\begin{equation}
\begin{aligned}
    \ddot x
+
\left(
\omega_m^2+\Omega_\text{rel}^2
\right)x
+
\left(
\Omega_1^2-\Omega_2^2
\right)x_\text{cm}
&=
0, \\
\ddot x_\text{cm}
+
\Omega_\text{cm}^2x_\text{cm}
+
\frac{m}{M}
\left(
\Omega_1^2-\Omega_2^2
\right)x
&=
0.
\end{aligned}
\end{equation}
Several features of the Hamiltonian in Eq.~\eqref{eq_hamiltonian_two_coupled_trapped}
are worth noting. First, the harmonic frequency of the relative coordinate is shifted due to the trapping potentials; the effective frequency of the relative mode is now given by $\omega_m^2+\Omega_\text{rel}^2$. Second, unlike in the only-coupled case, the center-of-mass coordinate is itself a harmonic oscillator with frequency $\Omega_\text{cm}$. Finally, the last term mixes the relative and center-of-mass coordinates. Therefore, unsurprisingly, in the generic trapped and coupled case, the transformation to the relative and center-of-mass coordinates does not fully diagonalize the Hamiltonian. 



\secpar{Two coupled untrapped particles}
Taking $\Omega_1 = \Omega_2 = 0$ in Eq.~\eqref{eq_hamiltonian_two_coupled_trapped_first}, the Hamiltonian reduces to
\begin{equation}
    H
=
\frac{p_1^2}{2m_1}
+
\frac{p_2^2}{2m_2}
+
\frac{k}{2}(x_1-x_2)^2 .
\end{equation}
This is the case we considered in Eq.~\eqref{eq_hamiltonian_two_coupled}. Or, to repeat for completeness, in terms of the relative and center-of-mass coordinates (cf.~Eq.~\eqref{eq_hamiltonian_two_coupled_trapped}), noting that $\Omega_\text{rel} = \Omega_\text{cm} = 0$, the Hamiltonian is
\begin{equation}
    H
=
\frac{p^2}{2m}
+
\frac{1}{2}m\omega_m^2x^2
+
\frac{p_\text{cm}^2}{2M}.
\end{equation}
Thus, the dynamics separates exactly into a harmonic oscillator for the relative coordinate and a free particle for the center of mass. The corresponding equations of motion are
\begin{equation}\label{eq_eom_for_x_xcm_uncoupled_eqs}
\begin{aligned}
    \ddot{x} + \omega_m^2 \, x &= 0, \\
    \ddot{x}_\text{cm} &=0.
\end{aligned}
\end{equation}
Since $\ddot{x}_\text{cm}=0$, the center-of-mass motion is inertial (cf.~discussion around Eq.~\eqref{eq_quadrupole_D_and_x}). Upon quantization, the full Hilbert space factorizes as $\mathcal H=\mathcal H_\text{rel}\otimes\mathcal H_\text{cm}$; the center-of-mass sector is diagonal in momentum eigenstates, whereas the relative sector is a harmonic oscillator. Introducing the ladder operators $b$ and $b^\dagger$ for the relative coordinate, we obtain,
\begin{equation}
    H = \hbar \, \omega_m \, b^\dagger b + \frac{p_\text{cm}^2}{2M},
\end{equation}
where we have omitted the zero-point energy.

\secpar{Two uncoupled trapped particles}
Taking $k=0$ in Eq.~\eqref{eq_hamiltonian_two_coupled_trapped_first}, the Hamiltonian becomes,
\begin{equation}
    H = \frac{p_1^2}{2m_1} + \frac{1}{2} m_1 \,\Omega_1^2 \, x_1^2 + \frac{p_2^2}{2 m_2} + \frac{1}{2} m_2 \, \Omega_2^2 \, x_2^2.
\end{equation}
This describes two independent trapped nanoparticles. Since there is no coupling term between them, the Hamiltonian is already diagonal in the original particle coordinates. For completeness, if this same uncoupled trapped system is rewritten in relative and center-of-mass coordinates, the Hamiltonian contains the mixing term,
\begin{equation}
    H = \frac{p^2}{2m} + \frac{1}{2}m\, \Omega_\text{rel}^2 \, x^2
    +
    \frac{p_\text{cm}^2}{2M}+ \frac{1}{2} M \,\Omega_\text{cm}^2 \, x_\text{cm}^2 + m \,x \, x_\text{cm} (\Omega_1^2 - \Omega_2^2)
\end{equation}
Therefore, for two uncoupled trapped particles with unequal trap frequencies, the relative and center-of-mass coordinates are not the natural normal coordinates. The natural basis is simply the original particle basis, $x_1$ and $x_2$. The full Hilbert space factorizes as $\mathcal H=\mathcal H_\text{1}\otimes\mathcal H_\text{2}$, for the first and second nanoparticle, respectively. Introducing ladder operators, we obtain
\begin{equation}
    H = \hbar \,\Omega_1 \, a_1^\dagger a_1 + \hbar \, \Omega_2 \, a_2^\dagger a_2.
\end{equation}


\secpar{Two coupled trapped particles with equal frequency}
A simple limiting case, without changing the essential physics, is obtained when the two nanoparticles are trapped with the same frequency, 
\begin{equation}
    \Omega_1=\Omega_2= \Omega.
\end{equation}
In this case, the mixing term in Eq.~\eqref{eq_hamiltonian_two_coupled_trapped} proportional to $\Omega_1^2-\Omega_2^2$ vanishes, so the relative and center-of-mass coordinate sectors decouple. The trap-induced frequencies also become equal, $\Omega_\text{rel}^2=\Omega_\text{cm}^2=\Omega^2$. Therefore, the Hamiltonian separates into two independent harmonic oscillators: one for the relative coordinate and one for the center-of-mass coordinate,
\begin{equation}
    H = \frac{p^2}{2m} + \frac{1}{2}m(\omega_m^2+\Omega^2) x^2 + \frac{p^2_\text{cm}}{2M} + \frac{1}{2} M\,\Omega^2 x_\text{cm}^2.
\end{equation}
The relative coordinate remains a harmonic oscillator, but its frequency is shifted from $\omega_m^2$ to an effective frequency $\omega_m^2+\Omega^2$. The center-of-mass coordinate, which was free in the untrapped case, now becomes a harmonic oscillator with frequency $\Omega$. Thus, in the equal-trap-frequency limit, the transformation to relative and center-of-mass coordinates fully diagonalizes the Hamiltonian. This decoupling means that the free matter Hamiltonian can be quantized as two independent oscillator sectors, since $\Hilbert = \Hilbert_\text{rel} \otimes \Hilbert_\text{cm}$. Introducing separate ladder operators for the relative and center-of-mass modes and omitting the corresponding zero-point energies, the Hamiltonian becomes
\begin{equation}
    H = \hbar \sqrt{\omega_m^2+\Omega^2}\, b^\dagger b + \hbar \, \Omega \, c^\dagger c.
\end{equation}
Here $b$ and $b^\dagger$ are the ladder operators associated with the relative mode, while $c$ and $c^\dagger$ are the ladder operators associated with the center-of-mass mode. Thus, in this limiting case, the trapped two-nanoparticle system is described by two independent quantized harmonic oscillators with frequencies $\sqrt{\omega_m^2 +\Omega^2}$ and $\Omega$, respectively. In this equal-trap-frequency limit, the center-of-mass and relative coordinates diagonalize the matter Hamiltonian into two independent harmonic oscillators. This decoupling, however, refers only to the free matter Hamiltonian. A gravitational interaction couples to the mass quadrupole, which contains both $M x_{\rm cm}^2$ and $m x^2$. Hence, in this case, both normal modes couple to gravitational waves.

\secpar{Physically relative}
Consider, once again, the case of $\Omega_1 = \Omega_2 = \Omega$. As was worked out above, this removed the mixing term. If we further impose
\begin{equation}
    \Omega \ll \omega_m,
\end{equation}
then we approximately recover the description where the particles were coupled but untrapped. Physically, we are requiring that over dynamics near the internal frequency $\omega_m$, the center-of-mass restoring term is negligible
\begin{equation}
    \frac{1}{2} M \Omega^2 x_\text{cm}^2 \ll \frac{1}{2} m \omega_m^2 x^2,
\end{equation}
in the sense that the common-mode restoring force is slow compared to the internal-mode dynamics. The center-of-mass mode is approximately untrapped. The above condition means that the external trap should be weak compared with the interparticle coupling. Then the pair of particles behaves like a weakly confined dimer: its common motion is nearly free, while its internal separation is harmonically trapped. This then gives the Hamiltonian
\begin{equation}\label{eq_effectively_coupled_Hamiltonian}
    H_\text{eff}
=
\frac{p^2}{2m}
+
\frac{1}{2}m\omega_m^2x^2
+
\frac{p_\text{cm}^2}{2M},
\end{equation}
which is exactly the Hamiltonian describing two coupled untrapped particles.





\section{Nonrelativistic matter and gravitons: Gravity Hamiltonian}\label{app_gravity_hamiltonian}

\subsection{From the Einstein-Hilbert action to the massless Fierz-Pauli action}\label{sec_from_einstein_hilbert_to_fierz_pauli}

The Einstein-Hilbert (EH) action in vacuum is given by \cite{einstein1916HamiltonPrincipleGeneral,hilbert1915FoundationsPhysics},
\begin{equation}\label{eq_einstein_hilbert_action}
    S_{\mathrm{EH}}
    = \frac{c^3}{16\pi G}
      \int d^4x\, \mathcal{L}_\text{EH}, \qquad \mathcal{L}_\text{EH} \equiv \sqrt{-g}\,R ,
\end{equation}
where $G$ denotes Newton's gravitational constant, $R$ denotes the Ricci scalar and $g$ denotes the nonvanishing determinant of the metric tensor $g_\mn$ \cite{carroll2014spacetimeandgeometry,wald1984generalrelativity,weinberg_1972_gravitation_and_cosmology_principles_and_applications}. In Wald's abstract index notation \cite{wald1984generalrelativity}, the latter enables us to always define the inverse $g^\mn$ via
\begin{equation}\label{eq_inverse_metric_relation}
    g_{\mu \rho} \,g^{\rho \nu} = \tensor{\delta}{_\mu^\nu}.
\end{equation}
We use the mostly-plus metric signature with Minkowski metric,
\begin{equation}
    \eta_\mn = \text{diag}(-1,1,1,1),
\end{equation}
and take the Cartesian spacetime coordinates to be
\begin{equation}
    x^\mu = (ct, \mathbf x),   
\end{equation}
so that $d^4 x = c \, dt \, d^3 x$ and $\partial_0 = c^{-1} \partial_t$. Varying the action with respect to the metric gives the Einstein's equation
\begin{equation}
    G_\mn = 0.
\end{equation}
The Einstein tensor is given by $G_\mn = R_\mn + \frac{1}{2} g_\mn R$, where $R_\mn$ is the Ricci tensor and $R$ is the Ricci scalar.

For the weak-field limit, we expand around flat spacetime up to first order in the perturbation $h_\mn$ \cite{brizuela_martingarcia_marugan_2009_xpert_computer_algebra_for_metric},
\begin{equation}
    g_\mn(\epsilon) = \eta_\mn + \epsilon \, h_\mn,
\end{equation}
where $\epsilon$ denotes the dimensionless perturbation parameter.
Up to quadratic order, the inverse relation \eqref{eq_inverse_metric_relation} then allows us to identify the inverse metric
\begin{equation}
    g^\mn = \eta^\mn - \epsilon\, h^\mn + \epsilon^2 \, \tensor{h}{^\mu_\alpha} \tensor{h}{^\alpha^\nu}.
\end{equation}
Up to the same order, we obtain,
\begin{equation}
    \sqrt{-g} = 1 + \frac{\epsilon}{2} \, h+\frac{\epsilon^2}{8}(h^2 - 2 h_\mn h^\mn).
\end{equation}
Note the notation that we have used in the above equations
\begin{equation}
    \tensor{h}{^\mu_\nu} \equiv \eta^{\mu \lambda
    } h_{\lambda \nu},
    \qquad
    h \equiv \tensor{h}{_\alpha^\alpha}.
\end{equation}
We now apply the perturbative expansion to the Einstein-Hilbert action \eqref{eq_einstein_hilbert_action}. Generically, up to quadratic order, in terms of the perturbation parameter, for the Lagrangian we have \cite{brizuela_martingarcia_marugan_2009_xpert_computer_algebra_for_metric},
\begin{equation}\label{eq_expanded_lagrangian_eh}
    \mathcal L_{\mathrm{EH}}(\epsilon)
=
\mathcal L_{\mathrm{EH}}^{(0)}
+
\epsilon\,\mathcal L_{\mathrm{EH}}^{(1)}
+
\frac{\epsilon^2}{2!}\mathcal L_{\mathrm{EH}}^{(2)},
\qquad
 \qquad
\mathcal L_{\mathrm{EH}}^{(n)}
\equiv
\left.
\frac{d^n \mathcal L_{\mathrm{EH}}
\!\left[g_{\mu\nu}(\epsilon)\right]}{d\epsilon^n}
 \right|_{\epsilon=0}.
\end{equation}
Since the background spacetime is flat, the zeroth-order contribution from the background curvature vanishes,
\begin{equation}
    \mathcal{L}_\text{EH}^{(0)} = 0.
\end{equation}
At first order in the metric perturbation, we obtain
\begin{equation}
    \mathcal{L}_\text{EH}^{(1)} = \nabla_\alpha \nabla_\beta h^{\alpha \beta} - \nabla_\alpha \nabla^\alpha \, h = \nabla_\alpha (\nabla_\beta h^{\alpha \beta} - \nabla^\alpha \, h).
\end{equation}
But this is just a boundary term in the action; it can be written as a total divergence and we assume that it can be neglected. Note that the covariant derivative is associated with the background metric. 
At second order in the perturbation, we obtain
\begin{equation}
\begin{aligned}
\mathcal L_{\mathrm{EH}}^{(2)}
={}&
-h\,\nabla_\alpha\nabla^\alpha h
-\frac12\nabla_\alpha h\,\nabla^\alpha h
\\
&+
2\nabla^\alpha h\,\nabla_\beta h_\alpha{}^\beta
+h\,\nabla_\beta\nabla_\alpha h^{\alpha\beta}
+2h^{\alpha\beta}\nabla_\beta\nabla_\alpha h
\\
&
-2h^{\alpha\beta}\nabla_\beta\nabla_\kappa h_\alpha{}^\kappa
-2\nabla_\alpha h^{\alpha\beta}\,
  \nabla_\kappa h_\beta{}^\kappa
\\
&
-2h^{\alpha\beta}\nabla_\kappa\nabla_\beta h_\alpha{}^\kappa
-\nabla_\beta h_{\alpha\kappa}\,
 \nabla^\kappa h^{\alpha\beta}
\\
&+
2h^{\alpha\beta}\nabla_\kappa\nabla^\kappa h_{\alpha\beta}
+\frac32\nabla_\kappa h_{\alpha\beta}\,
 \nabla^\kappa h^{\alpha\beta}.
\end{aligned}
\end{equation}
We have conveniently grouped the terms line by line so that, within each line, they combine naturally under the corresponding integrations by parts into a bulk contribution and a total-divergence term. The combination is, of course, not unique; we integrate by parts such that the quadratic Lagrangian contains only products of first derivatives. Namely,
\begin{equation}
\begin{aligned}
\mathcal L_{\mathrm{EH}}^{(2)}
={}&
\frac{1}{2}\nabla_\alpha h \nabla^\alpha h + \nabla_\alpha(-h \nabla^\alpha h)
\\
&- \nabla_\alpha h \nabla_\beta h^{\alpha \beta} + \nabla_\alpha(h \nabla_\beta h^{\alpha \beta} + 2 h^{\alpha\beta} \nabla_\beta h)
\\
&
+ \nabla_\alpha(-2 h^{\alpha\beta} \nabla_\kappa \tensor{h}{_\beta^\kappa})
\\
&
+\nabla_\beta h_{\alpha\kappa} \nabla^\kappa h^{\alpha \beta} + \nabla_\alpha(-2 h^{\kappa\beta} \nabla_\beta \tensor{h}{_\kappa^\alpha})
\\
&
-\frac12 \nabla_\kappa h_{\alpha\beta} \nabla^\kappa h^{\alpha\beta} + \nabla_\alpha(2 h^{\kappa\beta} \nabla^\alpha h_{\kappa\beta}).
\end{aligned}
\end{equation}
Here we have employed the symmetry of the metric perturbation $h_\mn = h_{\nu\mu}$, together with the background metric compatibility. Note that no covariant derivatives have been commuted. Consequently, the integrations by parts displayed at this stage do not themselves require the background to be flat. Note that this statement applies only to the integrations by parts: in the present derivation, flatness has already been assumed by setting $\sqrt{-\eta}=1$ and by omitting terms that would be present on a curved background. Neglecting the boundary terms, we finally reach the massless Fierz-Pauli (FP) Lagrangian~\cite{ortin_2015_gravity_and_strings,fierz_pauli_1939_on_relativistic_wave_equations_for_particles},
\begin{equation}
\mathcal{L}_\text{FP} \equiv\mathcal L_{\mathrm{EH}}^{(2)}
=
\frac{1}{2}\nabla_\alpha h \nabla^\alpha h - \nabla_\alpha h \nabla_\beta h^{\alpha \beta} 
+\nabla_\beta h_{\alpha\kappa} \nabla^\kappa h^{\alpha \beta}
-\frac12 \nabla_\kappa h_{\alpha\beta} \nabla^\kappa h^{\alpha\beta}.
\end{equation}
Thus the Einstein-Hilbert action~\eqref{eq_einstein_hilbert_action}, in our flat background spacetime in Cartesian coordinates, now becomes the massless Fierz-Pauli action,
\begin{equation}\label{eq_action_pauli_fierz}
    S_\text{FP} = 
    \frac{c^3}{16\pi G}
    \frac{\epsilon^2}{2} \int d^4x \,\mathcal{L}_\text{FP},
    \qquad
    \mathcal{L}_\text{FP}=\frac{1}{2}\Big\{ 
    \partial_\alpha h \,\partial^\alpha h - 2\,\partial_\alpha h \,\partial_\beta h^{\alpha \beta} 
+2\,\partial_\beta h_{\alpha\kappa} \,\partial^\kappa h^{\alpha \beta}
- \partial_\kappa h_{\alpha\beta} \,\partial^\kappa h^{\alpha\beta}
    \Big\}.
\end{equation}
Note that the factor of $1/2$ in the action arises from the quadratic term in the perturbative expansion of the Lagrangian~\eqref{eq_expanded_lagrangian_eh}. Therefore, it is useful to record the formal relation between the Fierz-Pauli action and the second perturbation of the Einstein-Hilbert action, up to boundary terms,
\begin{equation}
S_{\mathrm{FP}}
\equiv
\frac{\epsilon^2}{2!}
\left.
\frac{d^2S_{\mathrm{EH}}
\!\left[g_{\mu\nu}(\epsilon)\right]}
{d\epsilon^2}
\right|_{\epsilon=0}.
\end{equation}
In summary, we have obtained the Fierz-Pauli action directly by expanding the Einstein-Hilbert action to second order in the metric perturbation. From this point onward, having made the perturbative order explicit, we can safely set the bookkeeping parameter $\epsilon =1$. 

One may instead begin with the most general Lorentz-invariant quadratic Lagrangian containing terms with two first-order derivatives of the metric perturbation, modulo total derivatives, and determine its coefficients by canonical normalization together with the requirement that the resulting equations of motion satisfy the off-shell Bianchi identity \cite{ortin_2015_gravity_and_strings}. The result above given in Eq.~\eqref{eq_action_pauli_fierz}, agrees exactly with this procedure elaborated in \cite[Eq.~3.84]{ortin_2015_gravity_and_strings}, up to an overall minus sign which is simply due to the opposite metric-signature convention adopted there.

\subsection{Equations of motion, the harmonic gauge, and the transverse-traceless (TT) gauge}\label{sec_equations_of_motion_TT_gauge}

To obtain the equation of motion, we vary the Fierz-Pauli action \eqref{eq_action_pauli_fierz} with respect to the perturbation. After integrating by parts and neglecting boundary terms, the variation is
\begin{equation}
    \delta S_\text{FP} = \frac{c^3}{64 \pi G}\int d^4x \, \delta h^\mn \,G_{\mu\nu}^{(1)},
    \qquad
    G_{\mu\nu}^{(1)}
\equiv
\Box h_{\mu\nu}
+\partial_\mu\partial_\nu h
-\partial_\mu\partial^\alpha h_{\alpha\nu}
-\partial_\nu\partial^\alpha h_{\alpha\mu}
-\eta_{\mu\nu}\Box h
+\eta_{\mu\nu}
\partial_\alpha\partial_\beta h^{\alpha\beta}.
\end{equation}
Here we have used the notation $\Box \equiv \partial_\alpha \partial^\alpha$. Then the condition $\delta S_\text{FP} = 0$, gives us exactly Einstein's equation to first order,
\begin{equation}\label{eq_vacuum_equations}
    G_\mn^{(1)} = 0.
\end{equation}
Both the Fierz-Pauli action and its equations of motion are invariant under the linearized diffeomorphism,
\begin{equation}\label{eq_linearized_diffeomorphism}
\begin{aligned}
    h_\mn &\longrightarrow h_\mn' = h_\mn + \partial_\mu \xi_\nu + \partial_\nu \xi_\mu, \\
    h &\longrightarrow h' = h+ 2 \partial_\mu \xi^\mu,
\end{aligned}
\end{equation}
for the coordinate transformation
\begin{equation}
    x^\mu \longrightarrow x'^\mu = x^\mu - \xi^\mu,
\end{equation}
where $\xi^\mu(x)$ is an arbitrary spacetime-dependent vector field of the same perturbative order as $h_\mn$. That is, the perturbation contains gauge redundancy. We first impose the harmonic gauge (de Donder gauge; the gravitational analogue of the Lorenz gauge in electromagnetism),
\begin{equation}\label{eq_harmonic_gauge_trace_reversed_perturbation}
    \partial^\mu\bar h_{\mu\nu}=0, \qquad \bar h_{\mu\nu}
\equiv h_{\mu\nu}-\frac12\eta_{\mu\nu}h.
\end{equation}
In terms of the trace-reversed perturbation, the vacuum equations \eqref{eq_vacuum_equations} read
\begin{equation}
   \Box\bar h_{\mu\nu}
-\partial_\mu\partial^\alpha\bar h_{\alpha\nu}
-\partial_\nu\partial^\alpha\bar h_{\alpha\mu}
+\eta_{\mu\nu}\partial_\alpha\partial_\beta
\bar h^{\alpha\beta}
=0.
\end{equation}
Thus, in the harmonic gauge \eqref{eq_harmonic_gauge_trace_reversed_perturbation}, the vacuum equations \eqref{eq_vacuum_equations} reduce to wave equations for the trace-reversed perturbation,
\begin{equation}\label{eq_wave_equation}
    \Box \bar{h}_\mn = 0.
\end{equation}

Before proceeding to solve the wave equation, we examine the gauge freedom that remains after imposing the harmonic gauge. Under the linearized diffeomorphism \eqref{eq_linearized_diffeomorphism}, the trace-reversed perturbation transforms as
\begin{equation}
    \bar h'_{\mu\nu}
=
\bar h_{\mu\nu}
+\partial_\mu\xi_\nu+\partial_\nu\xi_\mu
-\eta_{\mu\nu}\partial_\alpha\xi^\alpha.
\end{equation}
Taking its divergence gives
\begin{equation}
    \partial^\mu\bar h'_{\mu\nu}
=
\partial^\mu\bar h_{\mu\nu}
+\Box\xi_\nu.
\end{equation}
Therefore, if the original perturbation is in the harmonic gauge \eqref{eq_harmonic_gauge_trace_reversed_perturbation}, then the transformed perturbation remains in the harmonic gauge provided
\begin{equation}
    \Box \xi_\nu = 0.
\end{equation}
This is how we see that gauge freedom remains: the harmonic gauge does not require $\xi^\mu = 0$, but it only requires it to satisfy the homogeneous wave equation, which has many nonzero solutions.

Let us now go back to the wave equation \eqref{eq_wave_equation}. For radiative vacuum solutions, the remaining gauge freedom can be used to choose the transverse-traceless (TT) gauge, in which the perturbation is purely spatial, transverse to the direction of the propagation, and traceless. Unlike the harmonic gauge, the TT gauge is a noncovariant gauge choice, because it singles out a timelike direction, or equivalently, a preferred inertial frame. It is perhaps worthwhile to note that the TT gauge is appropriate for freely propagating vacuum waves, although it cannot generally be imposed on the entire perturbation in the presence of sources. 

To solve the wave equation \eqref{eq_wave_equation}, let us introduce an ansatz. Because the equation is linear, a general radiative solution can be decomposed into plane-wave modes. Let us consider one such mode,
\begin{equation}\label{eq_ansatz_mode}
    h_\mn = e_\mn e^{i k_\alpha x^\alpha}, 
    \qquad
    \bar{h}_\mn = \bar{e}_\mn e^{i k_\alpha x^\alpha}.
\end{equation}
We take the polarization tensor $e_\mn$  to be real, $e^*_\mn = e_\mn$, and its trace-reversed definition is $\bar e_{\mu\nu}=e_{\mu\nu}-\tfrac12\eta_{\mu\nu}\eta^{\alpha\beta}e_{\alpha\beta}$. The wave four-vector is $k^\mu = (\omega_k/c,\mathbf k)$, such that with our mostly-plus metric signature we have $i k_\alpha x^\alpha = -i(\omega_k t - \mathbf k \cdot \mathbf x)$. The real solution is understood to be obtained by adding the complex-conjugate mode. There are two consequences of the plane-wave ansatz: the null dispersion relation and the harmonic constraint on the polarization tensor. Substituting the ansatz \eqref{eq_ansatz_mode} into the wave equation \eqref{eq_wave_equation} gives the null dispersion relation
\begin{equation}
    k_\alpha k^\alpha = 0 \qquad \to \qquad \omega_k = c k, \qquad k \equiv|\mathbf k|.
\end{equation}
The harmonic gauge condition \eqref{eq_harmonic_gauge_trace_reversed_perturbation} for the ansatz \eqref{eq_ansatz_mode} gives,
\begin{equation}\label{eq_transversality_for_the_trace_reversed}
    k^\mu \bar{e}_\mn = 0.
\end{equation}
This can be thought of as the harmonic gauge in the Fourier space. Accordingly, the residual gauge parameter relevant to this Fourier mode can be chosen as
\begin{equation}\label{eq_xi_mode}
    \xi_\mu(x) = i a_\mu e^{ik_\alpha x^\alpha}.
\end{equation}
Here we keep $a_\mu$ real, similarly to the polarization tensor, and we add the $i$ such that the transformed polarization tensor also remains real. From the linearized diffeomorphism \eqref{eq_linearized_diffeomorphism}, for the mode under consideration using \eqref{eq_ansatz_mode} and \eqref{eq_xi_mode}, we then obtain
\begin{equation}
    e_\mn' = e_\mn - k_\mu a_\nu - k_\nu a_\mu.
\end{equation}
Its trace then transforms as
\begin{equation}
    e' = e- 2 k_\mu a^\mu.
\end{equation}
We can now use one combination of the four components of $a_\mu$ to choose $k^\mu a_\mu = e/2$. This gives
\begin{equation}\label{eq_tracelessnes}
    \eta^\mn e_\mn' = 0.
\end{equation}
This is the traceless condition. In turn, this makes the polarization tensor and its trace-reversed definition coincide, importantly in the transformed coordinates. Thus, from the transversality of the trace-reversed polarization tensor, we obtain the transversality of the polarization tensor in the transformed coordinates,
\begin{equation}\label{eq_transversality}
    k^\mu e_\mn' = 0.
\end{equation}
There is still gauge freedom remaining. Namely, the traceless condition \eqref{eq_tracelessnes} only fixed one combination of the four components of $a_\mu$. Thus the transformed polarization tensor still depends on three undetermined components of $a_\mu$. In the end, we want only two physical components to survive and to describe the deformation in the plane perpendicular to the wave's propagation. But a null vector $k^\mu$ alone does not define an ordinary spatial plane. Because it is null, the space orthogonal to it contains $k^\mu$ itself. To identify a physical two-dimensional transverse plane, we must also specify another independent reference vector. But note that we have already done this implicitly when we wrote down $k^\mu = (\omega_k/c,\mathbf k)$. In particular, we can write it as
\begin{equation}\label{eq_kmu_decomposition_timelike_spacelike_u_n}
    k^\mu =k(u^\mu + n^\mu)
\end{equation}
where $u^\mu$ and $n^\mu$ are unit vectors orthogonal to each other, satisfying
\begin{equation}
    u^\mu u_\mu = -1, \qquad n^\mu n_\mu = 1, \qquad u^\mu n_\mu = 0.
\end{equation}
In the observer's rest frame,
\begin{equation}
    u^\mu = (1, \mathbf 0), \qquad n^\mu = (0, \mathbf n).
\end{equation}
This gives us $k^\mu = k(1, \mathbf n) = (\omega_k/c, \mathbf k)$, with a  timelike $u^\mu$ and a spacelike $n^\mu$.
Then the transversality condition \eqref{eq_transversality} reads
\begin{equation}\label{eq_transversality_condition_umu_nmu}
    u^\mu e_\mn' = - n^\mu e_\mn'
\end{equation}
To place both physical polarizations entirely in the spatial plane perpendicular to the wave, we choose
\begin{equation}\label{eq_temporal_gauge}
    u^\mu e_\mn' = 0.
\end{equation}
This is the temporal gauge condition. Finally, we have no remnant gauge freedom. 
Together, the tracelessness condition \eqref{eq_tracelessnes}, the transversality condition \eqref{eq_transversality} and the temporal gauge condition \eqref{eq_temporal_gauge} define the transverse-traceless (TT) (radiation) gauge. More generally, for the gauge-transformed perturbation, the three conditions can be written as
\begin{equation}\label{eq_TT_gauge_perturbation_transformed}
\eta^{\mu\nu}h'_{\mu\nu}=0,
\qquad
\partial^\mu h'_{\mu\nu}=0,
\qquad
u^\mu h'_{\mu\nu}=0.
\end{equation}
On a Minkowski background, these conditions can be imposed on any freely propagating radiative vacuum perturbation that admits a Fourier decomposition. They therefore apply not only to a single plane wave, but also to superpositions of plane-wave modes, including, say, linearized cylindrical waves, with the residual gauge freedom fixed mode by mode relative to the same chosen $u^\mu$.

Most commonly, a metric perturbation that has been put into the TT gauge is written as $h_\mn^\text{TT}$. This notation, however, is also commonly used for the transverse-traceless part obtained by applying the TT projector to the spatial perturbation, and in principle, they are distinct \cite{ashtekar_bonga_2017_on_the_ambiguity_in_the_notion}. For the freely propagating vacuum solutions considered here, however, they exactly coincide: the residual gauge transformation removes precisely the longitudinal components annihilated by the TT projector, while the resulting spatial perturbation is already transverse and traceless and is therefore unchanged by that projector. To see this, let us generalize the spatial projector into a four-dimensional TT projector, with the help of $u^\mu$ and $n^\mu$,
\begin{equation}
    P_{\mu\nu}{}^{\rho\sigma}
=
\frac12
\left(
Q_\mu{}^\rho Q_\nu{}^\sigma
+
Q_\mu{}^\sigma Q_\nu{}^\rho
-
Q_{\mu\nu}Q^{\rho\sigma}
\right), \qquad Q_{\mu\nu}
=
\eta_{\mu\nu}
+u_\mu u_\nu
-n_\mu n_\nu.
\end{equation}
Thus we can define
\begin{equation}
    e^{\mathrm{TT}}_{\mu\nu} \equiv P_{\mu\nu}{}^{\rho\sigma}e_{\rho\sigma}, \qquad h^{\mathrm{TT}}_{\mu\nu} \equiv P_{\mu\nu}{}^{\rho\sigma}h_{\rho\sigma}.
\end{equation}
In the inertial frame where $u^\mu = (1, \mathbf 0)$, we have
\begin{equation}
    Q_{00}=Q_{0i}=0,
\qquad
Q_{ij}=\delta_{ij}-n_i n_j.
\end{equation}
This reduces to the usual projector acting on spatial indices,
\begin{equation}
    h^{\mathrm{TT}}_{ij}=P_{ij}{}^{kl}h_{kl}.
\end{equation}
For completeness, the spatial TT projector is defined as
\begin{equation}\label{eq_TT_projector_spatial}
    P_{ij}{}^{kl}
=
\frac12
\left(
Q_i{}^kQ_j{}^l
+
Q_i{}^lQ_j{}^k
-
Q_{ij}Q^{kl}
\right),
\qquad
Q_{ij}=\delta_{ij}-n_i n_j.
\end{equation}
Now note that for one plane-wave mode, we had the gauge transformation
\begin{equation}
    e_\mn' = e_\mn - k_\mu a_\nu - k_\nu a_\mu.
\end{equation}
Acting with the projector on both sides gives
\begin{equation}
P_{\mu\nu}{}^{\rho\sigma}e'_{\rho\sigma}
=
P_{\mu\nu}{}^{\rho\sigma}
e_{\rho\sigma}
-
P_{\mu\nu}{}^{\rho\sigma}(
k_\rho a_\sigma
+k_\sigma a_\rho).
\end{equation}
For the left-hand side we note that  $e_\mn'$ is already transverse and traceless, so $P_{\mu\nu}{}^{\rho\sigma}e'_{\rho\sigma}=e'_{\mu\nu}.$ The first term on the right-hand side becomes simply $P_{\mu\nu}{}^{\rho\sigma}
e_{\rho\sigma}=e_{\mu\nu}^\text{TT}$. The important part is that the second term in the right-hand side vanishes,
\begin{equation}
P_{\mu\nu}{}^{\rho\sigma}
(k_\rho a_\sigma+k_\sigma a_\rho)
=
\frac12Q_\mu{}^\rho Q_\nu{}^\sigma
(k_\rho a_\sigma+k_\sigma a_\rho)
+
\frac12Q_\mu{}^\sigma Q_\nu{}^\rho
(k_\rho a_\sigma+k_\sigma a_\rho)
-
\frac12Q_{\mu\nu}Q^{\rho\sigma}
(k_\rho a_\sigma+k_\sigma a_\rho) = 0,
\end{equation}
because every term contains a contraction $Q_\mn$ and $k^\mu$, which is always zero. Thus, indeed,
\begin{equation}\label{eq_polarization_tensor_prime_equal_TT}
    e'_{\mu\nu}=e_{\mu\nu}^{\mathrm{TT}}.
\end{equation}
Since the field equations and the TT projection are all linear, this equality extends mode by mode to any superposition of plane-wave solutions. Thus, for a general freely propagating vacuum solution, the gauge-transformed perturbation coincides with the TT projection of the original perturbation,
\begin{equation}\label{eq_TT_gauge_perturbation_equal_to_TT}
    h_\mn'=h_\mn^\text{TT}.
\end{equation}

In the TT gauge, in the massless Fierz-Pauli action \eqref{eq_action_pauli_fierz} the first two terms vanish because of tracelessness, while the third disappears from the action after integration by parts because of transversality. Neglecting the resulting boundary term, only the fourth term remains (recall that we have set $\epsilon = 1$). Namely, we end up with
\begin{equation}\label{eq_gravity_action_covariant_TT}
    S^{(g)}\equiv S_{\mathrm{FP}}^{\mathrm{TT}}
=
-\frac{c^3}{64\pi G}
\int d^4x\,
\partial_\rho h_{\mu\nu}^{\mathrm{TT}}
\partial^\rho h_{\mathrm{TT}}^{\mu\nu}.
\end{equation}
The temporal gauge condition implies $h_{0\mu}^\text{TT}$, so the contraction may equivalently be written solely in terms of the spatial components $h_{ij}^\text{TT}$, as is usually done in the literature \cite{maggiore_2007_gravitational_waves_vol}. For notational convenience, we henceforth suppress the superscript TT, with $h_\mn$ understood to denote the perturbation in the TT gauge. This is the quadratic gravitational action presented in the main text in Eq.~\eqref{eq_gravity_action}.

As remarked before, the most general real radiative classical (CL) solution can be expressed as a superposition of plane-wave modes. In particular, we write
\begin{equation}
    h_{ij}^{\mathrm{CL}}(t,\mathbf x)
=
\sum_\lambda
\int d^3k\,
A_{\mathbf k,\lambda}
e^\lambda_{ij}(\mathbf n)
e^{-i(\omega_k t-\mathbf k\cdot\mathbf x)}
+\mathrm{c.c.}
\end{equation}
Here $A_\klambda$ is the arbitrary complex classical amplitude, containing both amplitude and phase, and the complex conjugate (c.c.) makes the field real, and $e_{ij}^\lambda(\mathbf n)$ is the real basis polarization tensor (defined thoroughly in Section~\ref{sec_polarization_tensor}). While the wave equation fixes $\omega_k = ck$, it cannot determine $A_\klambda$, because those coefficients represent the initial conditions of the classical gravitational wave. Putting the action into canonical form does not determine these classical amplitudes either; rather, it fixes the normalization of the field used to describe them. With the polarization-tensor normalization defined in Eq.~\eqref{eq_orthonormality_basis_polarization_tensors}, the following rescaling brings the action for each independent polarization $\lambda$ to the conventional canonical form with an overall factor $1/2$,
\begin{equation}
    A_\klambda \longrightarrow \sqrt{\frac{c^3}{16\pi G}} A_\klambda.
\end{equation}
The amplitudes on the right-hand side are understood to be those of the canonically normalized field. The role of the action, therefore, is to identify the canonically normalized gravitational field.

\subsection{Quantization modes}\label{sec_quantization_modes_gravity}

With suppressed TT superscript, using $x^0 = ct$, the action \eqref{eq_gravity_action_covariant_TT} becomes
\begin{equation}
S^{(g)}
=
\int dt\,L^{(g)},
\qquad
L^{(g)}(t)
=
\int d^3x\,\mathcal L^{(g)}.
\qquad
\mathcal L^{(g)}
=
\frac{c^2}{64\pi G}
\left[
\dot h_{ij}\dot h^{ij}
-
c^2\partial_kh_{ij}\partial^kh^{ij}
\right].
\end{equation}
The momentum canonically conjugate to $h_{ij}$ is therefore
\begin{equation}
    \Pi^{ij}
\equiv
\frac{\partial\mathcal L^{(g)}}{\partial\dot h_{ij}}
=
\frac{c^2}{32\pi G}\dot h^{ij}.
\end{equation}
In terms of the canonical variables, the action reads
\begin{equation}
    S^{(g)}
=
\int dt\,d^3x
\left[
\Pi^{ij}\dot h_{ij}
-
\mathcal H^{(g)}
\right].
\end{equation}
Here, the gravitational Hamiltonian is defined by the Legendre transform,
\begin{equation}\label{eq_gravity_hamiltonian_definition}
    H^{(g)}
\equiv
\int d^3x\,\mathcal H^{(g)},
\qquad
\mathcal H^{(g)}
\equiv
\Pi^{ij}\dot h_{ij}-\mathcal L^{(g)}.
\end{equation}
The canonical term (the first term in the action above) identifies the Because we are already working on the reduced TT phase space, the appropriate Poisson bracket (PB) contains the spatial TT projector,
\begin{equation}
    \left\{
h_{ij}(t,\mathbf x),
\Pi^{kl}(t,\mathbf y)
\right\}_{\mathrm{PB}}
=
P_{ij}{}^{kl}\,
\delta^{(3)}(\mathbf x-\mathbf y),
\end{equation}
together with
\begin{equation}
    \{h_{ij},h_{kl}\}_{\mathrm{PB}}=0,
\qquad
\{\Pi^{ij},\Pi^{kl}\}_{\mathrm{PB}}=0.
\end{equation}
We now promote the gravitational perturbation and its conjugate momentum to operators. Canonical quantization promotes these brackets according to $\{\,\cdot,\cdot\,\}_{\mathrm{PB}}
\longrightarrow
\frac{1}{i\hbar}
[\,\cdot,\cdot\,].$ This gives 
\begin{equation}\label{eq_canonical_commutation_relation_h_Pi}
    \left[
 h_{ij}(t,\mathbf x),
\Pi^{kl}(t,\mathbf y)
\right]
=
i\hbar
P_{ij}{}^{kl}
\delta^{(3)}(\mathbf x-\mathbf y).
\end{equation}
Expanding the field operator in the complete basis of transverse-traceless plane-wave modes introduced above, we write
\begin{equation}
    h_{ij}(t,\mathbf x)
=
\sum_\lambda
\int d^3k\,
C_k\,e^\lambda_{ij}(\mathbf n)
\left[
 g_{\mathbf k,\lambda}
e^{-i(\omega_kt-\mathbf k\cdot\mathbf x)}
+
 g^\dagger_{\mathbf k,\lambda}
e^{i(\omega_kt-\mathbf k\cdot\mathbf x)}
\right],
\end{equation}
Here, $\hat g_{\mathbf k,\lambda}$ and $\hat g^\dagger_{\mathbf k,\lambda}$ are, respectively, the annihilation and creation operators associated with a graviton of wave vector $\mathbf k$ and polarization $\lambda$. Since the polarization tensors are chosen to be real, the second term is the Hermitian conjugate of the first and ensures that $\hat h_{ij}$ is Hermitian. The wave equation fixes the mode functions and the dispersion relation $\omega_k=ck$, but it does not determine the normalization $C_k$. This normalization is fixed by requiring the field operators to satisfy the canonical equal-time commutation relations and picking a convention for the commutator of the mode operators.

We use the Fourier-transform convention
\begin{equation}
f(\mathbf x)
=
\int d^3k\,f(\mathbf k)e^{i\mathbf k\cdot\mathbf x},
\qquad
f(\mathbf k)
=
\frac{1}{(2\pi)^3}
\int d^3x \,f(\mathbf x)e^{-i\mathbf k\cdot\mathbf x},
\end{equation}
so that
\begin{equation}
\delta^{(3)}(\mathbf x-\mathbf y)
=
\frac{1}{(2\pi)^3}
\int d^3k \,
e^{i\mathbf k\cdot(\mathbf x-\mathbf y)}.
\end{equation}

We expect that the commutator of the graviton annihilation and creation operator can be written as
\begin{equation}
    [g_{\mathbf k,\lambda},
g^\dagger_{\mathbf k',\lambda'}]
=
D_k\,
\delta^{(3)}(\mathbf k-\mathbf k')
\delta_{\lambda\lambda'},
\end{equation}
where $D_k$ is to be specified. Requiring $D_k = 1$, and using the commutation relation \eqref{eq_canonical_commutation_relation_h_Pi}, fixes $C_k^2$. Since the quantity under the square root is positive, $C_k$ is real. We conventionally choose the positive square root, since an overall minus sign can be absorbed into the definition of the mode operators thus obtaining,
\begin{equation}
    C_k
=
\sqrt{\frac{G\hbar}{\pi^2c^2\omega_k}}.
\end{equation}
Now that $C_k$ has been fixed, for convenience, we package the normalization and the basis polarization tensor into a single mode tensor
\begin{equation}
    \mathcal G_{ij}^{\mathbf k,\lambda}
\equiv
C_k e^\lambda_{ij}(\mathbf n)
=
\sqrt{\frac{G\hbar}{\pi^2c^2\omega_k}}\,
e^\lambda_{ij}(\mathbf n).
\end{equation}
Because $C_k$ and $e_{ij}^\lambda(\mathbf n)$ are both real, so is the above. Then the field expansion can be written as
\begin{equation}
    h_{ij}(t,\mathbf x)
=
\sum_\lambda
\int d^3k\,
\mathcal G_{ij}^{\mathbf k,\lambda}
g_{\mathbf k,\lambda}
e^{-i(\omega_kt-\mathbf k\cdot\mathbf x)}
+\text{h.c.}.
\end{equation}
This is exactly Eq.~\eqref{eq_hij_in_modes} referred to in the main text. Substituting this into the Hamiltonian \eqref{eq_gravity_hamiltonian_definition}, gives
\begin{equation}
    H^{(g)}
=
\sum_\lambda
\int d^3k\,
\hbar\omega_k\,
g^\dagger_{\mathbf k,\lambda}g_{\mathbf k,\lambda},
\end{equation}
where we have omitted the zero-point contribution. This is exactly Eq.~\eqref{eq_graviton_hamiltonian} referred to in the main text.

\secpar{Box quantization counterpart}
For completeness, we also provide the expressions obtained via box quantization, which we could have started with. Taking a cubic box of side length $L$ and volume $V = L^3$, for the wave vectors we have,
\begin{equation}
    \mathbf k = \frac{2\pi}{L} \mathbf m, \qquad m \in \mathbb{Z}^3.
\end{equation}
The integral and delta function are replaced according to
\begin{equation}
    \int d^3 k \longrightarrow \frac{(2\pi)^3}{V} \sum_k,
    \qquad
    \delta^{(3)}(\mathbf k-\mathbf k')
\longrightarrow
\frac{V}{(2\pi)^3}
\delta_{\mathbf k,\mathbf k'}.
\end{equation}
The (same) Fourier-transform convention, in this discrete case would read
\begin{equation}
    f(\mathbf x)
=
\sum_{\mathbf k}f_{\mathbf k}e^{i\mathbf k\cdot\mathbf x},
\qquad
f_{\mathbf k}
=
\frac1V
\int_V d^3x\,f(\mathbf x)e^{-i\mathbf k\cdot\mathbf x},
\end{equation}
so that
\begin{equation}
    \delta_V^{(3)}(\mathbf x-\mathbf y)
=
\frac1V
\sum_{\mathbf k}
e^{i\mathbf k\cdot(\mathbf x-\mathbf y)}.
\end{equation}
The box-normalized field expansion is then
\begin{equation}
    h_{ij}(t,\mathbf x)
=
\sum_{\mathbf k,\lambda}
\mathcal G_{ij}^{\mathbf k,\lambda;V}
g_{\mathbf k,\lambda}^{V}
e^{-i(\omega_kt-\mathbf k\cdot\mathbf x)}
+\mathrm{h.c.}.
\end{equation}
Here, the box-normalized single mode tensor is related to the continuum counterpart via
\begin{equation}
    \mathcal G_{ij}^{\mathbf k,\lambda;V}
=
\sqrt{\frac{(2\pi)^3}{V}}\,
\mathcal G_{ij}^{\mathbf k,\lambda}.
\end{equation}
Accordingly, the box-normalized mode operator is related to the continuum counterpart via
\begin{equation}
g^V_{\mathbf k,\lambda}
=
\sqrt{\frac{(2\pi)^3}{V}}\,
g_{\mathbf k,\lambda}.
\end{equation}
Using
\begin{equation}
    [g_{\mathbf k,\lambda}^V,
g^{V\dagger}_{\mathbf k',\lambda'}]
=
\delta_{\mathbf k,\mathbf k'}
\delta_{\lambda,\lambda'},
\end{equation}
the discrete Hamiltonian becomes
\begin{equation}
    H^{(g)}
=
\sum_{\mathbf k,\lambda}
\hbar\omega_k
\left(
g^\dagger_{\mathbf k,\lambda}g_{\mathbf k,\lambda}
+\frac12
\right).
\end{equation}
where we have included the zero-point contribution, which can be discarded.

In open quantum systems, the box quantization should be understood as an intermediate regularization rather than as a physical restriction on the gravitational field. The finite quantization volume is introduced only as an intermediate regularization. For a gravitational field propagating in unbounded space, the limit $V \to\infty $, renders the mode spectrum continuous and converts the discrete sum into an integral when evaluating the spectral density. The long-time limit (Markov approximation) of the gravitational correlation function produces the resonance delta function $\delta(\omega_k- 2\omega_m)$. For a finite box, the wave vectors are discrete, so the corresponding expression would be a sum over isolated frequencies; unless one of those frequencies were exactly equal to $2\omega_m$, the delta function would select nothing. The continuum limit, on the other hand, allows the resonance delta function to select the continuous shell of modes satisfying $\omega_k = 2\omega_m$ (cf. Eq.~\eqref{eq_spectral_density_gravitons_thesumlambda_integralk}). Physically, a finite environment has a discrete mode spectrum. An infinite, or effectively infinite, environment instead possesses a continuous spectrum into which excitations can disperse, giving decaying correlation functions and effectively irreversible dynamics. This is the standard continuum-bath treatment used in open quantum systems.

\subsection{Polarization tensor}\label{sec_polarization_tensor}

After fixing the residual gauge freedom, we found that (Eq.~\eqref{eq_polarization_tensor_prime_equal_TT}),
\begin{equation}
    e'_{\mu\nu}=e_{\mu\nu}^{\mathrm{TT}}.
\end{equation}
Suppressing the prime (and the TT superscript), $e_\mn$ denotes a tensor satisfying
\begin{equation}\label{eq_polarization_tensor_three_conditions_kmu_umu}
    \eta^{\mu\nu}e_{\mu\nu}=0,
\qquad
k^\mu e_{\mu\nu}=0,
\qquad
u^\mu e_{\mu\nu}=0.
\end{equation}
For a fixed propagation direction, the space of symmetric tensors satisfying these conditions is two-dimensional. We therefore choose two basis tensors and label them by
\begin{equation}
    \lambda \in \{+,\times\},
\end{equation}
commonly referred to as the plus and cross polarizations. We denote these basis polarization tensors by $e_\mn^\lambda(\mathbf n)$, where $\mathbf n = \mathbf k/k$. Note that the label $\lambda$ is not a spacetime index; it only identifies the two independent polarization states, so we always write the sum over $\lambda$ explicitly. The general TT polarization amplitude can then be decomposed as
\begin{equation}
    e_\mn(\mathbf k) = \sum_{\lambda = +,\times} A_\klambda e_\mn^\lambda(\mathbf n).
\end{equation}
Thus, $e_\mn$ is the complete polarization amplitude for a given wave vector, whereas $e_\mn^\lambda$ are the two real basis tensors used to expand it. The complex coefficients $A_\klambda$ contain the amplitude and phase of each polarization. In the inertial frame defined by $u^\mu$, their temporal components vanish, $e_{0\mu}^\lambda = 0$, so it is sufficient to construct their spatial components, as we do below.

For a gravitational wave traveling in the $\mathbf n$-direction, the polarization tensor must be transverse with respect to $\mathbf n$,
\begin{equation}\label{eq_polten_transverse_conditions}
    n^i e_{ij}^\lambda(\mathbf n) = 0.
\end{equation}
This follows from the second and third equation in Eq.~\eqref{eq_polarization_tensor_three_conditions_kmu_umu} having in mind the decomposition \eqref{eq_kmu_decomposition_timelike_spacelike_u_n} (cf. Eq.~\eqref{eq_transversality_condition_umu_nmu}).
That is, the two basis vectors used to build the polarization tensor must lie in the plane perpendicular to $\mathbf n$.

Let us start with a general unit vector,
\begin{equation}\label{eq_polten_unit_vector}
    \mathbf n = (n_x,n_y,n_z), \quad n\equiv|\mathbf n|^2=n_x^2 + n_y^2+n_z^2 = 1.
\end{equation}
If we define $\theta$ as the angle from the $+z$-axis and $\phi$ as the angle from the $+x$-axis, then, in these spherical coordinates the components of the unit vector $\mathbf n$ are
\begin{equation}\label{eq_polten_spherical_coordinates_unit}
\begin{aligned}
    n_x &= \sin \theta \cos \phi,
    \\
    n_y &= \sin \theta \sin \phi,
    \\
    n_z &= \cos \theta.
\end{aligned}
\end{equation}
Now we can choose two transverse unit vectors $e_\theta \equiv \mathbf x= (x_1,x_2,x_3)$ and $e_\phi \equiv \mathbf y = (y_1,y_2,y_3)$ perpendicular to $\mathbf n$. In general,
\begin{equation}\label{eq_polarization_tensors_in_basis}
\begin{aligned}
    e^+_{ij} &= x_i x_j-y_iy_j, \\
    e^\times_{ij} &= x_i y_j + y_ix_j.
\end{aligned}
\end{equation}
Thus, in summary, they are symmetric, real, transverse, and traceless. For future reference, note that these definitions fix the orthogonality and normalization as
\begin{equation}\label{eq_orthonormality_basis_polarization_tensors}
    e^\lambda_{ij}e^{ij}_\lambdaprime
=
2\delta^\lambda_\lambdaprime.
\end{equation}
And their completeness relation is
\begin{equation}\label{eq_polarization_tensor_completeness_relation}
\sum_{\lambda=+,\times}
e^\lambda_{ij}e^{ kl}_\lambda
=2P_{ij}{}^{kl},
\end{equation}
where $P_{ij}{}^{kl}$ denotes the TT spatial projector given in Eq.~\eqref{eq_TT_projector_spatial}.

At first sight, one seems to have 9 equations for 6 variables. Namely, 6 equations (three for each $\lambda$) from the transverse condition \eqref{eq_polten_transverse_conditions} and 3 more from the orthonormality conditions,
\begin{equation}\label{eq_polten_orthonormality_conditions}
    x_ix^i = 1, \quad y_i y^i = 1, \quad x_i y^i = 0.
\end{equation}
However, from the transverse condition, only 2 equations are independent,
\begin{equation}
    n^i x_i = 0, \quad n^i y_i = 0.
\end{equation}
These two equations which can be derived using the orthonormality conditions \eqref{eq_polten_orthonormality_conditions} and contracting with $x_j$ and $y_j$, satisfy all six equations \eqref{eq_polten_transverse_conditions}. Therefore, together with the orthonormality conditions \eqref{eq_polten_orthonormality_conditions}, we get 5 equations for 6 variables, which is an underdetermined system. Let us make the choice
\begin{equation}
\begin{aligned}
    \mathbf x = e_\theta &= (\cos \theta \cos \phi, \cos \theta \sin \phi, - \sin \theta), \\
    \mathbf y =e_\phi &= (- \sin \phi, \cos \phi, 0).
\end{aligned}
\end{equation}
In this case, for the $(xx)$ component of the polarization tensor \eqref{eq_polarization_tensors_in_basis} for $\lambda = \{+,\times \}$, we obtain,
\begin{equation}\label{eq_polten_e_xx_in_n_direction_spherical}
\begin{aligned}
    e_{xx}^+(\mathbf n) &=\cos^2 \theta \,\cos^2 \phi - \sin^2 \phi, \\
    e_{xx}^\times(\mathbf n) &= - \cos \theta \,\sin 2 \phi.
\end{aligned}
\end{equation}
The chosen basis satisfies all the conditions given by Eqs.~\eqref{eq_polten_transverse_conditions} and \eqref{eq_polten_orthonormality_conditions}. In the case of $z$-direction propagation, i.e.~$\theta = 0$ and $\phi = 0$, we have $\mathbf n = (0,0,1)$ and
\begin{equation}
\begin{aligned}
    \mathbf x  &= (1, 0, 0), \\
    \mathbf y &= (0, 1, 0),
\end{aligned}
\end{equation}
such that
\begin{equation}
\begin{aligned}
    e_{xx}^+(\mathbf n) &=1, \\
    e_{xx}^\times(\mathbf n) &= 0.
\end{aligned}
\end{equation}

\section{Nonrelativistic matter and gravitons: Interaction Hamiltonian}\label{app_interaction_hamiltonian}

Here we provide the expression for the energy-momentum tensor for a point particle after the $\tau$-integration, and three alternative derivations for the interaction Hamiltonian given in Eq.~\eqref{eq_interaction_Hamiltonian_derived}.

\subsection{The energy-momentum tensor for a point particle}\label{sec_energy_momentum_tensor_specialization}

We describe the internal mode as an effective point particle of invariant rest mass $m$. Its covariant energy-momentum tensor in the main text was written as
\begin{equation}\label{eq_app_energy_momentum_covariant}
    T^{\mu\nu}(x)
    =
    c^2\int d\tau\,
    \frac{
        p^\mu(\tau)p^\nu(\tau)
    }{
        \sqrt{-p_\lambda p^\lambda}
    }
    \delta^{(4)}(x-z(\tau)),
\end{equation}
where $z^\mu(\tau)$ is the worldline, $\tau$ is proper time, $u^\mu=dz^\mu/d\tau$, and $p^\mu=m u^\mu$. We use the normalization $\sqrt{-p_\lambda p^\lambda}=m c$ for the invariant mass in this intermediate expression. To reduce Eq.~\eqref{eq_app_energy_momentum_covariant} to the usual form, we split the spacetime delta function as
\begin{equation*}
    \delta^{(4)}(x-z(\tau))
    =
    \delta\!\left(x^0-z^0(\tau)\right)
    \delta^{(3)}\!\left(\mathbf{x}-\mathbf{z}(\tau)\right).
\end{equation*}
For a future-directed worldline $z^0(\tau)$ is monotonic, so for each fixed $x^0$ there is a unique proper time $\tau_{x^0}$ such that $z^0(\tau_{x^0})=x^0$. Using the distribution identity
\begin{equation*}
    \delta\!\left(f(\tau)\right)
    =
    \sum_{\tau_*}
    \frac{\delta(\tau-\tau_*)}{|f'(\tau_*)|},
\end{equation*}
with $f(\tau)=x^0-z^0(\tau)$, we obtain
\begin{equation*}
    \delta\!\left(x^0-z^0(\tau)\right)
    =
    \frac{\delta(\tau-\tau_{x^0})}{d z^0/d\tau|_{\tau_{x^0}}}
    =
    \frac{m}{p^0(\tau_{x^0})}
    \delta(\tau-\tau_{x^0}),
\end{equation*}
where we used $dz^0/d\tau=u^0=p^0/m$ and dropped the absolute value for a future-directed particle. Substituting this into Eq.~\eqref{eq_app_energy_momentum_covariant} gives the proper-time integral explicitly,
\begin{align*}
    T^{\mu\nu}(x)
    &=
    \int d\tau\,
    \frac{p^\mu(\tau)p^\nu(\tau)}{m}
    \delta\!\left(x^0-z^0(\tau)\right)
    \delta^{(3)}\!\left(\mathbf{x}-\mathbf{z}(\tau)\right)
    \\
    &=
    \int d\tau\,
    \frac{p^\mu(\tau)p^\nu(\tau)}{m}
    \frac{m}{p^0(\tau_{x^0})}
    \delta(\tau-\tau_{x^0})
    \delta^{(3)}\!\left(\mathbf{x}-\mathbf{z}(\tau)\right)
    \\
    &=
    \frac{p^\mu(\tau_{x^0})p^\nu(\tau_{x^0})}{p^0(\tau_{x^0})}
    \delta^{(3)}\!\left(\mathbf{x}-\mathbf{z}(\tau_{x^0})\right).
\end{align*}
Finally, denoting the spatial trajectory in the detector frame by $\mathbf{q}(t)$ (to distinguish from $\mathbf x$ which is a dummy variable in the minimal-coupling action), we write~\cite{weinberg_1972_gravitation_and_cosmology_principles_and_applications,bose_mazumdar_schut_ea_2022_mechanism_for_the_quantum_natured}
\begin{equation}\label{eq_app_energy_momentum_equal_time}
    T^{\mu\nu}(t,\mathbf x)
    =
    \frac{p^\mu p^\nu}{p^0}
    \delta^{(3)}
    \!\left(
        \mathbf{x}
        -
        \mathbf{q}(t)
    \right).
\end{equation}
Thus the spatial part of the worldline $\mathbf q(t)$ is taken to be the relative displacement of the two masses in the detector frame.

\subsection{Interaction Hamiltonian from minimal coupling: from TT to FNC}\label{sec_interaction_Hamiltonian_TT_to_FNC}

Let us start with the universal linearized-gravity coupling,
\begin{equation}
    S_I = \frac{1}{2} \int d^4x \, h_\mn T^\mn.
\end{equation}
For a point particle,
\begin{equation}
    T^\mn(t,\mathbf x)= \frac{p^\mu p^\nu}{p^0} \delta^{(3)}(\vec{x}-\vec{q}(t)),
\end{equation}
where we have denoted $p^\mu = m u^\mu$. Evaluating the spatial integral and taking the nonrelativistic limit $u^\mu = (1, \dot{x}^i)$, gives the Lagrangian in terms of the metric perturbation. We then specialize to the TT gauge, where $h_{00}= h_{0i}=0$ and $h_{ij}\neq 0$, and we switch to common notation $q^i \rightarrow x^i$, which yields the Lagrangian,
\begin{equation}
    L_I = \frac{m}{2} h_{ij} \dot{x}^i \dot{x}^j.
\end{equation}
We can eliminate the velocity coupling by adding total derivatives, and using the zeroth-order equation of motion $\ddot{x}_0^j = 0$; since we work only to linear order in the perturbation, any factor of $\ddot{x}^j$ multiplied by $h_{ij}$ is already a second-order contribution. Using the symmetry properties of the perturbation, the remaining mixed terms cancel, leaving the tidal piece
\begin{equation}
    L_I = \frac{m}{4} \ddot{h}_{ij} \,  x^i x^j.
\end{equation}
At this point, we move to Fermi's normal coordinates, evaluating the perturbation at the reference worldline which we choose as $x^i = 0$. Thus, ${h}_{ij}(t,x^i) = {h}_{ij}(t,0)$. Specializing for motion along the $x$-direction, and using $L_I = - H_I$, we finally obtain,
\begin{equation}
    H_I = - \frac{m}{4} \ddot{h}_{xx}(t,0) \,x^2.
\end{equation}
This denotes the interaction Hamiltonian of our theory.

\subsection{Interaction Hamiltonian from minimal coupling: from FNC to TT}

Let us start with the universal linearized-gravity coupling,
\begin{equation}
    S_I = \frac{1}{2} \int d^4x \, h_\mn T^\mn.
\end{equation}
For a point particle,
\begin{equation}
    T^\mn(t,\mathbf x) = \frac{p^\mu p^\nu}{p^0} \delta^{(3)}(\vec{x}-\vec{q}(t)),
\end{equation}
where we have denoted $p^\mu = m u^\mu$. We now take the nonrelativistic limit and go to the FNC coordinates, keeping factors of $v^i \equiv \dot{x}^i $ and $c$ explicit in order to see the leading-order contributions, and we also switch to common notation $q^i \rightarrow x^i$. In the nonrelativistic limit,
\begin{equation}
\begin{aligned}
    T^{00} &= m c^2 \, \delta^{(3)}(\vec{x} - \vec{q}(t)), \\
    T^{0i} &=  m c  v^i \, \delta^{(3)}(\vec{x} - \vec{q}(t)), \\
    T^{ij} &= m v^i v^j \, \delta^{(3)}(\vec{x} - \vec{q}(t)).
\end{aligned}
\end{equation}
We have taken into account that $\gamma= 1/\sqrt{1-v^2/c^2} = 1 +\mathcal{O}(v^2/c^2)$. We construct the Fermi normal coordinates around the reference geodesic, which we take to be $x^i = 0$. This results in the perturbations
\begin{equation}
\begin{aligned}
    h_{00} &= - R_{0i0j} x^i x^j, \\
    h_{0i} &= - \frac{2}{3}R_{0kij}x^k x^j, \\
    h_{ij} &= - \frac{1}{3} R_{ikjl} x^k x^l. 
\end{aligned}
\end{equation}
Plugging the energy-momentum components in the action, after evaluating the spatial integrals we end up with the Lagrangian
\begin{equation}
    L_I = \frac{1}{2} mc^2\left( h_{00} + \frac{v^i}{c} h_{0i} + \frac{v^i v^j}{c^2} h_{ij}   \right).
\end{equation}
This makes it obvious that the leading-order contribution is given by the first term. Using the perturbations in the Fermi normal coordinates, we have
\begin{equation}
    L_I = - \frac{m}{2} R_{0i0j} x^i x^j + \mathcal{O}\left(\frac{v^2}{c^2}\right).
\end{equation}
It is at this point that we consider the TT gauge. Specifically, the Riemann tensor components evaluated along the reference worldline are related to the TT-gauge metric perturbation by
\begin{equation}
R_{0i0j} = -\frac{1}{2c^2} \partial_t^2 h_{ij}.
\end{equation}
This replaces the curvature tensor in the interaction Lagrangian with the metric perturbation. Specializing to motion along the $x$-direction and using $H_I = - L_I$, we finally obtain
\begin{equation}
    H_I = - \frac{m}{4} \ddot{h}_{xx}(ct,0) x^2.
\end{equation}
This denotes the interaction Hamiltonian of our theory.

\subsection{Interaction Hamiltonian from the general-relativistic point-particle Lagrangian}\label{sec_interaction_Hamiltonian_general_relativistic_point_particle}

Let us start with the reparametrization-invariant action of a free relativistic point particle
\begin{equation}
    S=-mc\!\int ds,\qquad ds^2=g_{\mu\nu}\,dx^\mu dx^\nu.
\end{equation}
We choose our coordinates as $x^\mu = (ct, x^i)$ and parametrize the curve by a monotonic affine parameter, which we take to be $\lambda = t$. Switching to proper time $d\tau^2 = -ds^2$, we get the Lagrangian,
\begin{equation}
    L = - mc \sqrt{- g_\mn \dot{x}^\mu \dot{x}^\nu}.
\end{equation}
The dot denotes derivatives with respect to $t$. Let us define $v^i \equiv \dot{x}^i$. Using the Fermi normal coordinates, in the weak-field, slow-motion approximation, we may expand the Lagrangian perturbatively in powers of $v^i/c$ and $h_\mn$. We have
\begin{equation}
    L = - mc^2 \sqrt{-\left(g_{00} + g_{0i} \frac{v^i}{c} + g_{ij} \frac{v^i v^j}{c^2}\right)},
\end{equation}
where the metric components are given by
\begin{equation}
\begin{aligned}
    g_{00} &= -1 - R_{0i0j} x^i x^j, \\
    g_{0i} &= \frac{2}{3}R_{0jik} x^j x^k, \\
    g_{ij} &= \delta_{ij} - \frac{1}{3}R_{i k jl} x^k x^l,
\end{aligned}  
\end{equation}
and the corresponding Riemann tensor components in the TT-gauge are given by
\begin{equation}
\begin{aligned}
    R_{0i0j} &= -\frac{1}{2c^2} \partial_t^2 h_{ij}, \\
    R_{0jik} &= \frac{1}{2c} \left( \partial_t \partial_k h_{ij} - \partial_t \partial_j h_{ik}    \right), \\
    R_{i k jl} &= \frac{1}{2} \left( \partial_k \partial_l h_{ij} - \partial_i \partial_j h_{kl} - \partial_i \partial_l h_{kj} - \partial_k \partial_j h_{il}   \right).
\end{aligned}
\end{equation}
In Fermi's normal coordinates, it is understood that all curvature tensors are evaluated on the reference worldline, which in our case we choose as $x^i = 0$. Upon inspection, one can see that the leading-order contributions in the Lagrangian consist of the rest-energy term, the nonrelativistic kinetic term, and the weak curvature corrections with the dominant curvature contribution originating from the quadratic term in $g_{00}$. The latter represents the interaction Lagrangian $L_I = - H_I$, from which the corresponding interaction Hamiltonian directly follows. By taking the metric perturbation to represent a plane gravitational wave propagating along the $x$-direction, $h_{ij}(ct, x) = h_{ij}(ct - x)$, we finally obtain
\begin{equation}
    H_I = - \frac{m}{4}  \ddot{h}_{xx}(ct,0) \, x^2
\end{equation}
This is the term that captures the interaction induced by the passing gravitational wave, describing the relative geodesic deviation  with respect to the geodesic reference worldline.

\section{State of gravitons: Expectation values and dissipators}\label{eq_expectation_values_dissipators}

For technical reasons, we distinguish between a single-mode state and a single-mode state defined within a multi-mode system. For brevity, we refer to the latter as a single multi-mode state. This terminology is formal: although the system is formally treated as multi-mode, with all modes labeled by $(k, \lambda)$, only a single mode $(k_0,\lambda_0)$ is occupied, while all others remain in the vacuum.
Such a monochromatic occupation should be understood as an idealized limit of a wave-packet state, which provides the more physical description.
Nevertheless, this formal construction is useful here because retaining the multi-mode indices allows us to track operator labels and delta functions in subsequent expressions, ensuring clear bookkeeping of mode labels. For clarity of notation and convenience, we reserve the mode label $(k_0,\lambda_0)$ for the state, and use primed labels for the operators.
With this convention, we consider a comprehensive set of graviton states and derive the corresponding expectation values. In addition, for the multi-mode states, we additionally derive the resulting master equations.

The graviton annihilation and creation operators retain their explicit mode labels and satisfy the standard bosonic commutation relations (Section~\ref{app_gravity_hamiltonian}),
\begin{equation}\label{eq_comm_rel}
\begin{aligned}
    \comm*{g_\klambdaprime}{g_\klambdadoubleprime^\dagger}
    &= \deltathree(\kprime-\kdoubleprime)
    \delta_\lambdaprime^\lambdadoubleprime, \\
    \comm*{g_\klambdaprime}{g_\klambdadoubleprime} &= 0, \\
    \comm*{g_\klambdaprime^\dagger}{g_\klambdadoubleprime^\dagger} &= 0.
\end{aligned}
\end{equation}
The Kronecker and Dirac deltas ensure that only operators associated with the same mode fail to commute.

\subsection{Vacuum state}

The vacuum state is defined by
\begin{equation}
    g_\klambdaprime \ket{0} = 0.
\end{equation}
Using the above, together with its Hermitian conjugate, we quickly note that the one-point correlation functions vanish,
\begin{subequations}\label{eq_vac_exp_vals_onepoint}
\begin{align}
    \bra{0} g_\klambdaprime\ket{0} &= 0, \\
    \bra{0}g_\klambdaprime^\dagger\ket{0} &=  0 .
\end{align}    
\end{subequations}
Employing, in addition, the commutation relations \eqref{eq_comm_rel}, for the two-point correlation functions we immediately find
\begin{subequations}\label{eq_vac_exp_vals}
\begin{align}
    \bra{0}g_\klambdaprime^\dagger g_\klambdadoubleprime\ket{0} &= 0, \\
    \bra{0}g_\klambdaprime g_\klambdadoubleprime^\dagger \ket{0} &=  \deltathree(\kprime - \kdoubleprime) \, \delta_\lambdaprime^\lambdadoubleprime,  \\
    \bra{0}g_\klambdaprime g_\klambdadoubleprime \ket{0} &= 0, \\
    \bra{0}g_\klambdaprime^\dagger g_\klambdadoubleprime^\dagger \ket{0} &= 0.
\end{align}
\end{subequations}

The two-point bath correlator \eqref{eq_two_point_bath_conn_or_not_conn}, is therefore
\begin{equation}
    D^{(g)}(\tau) = \sum_\lambda\int d^3k\,
    (G_{\mathbf{k}}^\lambda)^2
    e^{-i\omega_k\tau}.
\end{equation}

Inserting the above into Eq.~\eqref{eq_master_equation_D}, after the secular approximation (Section~\ref{sec_secular_approximation}) and the renormalization procedure (Section~\ref{sec_lamb_shift_Hamiltonian_renormalized}), and transforming back to the Schrödinger picture, the Lindblad master equation for vacuum ($\mathcal{V}$) state gravitons reads
\begin{equation}
    \dot{\rho}^\order{m}_\mathcal{V}(t) =-\frac{i}{\hbar} [H^\order{m}, \rho_\mathcal{V}^\order{m}(t)]
+\gamma_{\mathcal{V}} \,\mathcal{D}[L_{\mathcal{V}}]\,\rho_\mathcal{V}^\order{m}(t)
\end{equation}
with
\begin{alignat}{2}
\gamma_{\mathcal V}
&=\frac{32}{15}t_P^2 \omega_m^3
&\qquad
L_{\mathcal V}
&=
b^2.
\end{alignat}
We have denoted the Planck time by
\begin{equation}
    t_P = \sqrt{\frac{\hbar G}{c^5}}.
\end{equation}

\subsection{Single multi-mode number state}

For an orthonormal single multi-mode number state $\ket{n_\klambdazero}$, the annihilation operator acts as
\begin{align}
    g_\klambdaprime \ket{n_\klambdazero} &= \sqrt{n_\klambdazero} \, \deltathree(\kzero-\kprime)\, \delta_\lambdazero^\lambdaprime \,\ket{n_\klambdazero - 1}.
\end{align}
This, together with its Hermitian conjugate and the commutation relations \eqref{eq_comm_rel}, give the expectation values
\begin{subequations}
\begin{align}
    \bra{n_\klambdazero}g_\klambdaprime^\dagger g_\klambdadoubleprime\ket{n_\klambdazero} &= n_\klambdazero \, \deltathree(\kprime-\kzero) \,\delta_\lambdaprime^\lambdazero \, \deltathree(\kdoubleprime-\kzero) \,\delta_\lambdadoubleprime^\lambdazero,  \\
    \bra{n_\klambdazero}g_\klambdaprime g_\klambdadoubleprime^\dagger \ket{n_\klambdazero} &=  \deltathree(\kprime - \kdoubleprime) \, \delta_\lambdaprime^\lambdadoubleprime  + n_\klambdazero \,\deltathree(\kprime-\kzero) \,\delta_\lambdaprime^\lambdazero \, \deltathree(\kdoubleprime-\kzero) \,\delta_\lambdadoubleprime^\lambdazero , \\
    \bra{n_\klambdazero}g_\klambdaprime g_\klambdadoubleprime \ket{n_\klambdazero} &= 0, \\
    \bra{n_\klambdazero}g_\klambdaprime^\dagger g_\klambdadoubleprime^\dagger \ket{n_\klambdazero} &= 0.
\end{align}    
\end{subequations}
For completeness, we note that the one-point correlation functions vanish,
\begin{subequations}\label{eq_sm_number_onepoint}
\begin{align}
    \bra{n_\klambdazero} g_\klambdaprime\ket{n_\klambdazero} &= 0, \\
    \bra{n_\klambdazero}g_\klambdaprime^\dagger\ket{n_\klambdazero} &=  0.  
\end{align}    
\end{subequations}

\subsection{Multi-mode number state}

For an orthonormal multi-mode number state $\ket{\underline{n}} \equiv \bigotimes_{(\klambdazero)} \ket{n_\klambdazero}$, the annihilation operator acts as
\begin{align}
    g_\klambdaprime \ket{\underline{n}} &= \sqrt{n_\klambdaprime} \,  \ket{\;\cdots \; n_\klambdaprime - 1 \;\cdots\;} .
\end{align}
This, together with its Hermitian conjugate and the commutation relations \eqref{eq_comm_rel}, give the two-point correlation functions
\begin{subequations}
\begin{align}
    \bra{\underline{n}}g_\klambdaprime^\dagger g_\klambdadoubleprime\ket{\underline{n}} &= n_\klambdaprime \, \deltathree(\kprime-\kdoubleprime) \,\delta_\lambdaprime^\lambdadoubleprime  ,\\
    \bra{\underline{n}}g_\klambdaprime g_\klambdadoubleprime^\dagger \ket{\underline{n}} &=  \deltathree(\kprime-\kdoubleprime) \, \delta_\lambdaprime^\lambdadoubleprime + n_\klambdaprime \, \deltathree(\kprime-\kdoubleprime) \,\delta_\lambdaprime^\lambdadoubleprime  ,\\
    \bra{\underline{n}}g_\klambdaprime g_\klambdadoubleprime \ket{\underline{n}} &= 0, \\
    \bra{\underline{n}}g_\klambdaprime^\dagger g_\klambdadoubleprime^\dagger \ket{\underline{n}} &= 0.
\end{align}    
\end{subequations}
For completeness, we note that the one-point correlation functions vanish,
\begin{subequations}\label{eq_mm_number_onepoint}
\begin{align}
    \bra{\underline{n}} g_\klambdaprime\ket{\underline{n}} &= 0 ,\\
    \bra{\underline{n}}g_\klambdaprime^\dagger\ket{\underline{n}} &=  0.  
\end{align}    
\end{subequations}

The one-point bath correlator \eqref{eq_onepoint_correlation_function} vanishes. The two-point bath correlator \eqref{eq_two_point_bath_conn_or_not_conn}, is therefore
\begin{equation}
    D^{(g)}(\tau) = \sum_\lambda\int d^3k\,
    (G_{\mathbf{k}}^\lambda)^2
    \left[
        \left(1+n_\klambda\right)e^{-i\omega_k\tau}
        +
        n_\klambda e^{i\omega_k\tau}
    \right].
\end{equation}

Let us suppose that the graviton occupation number is isotropic and polarization independent, $n_\klambda = n(\omega_k)$. Then, inserting the above into Eq.~\eqref{eq_master_equation_D}, after the secular approximation (Section~\ref{sec_secular_approximation}) and the renormalization procedure (Section~\ref{sec_lamb_shift_Hamiltonian_renormalized}), and transforming back to the Schrödinger picture, the Lindblad master equation for multi-mode number ($\mathcal{N}$) state gravitons reads
\begin{equation}
    \dot{\rho}^\order{m}_\mathcal{N}(t) =-\frac{i}{\hbar} [H^\order{m}, \rho_\mathcal{N}^\order{m}(t)]
+\gamma_{\mathcal{N},\downarrow} \,\mathcal{D}[L_{\mathcal{N},\downarrow}]\,\rho_\mathcal{N}^\order{m}(t)
+ \gamma_{\mathcal{N},\uparrow} \,\mathcal{D}[L_{\mathcal{N},\uparrow}]\,\rho_\mathcal{N}^\order{m}(t).
\end{equation}
with
\begin{alignat}{2}
\gamma_{\mathcal N,\downarrow}
&=
\gamma_{\mathcal V}\left[1+n(2\omega_m)\right],
&\qquad
L_{\mathcal N,\downarrow}
&=
b^2,
\\
\gamma_{\mathcal N,\uparrow}
&=
\gamma_{\mathcal V}n(2\omega_m),
&
L_{\mathcal N,\uparrow}
&=
b^{\dagger 2}.
\end{alignat}
Here $n(2\omega)$ is the dimensionless Fock occupation number of the resonant graviton modes, i.e.~the value of the occupation number $n(\omega_k)$ evaluated on shell at $\omega_k = 2 \omega_m$.

\subsection{Single multi-mode coherent state}

The coherent state can be defined in multiple ways. For a single multi-mode coherent state we write
\begin{equation}
    \ket{\alpha_\klambdazero} = D_\klambdazero \ket{0},
\end{equation}
where the displacement operator is given by
\begin{equation}\label{eq_displacement_operator}
    D_\klambdazero = e^{\alpha_\klambdazero \,g_\klambdazero^\dagger - \alpha_\klambdazero^* \, g_\klambdazero}.
\end{equation}
Note the unitarity of the displacement operator,
\begin{equation}\label{eq_D_is_unitary}
    D_\klambdazero^\dagger D_\klambdazero = D_\klambdazero D_\klambdazero^\dagger = \onematrix.
\end{equation}
In the number-state basis, the state can be written as
\begin{equation}
    \ket{\alpha_\klambdazero} = e^{-\frac{1}{2}\abs{\alpha_\klambdazero}^2}
    \sum_{n_\klambdazero = 0}^\infty \frac{\alpha_\klambdazero^n}{\sqrt{n_\klambdazero!}} \ket{n_\klambdazero}.
\end{equation}
Importantly, a coherent state is an eigenstate of the annihilation operator associated with its mode. More generally, for a single multi-mode coherent state $\ket{\alpha_\klambdazero}$, the annihilation operator $g_\klambdaprime$ acts as
\begin{equation}
    g_\klambdaprime \ket{\alpha_\klambdazero} = \alpha_\klambdazero \, \deltathree(\kprime - \kzero)\, \delta_\lambdaprime^\lambdazero \ket{\alpha_\klambdazero}.
\end{equation}
Using this, together with its Hermitian conjugate and the commutation relations \eqref{eq_comm_rel}, we obtain
\begin{subequations}\label{eq_sm_coherent_state_exp}
\begin{align}
    \bra{\alpha_\klambdazero}g_\klambdaprime^\dagger g_\klambdadoubleprime\ket{\alpha_\klambdazero} &= \abs{\alpha_\klambdazero}^2 \, \deltathree(\kprime-\kzero) \,\delta_\lambdaprime^\lambdazero \, \deltathree(\kdoubleprime-\kzero) \,\delta_\lambdadoubleprime^\lambdazero , \\
    \bra{\alpha_\klambdazero}g_\klambdaprime g_\klambdadoubleprime^\dagger \ket{\alpha_\klambdazero} &=  \deltathree(\kprime-\kdoubleprime) \,\delta_\lambdaprime^\lambdadoubleprime  + \abs{\alpha_\klambdazero}^2 \, \deltathree(\kprime-\kzero) \,\delta_\lambdaprime^\lambdazero \, \deltathree(\kdoubleprime-\kzero) \,\delta_\lambdadoubleprime^\lambdazero  ,\\
    \bra{\alpha_\klambdazero}g_\klambdaprime g_\klambdadoubleprime \ket{\alpha_\klambdazero} &= \alpha_\klambdazero^2 \, \deltathree(\kprime-\kzero) \,\delta_\lambdaprime^\lambdazero \, \deltathree(\kdoubleprime-\kzero) \,\delta_\lambdadoubleprime^\lambdazero  ,\\
    \bra{\alpha_\klambdazero}g_\klambdaprime^\dagger g_\klambdadoubleprime^\dagger \ket{\alpha_\klambdazero} &= \alpha_\klambdazero^{*2} \, \deltathree(\kprime-\kzero) \,\delta_\lambdaprime^\lambdazero \, \deltathree(\kdoubleprime-\kzero) \,\delta_\lambdadoubleprime^\lambdazero .
\end{align}    
\end{subequations}
Equivalently, one can derive the expectation values using the similarity transformation,
\begin{equation}\label{eq_sm_coherent_similarity_transformation}
    D^\dagger_\klambdazero g_\klambdaprime D_\klambdazero = g_\klambdaprime + \alpha_\klambdazero \, \deltathree(\kprime - \kzero)\,\delta_\lambdaprime^\lambdazero.
\end{equation}
This follows from the definition of the displacement operator \eqref{eq_displacement_operator} via the Baker–Campbell–Hausdorff formula, evaluated using the canonical commutation relations \eqref{eq_comm_rel}. Using this, its Hermitian conjugate, the unitarity property \eqref{eq_D_is_unitary} and the vacuum results \eqref{eq_vac_exp_vals}, one arrives at the same expectation values \eqref{eq_sm_coherent_state_exp}.

For completeness, we note that, importantly, one finds nonvanishing one-point correlation functions,
\begin{subequations}\label{eq_sm_coherent_onepoint}
\begin{align}
    \bra{\alpha_\klambdazero} g_\klambdaprime\ket{\alpha_\klambdazero} &= \alpha_\klambdazero \, \deltathree(\kprime-\kzero)\,\delta_\lambdaprime^\lambdazero ,\\
    \bra{\alpha_\klambdazero}g_\klambdaprime^\dagger\ket{\alpha_\klambdazero} &=  \alpha_\klambdazero^* \, \deltathree(\kprime-\kzero)\,\delta_\lambdaprime^\lambdazero  .
\end{align}    
\end{subequations}

The one-point correlation function \eqref{eq_onepoint_correlation_function}, is therefore
\begin{equation}\label{eq_one_point_single_coherent}
    C^{(g)}(t) = G_{\mathbf k_0}^{\lambda_0}(\alpha_\klambdazero e^{- i \omega_\kzero t} + \alpha_\klambdazero^* e^{+i \omega_\kzero t}).
\end{equation}
We can also parametrize (we also omit the indices to avoid clutter)
\begin{equation}
    \alpha_\klambdazero = |\alpha_\klambdazero| e^{i \phi_\klambdazero} \equiv |\alpha|e^{i\phi_\alpha}.
\end{equation}
Then
\begin{equation}\label{eq_one_point_parametrized_coherent_single}
    C^{(g)}(t) = 2 G_\kzero^\lambdazero |\alpha| \cos(\omega_{k_0} t - \theta).
\end{equation}
The (connected) two-point bath correlation function is identical to the one in the vacuum case.

\subsection{Multi-mode coherent state}

For a multi-mode coherent state $\ket{\underline{\alpha}}\equiv\bigotimes_{(\klambdazero)} \ket{\alpha_\klambdazero}$, the annihilation operator acts as
\begin{equation}
    g_\klambdaprime \ket{\underline{\alpha}} = \alpha_\klambdaprime \ket{\underline{\alpha}}.
\end{equation}
Using this, its Hermitian conjugate, and the commutation relations \eqref{eq_comm_rel}, we obtain
\begin{subequations}\label{eq_mm_coherent_state_exp}
\begin{align}
    \bra{\underline{\alpha}}g_\klambdaprime^\dagger g_\klambdadoubleprime\ket{\underline{\alpha}} &= \alpha_\klambdaprime^* \, \alpha_\klambdadoubleprime  ,\\
    \bra{\underline{\alpha}}g_\klambdaprime g_\klambdadoubleprime^\dagger \ket{\underline{\alpha}} &=  \deltathree(\kprime-\kdoubleprime) \delta_\lambdaprime^\lambdadoubleprime  + \alpha_\klambdaprime \, \alpha_\klambdadoubleprime^*,  \\
    \bra{\underline{\alpha}}g_\klambdaprime g_\klambdadoubleprime \ket{\underline{\alpha}} &= \alpha_\klambdaprime \, \alpha_\klambdadoubleprime,  \\
    \bra{\underline{\alpha}}g_\klambdaprime^\dagger g_\klambdadoubleprime^\dagger \ket{\underline{\alpha}} &= \alpha_\klambdaprime^* \, \alpha_\klambdadoubleprime^*.
\end{align}    
\end{subequations}
Equivalently, one can derive the expectation values using the (simpler) similarity transformation (same mode),
\begin{equation}\label{eq_mm_coherent_similarity_transformation}
    D^\dagger_\klambdaprime g_\klambdaprime D_\klambdaprime = g_\klambdaprime + \alpha_\klambdaprime.
\end{equation}
Using this, its Hermitian conjugate, the unitarity property \eqref{eq_D_is_unitary} and the vacuum results \eqref{eq_vac_exp_vals}, one arrives at the same expectation values \eqref{eq_mm_coherent_state_exp}.

For completeness, we note that, importantly, one finds nonvanishing one-point correlation functions,
\begin{subequations}\label{eq_mm_coherent_onepoint}
\begin{align}
    \bra{\underline{\alpha}} g_\klambdaprime\ket{\underline{\alpha}} &= \alpha_\klambdaprime, \\
    \bra{\underline{\alpha}}g_\klambdaprime^\dagger\ket{\underline{\alpha}} &=  \alpha_\klambdaprime^*.  
\end{align}    
\end{subequations}

The one-point correlation function \eqref{eq_onepoint_correlation_function}, is therefore
\begin{equation}\label{eq_one_point_multi_mode_coherent}
C^{(g)}(t)
=
\sum_\lambda \int d^3 k \,
G_{\mathbf k}^\lambda
\left(
\alpha_{\mathbf k,\lambda} e^{-i\omega_k t}
+
\alpha_{\mathbf k,\lambda}^* e^{+i\omega_k t}
\right).
\end{equation}
It is worthwhile noting that for a coherent amplitude profile supported only on one mode, $\alpha_\klambda = \alpha_\klambdazero \deltathree(\mathbf k - \kzero) \delta_{\lambda_0}^\lambda$, this formally reduces to the one-point correlation function for a single multi-mode coherent state given in Eq.~\eqref{eq_one_point_single_coherent}.

The (connected) two-point bath correlation function is identical to the one in the vacuum case.

As explained in Footnote~\ref{footnote_one_point_function}, the nonvanishing one-point function produces an additional coherent Hamiltonian contribution, while the connected two-point bath correlator coincides with the vacuum result in Eq.~\eqref{eq_two_point_bath_conn_or_not_conn}. Thus, inserting the above results for the one-point correlation functions and the (connected) two-point correlation function, the Lindblad master equation for the multi-mode coherent ($\mathcal{C}$) state gravitons reads,
\begin{equation}
\dot{\rho}_{\mathcal C}^{(m)}(t)
=
-\frac{i}{\hbar}
\left[
H^{(m)}+H_{\mathrm{CS}}(t),
\rho_{\mathcal C}^{(m)}(t)
\right]
+
\gamma_{\mathcal C}
\mathcal D[L_{\mathcal C}]
\rho_{\mathcal C}^{(m)}(t),
\qquad
\gamma_{\mathcal C}=\gamma_{\mathcal V},
\qquad
L_{\mathcal C}=b^2.
\end{equation}
Here, the coherent-shift Hamiltonian is given by
\begin{equation}\label{eq_coherent_shift_Hamiltonian_multimode}
\begin{aligned}
    H_\text{CS}(t) &\equiv H_I^{(m)} C^{(g)}(t) \\
    &=H_I^{(m)} \bra{\underline{\alpha}}
H_{II}^{(g)}(t)
\ket{\underline{\alpha}} \\
&=
(b+b^\dagger)^2
\sum_\lambda\int d^3k\,
G_{\mathbf k}^{\lambda}
\left(
\alpha_{\mathbf k,\lambda}e^{-i\omega_kt}
+
\alpha_{\mathbf k,\lambda}^*e^{i\omega_kt}
\right).
\end{aligned}
\end{equation}

\subsection{Single multi-mode squeezed vacuum state}

A single multi-mode squeezed vacuum state is defined as
\begin{equation}
    \ket{\xi_\klambdazero} = S_\klambdazero \ket{0},
\end{equation}
where the squeezing operator is given by
\begin{equation}\label{eq_squeezing_operator}
    S_\klambdazero = e^{\frac{1}{2} \,\xi_\klambdazero^*  \, g_\klambdazero^2 - \frac{1}{2}\, \xi_\klambdazero \, g_\klambdazero^{\dagger2} }.
\end{equation}
Note the unitarity of the squeezing operator,
\begin{equation}\label{eq_S_is_unitary}
    S_\klambdazero^\dagger S_\klambdazero = S_\klambdazero S_\klambdazero^\dagger = \onematrix.
\end{equation}
As is customary, we will use the parametrization
\begin{equation}\label{eq_squeezing_parametrization}
    \xi_\klambdazero = r_\klambdazero e^{i  \phi_\klambdazero}.
\end{equation}
To compute expectation values in the squeezed vacuum state, we reduce the problem to evaluating vacuum expectation values in the vacuum, by making use of the similarity transformation
\begin{align}\label{eq_sm_squeezing_similarity_transformation}
    S_\klambdazero^{\dagger}\,g_{\klambdaprime}\,S_\klambdazero
&=
g_\klambdaprime\,\left[1-\delta(\kprime-\kzero) \delta_{\lambda'}^\lambdazero\right] +
\Bigl(g_\klambdaprime \,u_\klambdazero
+g_{k'\lambda'}^{\dagger} \, v_\klambdazero\Bigr)\,
\delta(\kprime-\kzero) \delta_{\lambda'}^\lambdazero,
\end{align}
where we have defined
\begin{subequations}\label{eq_u_and_v}
\begin{align}
    u_\klambdazero &\equiv \cosh r_\klambdazero ,\\
    v_\klambdazero &\equiv -e^{i \phi_\klambdazero} \sinh r_\klambdazero.
\end{align}
\end{subequations}
The similarity transformation \eqref{eq_sm_squeezing_similarity_transformation} follows from the definition of the squeezing operator \eqref{eq_squeezing_operator} with the parametrization \eqref{eq_squeezing_parametrization} via the Baker–Campbell–Hausdorff formula, evaluated using the canonical commutation relations \eqref{eq_comm_rel}. Alternatively, it can be derived using the Bogoliubov transformations. Using this, its Hermitian conjugate, the unitarity property \eqref{eq_S_is_unitary}, and the vacuum results \eqref{eq_vac_exp_vals}, we obtain
\begin{subequations}
\begin{align}
    \bra{\xi_\klambdazero}g_\klambdaprime^\dagger g_\klambdadoubleprime\ket{\xi_\klambdazero} &=N_\klambdazero \, \deltathree(\kprime-\kzero) \,\delta_\lambdaprime^\lambdazero \, \deltathree(\kdoubleprime-\kzero) \, \delta_\lambdadoubleprime^\lambdazero \,\deltathree(\kprime -\kdoubleprime) \, \delta_\lambdaprime^\lambdadoubleprime ,\\
    \bra{\xi_\klambdazero}g_\klambdaprime g_\klambdadoubleprime^\dagger \ket{\xi_\klambdazero} &=  \deltathree(\kprime-\kdoubleprime) \,\delta_\lambdaprime^\lambdadoubleprime  \\&\;\;\;\;\;+ N_\klambdazero \, \deltathree(\kprime-\kzero) \, \delta_\lambdaprime^\lambdazero \, \deltathree(\kdoubleprime-\kzero) \,\delta_\lambdadoubleprime^\lambdazero \, \deltathree(\kprime -\kdoubleprime)  \,\delta_\lambdaprime^\lambdadoubleprime \nonumber,\\
    \bra{\xi_\klambdazero}g_\klambdaprime g_\klambdadoubleprime \ket{\xi_\klambdazero} &= M_\klambdazero \, \deltathree(\kprime-\kzero) \, \delta_\lambdaprime^\lambdazero \, \deltathree(\kdoubleprime-\kzero) \, \delta_\lambdadoubleprime^\lambdazero  \, \deltathree(\kprime-\kdoubleprime) \, \delta_\lambdaprime^\lambdadoubleprime  ,\\
    \bra{\xi_\klambdazero}g_\klambdaprime^\dagger g_\klambdadoubleprime^\dagger \ket{\xi_\klambdazero} &= M_\klambdazero^* \, \deltathree(\kprime-\kzero) \, \delta_\lambdaprime^\lambdazero \, \deltathree(\kdoubleprime-\kzero) \, \delta_\lambdadoubleprime^\lambdazero  \, \deltathree(\kprime-\kdoubleprime) \, \delta_\lambdaprime^\lambdadoubleprime.
\end{align}    
\end{subequations}
where we have defined, following standard literature,
\begin{subequations}\label{eq_N_and_M}
\begin{align}
    N_\klambdazero &\equiv \sinh^2 r_\klambdazero, \\
    M_\klambdazero &\equiv - e^{i \phi_\klambdazero} \sinh r_\klambdazero \cosh r_\klambdazero.
\end{align}
\end{subequations}
Note the relations between \eqref{eq_u_and_v} and \eqref{eq_N_and_M},
\begin{subequations}\label{eq_N_M_u_v_relations}
\begin{align}
    N_\klambdazero &= \abs{v_\klambdazero}^2, \\
    M_\klambdazero &= v_\klambdazero \, u_\klambdazero.
\end{align}
\end{subequations}
For completeness, we note that the one-point correlation functions vanish,
\begin{subequations}\label{eq_sm_squeezed_onepoint}
\begin{align}
    \bra{\xi_\klambdazero} g_\klambdaprime\ket{\xi_\klambdazero} &= 0 ,\\
    \bra{\xi_\klambdazero}g_\klambdaprime^\dagger\ket{\xi_\klambdazero} &=  0.  
\end{align}    
\end{subequations}

\subsection{Multi-mode squeezed vacuum state}

For a multi-mode squeezed vacuum state $\ket{\underline{\xi}} \equiv \bigotimes_{(\klambdazero)} \ket{\xi_\klambdazero}$, one finds the (simpler) similarity transformation (same mode),
\begin{align}
    S_\klambdaprime^{\dagger}\,g_{\klambdaprime}\,S_\klambdaprime
&=
g_\klambdaprime\,u_\klambdaprime
+g_{k'\lambda'}^{\dagger}\,v_\klambdaprime,
\end{align}
where $u_\klambdaprime$ and $v_\klambdaprime$ are defined in \eqref{eq_u_and_v}. Using the above, its Hermitian conjugate, the unitarity property \eqref{eq_S_is_unitary}, and the vacuum results \eqref{eq_vac_exp_vals}, we obtain,
\begin{subequations}
\begin{align}
    \bra{\underline{\xi}}g_\klambdaprime^\dagger g_\klambdadoubleprime\ket{\underline{\xi}} &=N_\klambdaprime \, \deltathree(\kprime -\kdoubleprime) \, \delta_\lambdaprime^\lambdadoubleprime, \\
    \bra{\underline{\xi}}g_\klambdaprime g_\klambdadoubleprime^\dagger \ket{\underline{\xi}} &=  \deltathree(\kprime-\kdoubleprime) \,\delta_\lambdaprime^\lambdadoubleprime  + N_\klambdaprime \, \deltathree(\kprime -\kdoubleprime) \, \delta_\lambdaprime^\lambdadoubleprime ,\\
    \bra{\underline{\xi}}g_\klambdaprime g_\klambdadoubleprime \ket{\underline{\xi}} &= M_\klambdaprime \, \deltathree(\kprime-\kdoubleprime) \, \delta_\lambdaprime^\lambdadoubleprime  ,\\
    \bra{\underline{\xi}}g_\klambdaprime^\dagger g_\klambdadoubleprime^\dagger \ket{\underline{\xi}} &= M_\klambdaprime^* \,  \deltathree(\kprime-\kdoubleprime) \, \delta_\lambdaprime^\lambdadoubleprime.
\end{align}    
\end{subequations}
where $N_\klambdaprime$ and $M_\klambdaprime$ are given by \eqref{eq_N_and_M}.

For completeness, we note that the one-point correlation functions vanish,
\begin{subequations}\label{eq_mm_squeezed_onepoint}
\begin{align}
    \bra{\underline{\xi}} g_\klambdaprime\ket{\underline{\xi}} &= 0 ,\\
    \bra{\underline{\xi}}g_\klambdaprime^\dagger\ket{\underline{\xi}} &=  0.  
\end{align}    
\end{subequations}

The two-point bath correlator \eqref{eq_two_point_bath_conn_or_not_conn}, is therefore
\begin{equation}
    D^{(g)} (t, t-\tau) = \sum_\lambda \int d^3 k\, (G_\mathbf k^\lambda)^2[(1 + N_\klambda)e^{- i \omega_k \tau} + N_\klambda e^{+ i \omega_k \tau} + M_\klambda e^{-2 i \omega_k t}e^{+i \omega_k \tau} + M_\klambda^* e^{+ 2 i \omega_k t}e^{- i \omega_k \tau}].
\end{equation}

Let us suppose that the graviton normal and anomalous contributions are isotropic and polarization independent, such that $N_\klambda = N(\omega_k)$ and $M_\klambda = M(\omega_k)$. In addition, let us express them in terms of the squeezing strength $r(\omega_k)$ and the squeezing phase $\phi(\omega_k)$ via the relations given in Eq.~\eqref{eq_N_and_M}. Then, inserting the above into Eq.~\eqref{eq_master_equation_D}, after the secular approximation (Section~\ref{sec_secular_approximation}), the renormalization procedure (Section~\ref{sec_lamb_shift_Hamiltonian_renormalized}), and diagonalization of the Kossakowski matrix (Section~\ref{sec_secular_approximation}), and transforming back to the Schrödinger picture, the Lindblad master equation for multi-mode squeezed ($\mathcal{S}$) state gravitons reads
\begin{equation}
    \dot{\rho}^\order{m}_\mathcal{S}(t) =-\frac{i}{\hbar} [H^\order{m}, \rho_\mathcal{S}^\order{m}(t)]
+\gamma_{\mathcal{S}} \,\mathcal{D}[L_{\mathcal{S}}]\,\rho_\mathcal{S}^\order{m}(t)
\end{equation}
with
\begin{alignat}{2}
\gamma_{\mathcal S}
&=\gamma_\mathcal{V} =\frac{32}{15}t_P^2 \omega_m^3
&\qquad
L_{\mathcal S}
&=
\cosh r(2\omega_m)\,b^2 - e^{i\phi(2\omega_m)}\, \sinh r(2\omega_m)\,b^{\dagger2}.
\end{alignat}
Here $r(2\omega_m)$ and $\phi(2\omega_m)$ are, respectively, the squeezing parameter and squeezing phase of the resonant graviton modes, i.e.~the functions $r(\omega_k)$ and $\phi(\omega_k)$ evaluated on shell at $\omega_k = 2\omega_m$.

\subsection{Single and multi-mode thermal state}

A thermal state cannot, of course, be defined in terms of a ket state. It is, by definition, a mixture of pure states and it can only be defined via the density operator. Let us directly consider the multi-mode thermal state. The multi-mode thermal state is fundamentally the Gibbs state of the complete free gravitational Hamiltonian \eqref{eq_graviton_hamiltonian},
\begin{equation}
    \rho^\beta
\equiv
\frac{1}{Z_g}e^{-\beta H^{(g)}},
\qquad
Z_g
\equiv
\operatorname{tr}_g
\left(e^{-\beta H^{(g)}}\right),
\end{equation}
where $\beta= 1/(k_B T)$ denotes the inverse temperature with the Boltzmann's constant denoted by $k_B$. However, the free gravitational Hamiltonian is a sum of independent mode Hamiltonians (one may temporarily think of $k$ as being discrete and only later take the continuum limit),
\begin{equation}
    H^{(g)}
=
\sum_{\lambda_0}
\int d^3k_0\,H^{(g)}_{\klambdazero}
,
\qquad
H^{(g)}_{\klambdazero}=
\hbar\omega_{k_0}
g_{\mathbf k_0,\lambda_0}^\dagger
g_{\mathbf k_0,\lambda_0}
\end{equation}
Consequently, the Gibbs state factorizes mode by mode
\begin{equation}
    \rho^\beta
=
\bigotimes_{(\mathbf k_0,\lambda_0)}
\rho_{\mathbf k_0,\lambda_0}^\beta.
\end{equation}
A single multi-mode thermal state is then defined as
\begin{equation}
    \rho_{\mathbf k_0,\lambda_0}^{\beta}
\equiv
\frac{
1
}{
Z_{\omega_{k_0}}
} e^{-\beta H_{\mathbf k_0,\lambda_0}^{(g)}},
\end{equation}
where
\begin{equation}
    \begin{aligned}
Z(\omega_{k_0})\equiv Z_{\mathbf k_0,\lambda_0}
&=
\operatorname{tr}_{g;\mathbf k_0,\lambda_0}
\left(
e^{-\beta H_{\mathbf k_0,\lambda_0}^{(g)}}
\right)
\\
&=
\sum_{n=0}^{\infty}
e^{-\beta\hbar\omega_{k_0}n}
=
\frac{1}{1-e^{-\beta\hbar\omega_{k_0}}}.
\end{aligned}
\end{equation}
Here, we have used the notation
\begin{equation}
    \operatorname{tr}_{\mathbf k_0,\lambda_0}(O)
\equiv
\sum_{n=0}^{\infty}
\bra{n_{\mathbf k_0,\lambda_0}}
O
\ket{n_{\mathbf k_0,\lambda_0}}.
\end{equation}
It is indeed the gravitational trace, but only over the selected graviton single multi-mode number state. The occupation of an arbitrary mode is then defined as
\begin{equation}\label{eq_occupation_arbitrary_mode_nbeta}
    n_{\mathbf k,\lambda}^{\beta}
=
\operatorname{tr}_{g}
\left(
\rho^\beta
g_{\mathbf k,\lambda}^{\dagger}
g_{\mathbf k,\lambda}
\right).
\end{equation}
Here, we have used the notation 
\begin{equation}
    \operatorname{tr}_g(O)
=
\sum_{\underline n}
\bra{\underline n}O\ket{\underline n}.
\end{equation}
This is the gravitational trace over the full multi-mode graviton Hilbert space (note the notation for the graviton multi-mode number state defined earlier). Thus for the mean thermal occupation number \eqref{eq_occupation_arbitrary_mode_nbeta} we obtain
\begin{equation}
    n_\beta(\omega_k)\equiv
n_{\mathbf k,\lambda}^{\beta}
=
\frac{1}{e^{\beta\hbar\omega_k}-1}.
\end{equation}
Since the two polarization modes have the same frequency and the thermal state is polarization independent, the mean occupation number depends on $\omega_k$ but not on $\lambda$. We therefore denote it by $n_\beta(\omega_k)$.

Then, the expectation values are given by
\begin{subequations}
\begin{align}
    \operatorname{tr}_g\!\left(
\rho^\beta
g_{\mathbf k',\lambda'}^\dagger
g_{\mathbf k'',\lambda''}
\right)
&=
n_\beta(\omega_{k'})
\delta^{(3)}(\mathbf k'-\mathbf k'')
\delta_{\lambda'}^{\lambda''}, \\
\operatorname{tr}_g\!\left(
\rho^\beta
g_{\mathbf k',\lambda'}
g_{\mathbf k'',\lambda''}^\dagger
\right)
&=
\delta^{(3)}(\mathbf k'-\mathbf k'')
\delta_{\lambda'}^{\lambda''} +n_\beta(\omega_{k'})\,
\delta^{(3)}(\mathbf k'-\mathbf k'')
\delta_{\lambda'}^{\lambda''},\\
\operatorname{tr}_g\!\left(
\rho^\beta
g_{\mathbf k',\lambda'}
g_{\mathbf k'',\lambda''}
\right)
&=0,\\
\operatorname{tr}_g\!\left(
\rho^\beta
g_{\mathbf k',\lambda'}^\dagger
g_{\mathbf k'',\lambda''}^\dagger
\right)
&=0.
\end{align}
\end{subequations}
For completeness, we note that the one-point correlation functions vanish,
\begin{subequations}
\begin{align}
    \operatorname{tr}_g\!\left(
\rho^\beta g_{\mathbf k',\lambda'}
\right)&=0, \\
\operatorname{tr}_g\!\left(
\rho^\beta g_{\mathbf k',\lambda'}^\dagger
\right)&=0.
\end{align}
\end{subequations}

The one-point bath correlator \eqref{eq_onepoint_correlation_function} vanishes. The two-point bath correlator \eqref{eq_two_point_bath_conn_or_not_conn}, is therefore
\begin{equation}
    D^{(g)}(\tau)
=
\sum_\lambda\int d^3k\,
\left(G_{\mathbf k}^{\lambda}\right)^2
\left[
\left(1+n_\beta(\omega_k)\right)e^{-i\omega_k\tau}
+
n_\beta(\omega_k)e^{i\omega_k\tau}
\right].
\end{equation}

Unlike a general multi-mode number state, the thermal state requires no additional assumptions of isotropy or polarization independence; they follow from the properties of the canonical Gibbs state. The derivation therefore proceeds exactly as in the multi-mode number-state case, with the occupation profile replaced by the distribution $n_\beta(\omega_k)$. Inserting the above into Eq.~\eqref{eq_master_equation_D}, after the secular approximation (Section~\ref{sec_secular_approximation}) and the renormalization procedure (Section~\ref{sec_lamb_shift_Hamiltonian_renormalized}), and transforming back to the Schrödinger picture, the Lindblad master equation for multi-mode thermal ($\mathcal{T}$) state gravitons reads
\begin{equation}
    \dot{\rho}^\order{m}_\mathcal{T}(t) =-\frac{i}{\hbar} [H^\order{m}, \rho_\mathcal{T}^\order{m}(t)]
+\gamma_{\mathcal{T},\downarrow} \,\mathcal{D}[L_{\mathcal{T},\downarrow}]\,\rho_\mathcal{T}^\order{m}(t)
+ \gamma_{\mathcal{T},\uparrow} \,\mathcal{D}[L_{\mathcal{T},\uparrow}]\,\rho_\mathcal{T}^\order{m}(t).
\end{equation}
with
\begin{alignat}{2}
\gamma_{\mathcal T,\downarrow}
&=
\gamma_{\mathcal V}\left[1+n_\beta(2\omega_m)\right],
&\qquad
L_{\mathcal T,\downarrow}
&=
b^2,
\\
\gamma_{\mathcal T,\uparrow}
&=
\gamma_{\mathcal V}n_\beta(2\omega_m),
&
L_{\mathcal T,\uparrow}
&=
b^{\dagger 2}.
\end{alignat}
Here $n_\beta(2\omega)$ denotes the Bose-Einstein thermal mean occupation number with inverse temperature $\beta$.

\section{State of gravitons: The rotating-wave approximation, secular approximation, and Lamb-shift Hamiltonian renormalization}

\subsection{Rotating-wave approximation (RWA) for the interaction Hamiltonian}

For completeness, we note how the interaction Hamiltonian would look if one performed a rotating-wave approximation directly at the Hamiltonian level. This approximation keeps only the near-resonant terms in the interaction-picture Hamiltonian, here corresponding to graviton modes with $\omega_k\simeq 2\omega_m$, and discards the rapidly oscillating counter-rotating terms.

The full interaction Hamiltonian in the interaction picture \eqref{eq_interaction_Hamiltonian_interaction_picture}, expanded, reads
\begin{equation}
    \begin{aligned}
H_{II}(t)
=
\sum_\lambda \int d^3 k\,
G_{\mathbf k}^{\lambda}
\Big[
&
b^2 \otimes g_{\mathbf k\lambda}\,
e^{-i(2\omega_m+\omega_k)t}
+
b^2 \otimes g_{\mathbf k\lambda}^{\dagger}\,
e^{-i(2\omega_m-\omega_k)t}
\\
&+
b^{\dagger 2} \otimes g_{\mathbf k\lambda}\,
e^{i(2\omega_m-\omega_k)t}
+
b^{\dagger 2} \otimes g_{\mathbf k\lambda}^{\dagger}\,
e^{i(2\omega_m+\omega_k)t}
\\
&+
(b^\dagger b + b b^\dagger)\otimes g_{\mathbf k\lambda}\,
e^{-i\omega_k t}
+
(b^\dagger b + b b^\dagger)\otimes g_{\mathbf k\lambda}^{\dagger}\,
e^{i\omega_k t}
\Big].
\end{aligned}
\end{equation}
Thus the frequency content of the full interaction Hamiltonian is
\begin{equation}
    \pm \omega_k,
\qquad
\pm(2\omega_m+\omega_k),
\qquad
\pm(2\omega_m-\omega_k).
\end{equation}
For a graviton bath with positive-frequency modes and a resonance near
\begin{equation}
    \omega_k \sim 2\omega_m,
\end{equation}
the rotating-wave approximation (RWA) drops the terms with explicit factors
\begin{equation}
    e^{\pm \omega_k},
\qquad
e^{\pm(2\omega_m+\omega_k)}.
\end{equation}
We therefore end up with
\begin{equation}
H_{II}(t)
\overset{\text{RWA}}{=}
\sum_\lambda
\int d^3 k\,
G_{\mathbf k}^{\lambda}
\left[
b^2\otimes g_{\mathbf k\lambda}^{\dagger}\,
e^{-i(2\omega_m-\omega_k)t}
+
b^{\dagger 2}\otimes g_{\mathbf k\lambda}\,
e^{+i(2\omega_m-\omega_k)t}
\right].
\end{equation}
We emphasize that this Hamiltonian-level RWA is presented only for comparison. In the derivation used in the main text, we do not make use of this approximation. Rather, we retain the full interaction Hamiltonian and apply the secular approximation at the master-equation level, where terms are organized according to their Bohr frequencies.

\subsection{Secular approximation for the matter operators}\label{sec_secular_approximation}
The evaluation requires the matter part of interaction-picture interaction Hamiltonian at the two times $t$ and $t-\tau$. These are
\begin{equation}
    H_{II}^{(m)}(t)
=
b^2 e^{-2i\omega_m t}
+
b^{\dagger 2} e^{2i\omega_m t}
+
b^\dagger b + b b^\dagger ,
\end{equation}
and
\begin{equation}
    H_{II}^{(m)}(t-\tau)
=
b^2 e^{-2i\omega_m(t-\tau)}
+
b^{\dagger 2} e^{2i\omega_m(t-\tau)}
+
b^\dagger b + b b^\dagger .
\end{equation}
For notational convenience, in this section, we henceforth suppress the explicit interaction-picture and time arguments of the matter density operator and write $\rho\equiv \rho_I^{(m)}(t)$. 

\secpar{Stationary two-point bath correlator}
Let us first consider the case when the bath two-point correlator is stationary $D^{(g)}(t,t-\tau)=D^\order{g}(\tau)$. Such is the case when the graviton bath is a vacuum state, a number state, a coherent state (when $D$ is calculated as a connected correlator), or a thermal state. Then the master equation \eqref{eq_master_equation_D} reads,
\begin{equation}
    \dot\rho
=
-\frac{1}{\hbar^2}
\int_0^\infty d\tau
\left(
D^\order{g}(\tau)
\left[
H_{II}^{(m)}(t),H_{II}^{(m)}(t-\tau)\rho
\right]
-
D^\order{g}(\tau)^*
\left[
H_{II}^{(m)}(t),\rho H_{II}^{(m)}(t-\tau)
\right]
\right),
\end{equation}
Expanding the first commutator gives
\begin{equation}
\begin{aligned}
\left[
H_{II}^{(m)}(t),H_{II}^{(m)}(t-\tau)\rho
\right]
=&\;
b^2 b^2 \rho \,e^{-4i\omega_m t}e^{2i\omega_m\tau}
+
b^2 b^{\dagger 2} \rho \,e^{-2i\omega_m\tau}
+
b^2(b^\dagger b + b b^\dagger) \rho \,e^{-2i\omega_m t}
\\
&+
b^{\dagger 2}b^2 \rho\, e^{2i\omega_m\tau}
+
b^{\dagger 2}b^{\dagger 2} \rho\,e^{4i\omega_m t}e^{-2i\omega_m\tau}
+
b^{\dagger 2}(b^\dagger b + b b^\dagger) \rho\,e^{2i\omega_m t}
\\
&+
(b^\dagger b + b b^\dagger)b^2 \rho\,e^{-2i\omega_m t}e^{2i\omega_m\tau}
+
(b^\dagger b + b b^\dagger)b^{\dagger 2} \rho\, e^{2i\omega_m t}e^{-2i\omega_m\tau}
\\
&+
(b^\dagger b + b b^\dagger)^2\rho 
\\
&-
b^2\rho \,b^2 \,e^{-4i\omega_m t}e^{2i\omega_m\tau}
-
b^2\rho \,b^{\dagger 2}e^{2i\omega_m\tau}
-
b^2\rho\,(b^\dagger b + b b^\dagger)e^{-2i\omega_m t}e^{2i\omega_m\tau}
\\
&-
b^{\dagger 2}\rho \,b^2 \,e^{-2i\omega_m\tau}
-
b^{\dagger 2}\rho \,b^{\dagger 2} \,e^{4i\omega_m t}e^{-2i\omega_m\tau}
-
b^{\dagger 2}\rho \,(b^\dagger b + b b^\dagger)\,e^{2i\omega_m t}e^{-2i\omega_m\tau}
\\
&-
(b^\dagger b + b b^\dagger)\,\rho \,b^2e^{-2i\omega_m t}
-
(b^\dagger b + b b^\dagger)\,\rho \,b^{\dagger 2}e^{2i\omega_m t}
\\
&-
(b^\dagger b + b b^\dagger)\,\rho\,(b^\dagger b + b b^\dagger).
\end{aligned}
\end{equation}
Expanding the second commutator gives
\begin{equation}
\begin{aligned}
\left[
H^\order{m}(t),\rho \,H^\order{m}(t-\tau)
\right]
=
&\;
b^2 \,\rho \,b^2e^{-4i\omega_m t}e^{2i\omega_m\tau}
+
b^2\rho \, b^{\dagger 2}e^{-2i\omega_m\tau}
+
b^2\rho\,(b^\dagger b + b b^\dagger)e^{-2i\omega_m t}
\\
&+
b^{\dagger 2}\rho \,b^2e^{2i\omega_m\tau}
+
b^{\dagger 2}\rho \,b^{\dagger 2}e^{4i\omega_m t}e^{-2i\omega_m\tau}
+
b^{\dagger 2}\rho\,(b^\dagger b + b b^\dagger)e^{2i\omega_m t}
\\
&+
(b^\dagger b + b b^\dagger)\,\rho \,b^2e^{-2i\omega_m t}e^{2i\omega_m\tau}
+
(b^\dagger b + b b^\dagger)\,\rho \, b^{\dagger 2}e^{2i\omega_m t}e^{-2i\omega_m\tau}
\\
&+
(b^\dagger b + b b^\dagger)\, \rho \,(b^\dagger b + b b^\dagger)
\\
&-
\rho \, b^2\,b^2 \, e^{-4i\omega_m t}e^{2i\omega_m\tau}
-
\rho \,b^2 \, b^{\dagger 2} \,e^{2i\omega_m\tau}
-
\rho \, b^2\, (b^\dagger b + b b^\dagger ) \, e^{-2i\omega_m t}e^{2i\omega_m\tau}
\\
&-
\rho \,b^{\dagger 2} \, b^2 \, e^{-2i\omega_m\tau}
-
\rho \, b^{\dagger 2} \, b^{\dagger 2} \, e^{4i\omega_m t}e^{-2i\omega_m\tau}
-
\rho \, b^{\dagger 2}\, (b^\dagger b + b b^\dagger) \, e^{2i\omega_m t}e^{-2i\omega_m\tau}
\\
&-
\rho\, (b^\dagger b + b b^\dagger)\,b^2\,e^{-2i\omega_m t}
-
\rho\, (b^\dagger b + b b^\dagger) \, b^{\dagger 2}e^{2i\omega_m t}
\\
&-
\rho\, (b^\dagger b + b b^\dagger)^2
\Big].
\end{aligned}
\end{equation}
The secular approximation (SA) effectively drops all terms with explicit time $t$-dependence, namely, in this case, terms with explicit factors
\begin{equation}
    e^{\pm2i\omega_m t},
\qquad
e^{\pm4i\omega_m t}.
\end{equation}
Therefore, collecting the frequencies, the surviving terms are
\begin{equation}
\begin{aligned}
\dot\rho
\overset{\text{SA}}{=}
-\frac{1}{\hbar^2}
\int_0^\infty d\tau
\Bigg\{
D^\order{g}(\tau)\Big[
&
(b^2 \, b^{\dagger 2} \rho \,  -
b^{\dagger 2}\,\rho \,b^2) e^{-2i\omega_m\tau}
\\
&
+
(b^{\dagger 2} \,b^2 \, \rho  -
b^2 \,\rho \, b^{\dagger 2}) e^{2i\omega_m\tau} \\
&+
(b^\dagger b + b b^\dagger)^2\,\rho -
(b^\dagger b + b b^\dagger) \, \rho \, (b^\dagger b + b b^\dagger) \Big]
&
\\
-
D^\order{g}(\tau)^*\Big[
&
(b^2 \,\rho \,b^{\dagger 2} -
\rho \, b^{\dagger 2} \, b^2) e^{-2i\omega_m\tau}
\\&+
(b^{\dagger 2} \, \rho \, b^2 -
\rho \,b^2 \, b^{\dagger 2}) e^{2i\omega_m\tau}
\\&+
(b^\dagger b + b b^\dagger) \, \rho \, (b^\dagger b + b b^\dagger) -
\rho \, (b^\dagger b + b b^\dagger)^2
\Big]
\Bigg\}.
\end{aligned}
\end{equation}
It is illuminating to write the above in terms of the so-called Bohr frequencies, obtaining a more compact form. Let us define the Bohr components
\begin{equation}
    A(2\omega_m) \equiv b^2,
    \qquad
    A(-2\omega_m) \equiv b^{\dagger 2},
    \qquad
    A(0) \equiv (b^\dagger b + b b^\dagger).
\end{equation}
And let us also define the $\tau$-integrated one-sided bath transform by
\begin{equation}
    \Gamma(\Omega) \equiv \frac{1}{\hbar^2} \int_0^\infty d\tau \, D^{(g)}(\tau) \,e^{i \Omega \tau} \equiv \frac{1}{2} \gamma(\Omega) + i S(\Omega),
    \qquad
    \Omega = \{0, \pm 2 \omega_m\}.
\end{equation}
Then the secular master equation can be written as
\begin{equation}
    \dot\rho
=
-\frac{i}{\hbar}
\left[
H_{\mathrm{LS}},\rho
\right]
+
\sum_{\Omega=0,\pm 2\omega_m}
\gamma(\Omega)
\mathcal D[A(\Omega)]\rho,
\end{equation}
where the Lamb-shift Hamiltonian is given by
\begin{equation}
H_{\mathrm{LS}}
=
\hbar
\sum_{\Omega=0,\pm 2\omega_m}
S(\Omega)
A^\dagger(\Omega)A(\Omega).
\end{equation}
In the case of vacuum, because of the specific properties of the graviton bath, we will find 
$\gamma(-2\omega_m)=\gamma(0)=0$, so that only the dissipative contribution
proportional to $\gamma(2\omega_m)$ remains. For the Lamb-shift Hamiltonian,
all frequency components contribute; however, these terms will be absorbed
through a renormalization procedure which we illustrate in detail in Section~\ref{sec_lamb_shift_Hamiltonian_renormalized}.

\secpar{Nonstationary two-point bath correlator: squeezed states}
The stationarity assumption of the two-point bath correlator does not hold, however, for a squeezed state, so the SA should be modified accordingly: one first multiplies the explicit time-dependent factors in $D$ with those from the matter operators, does the $\tau$-integration, and only then discards the remaining rapidly oscillating terms (containing time $t$-dependence) \cite{breuer_petruccione_2002_the_theory_of_open_quantuma}.

In the case of the multi-mode squeezed vacuum case, we have the two-point bath correlator of the form
\begin{align}
    D^{(g)}(t,t-\tau) = F^{(g)}(\tau) + G^{(g)}(t,t-\tau).
\end{align}
Therefore,  $F(\tau)$ is going to multiply the same matter operators as elaborated above in the stationary case. However the second term is $t$-dependent,
\begin{equation}
    G^{(g)}(t,t-\tau) = \sum_\lambda \int d^3 k \, (G_{\mathbf k}^\lambda)^2 \left[M_\klambda e^{-i 2 \omega_k t} e^{+i \omega_k \tau}
        +
        M_\klambda^*e^{i 2\omega_k t} e^{- i\omega_k\tau} \right]
\end{equation}
Multiplying the first commutator by the \(M_{\mathbf{k},\lambda}\)-part gives
\begin{align*}
&
M_{\mathbf{k},\lambda}
e^{-2i\omega_k t}e^{i\omega_k\tau}
\left[
H_{II}^{(m)}(t),
H_{II}^{(m)}(t-\tau)\rho
\right]
\\
=&\;
M_{\mathbf{k},\lambda}
\Big[
b^2 b^2 \rho \,
e^{-2i(\omega_k+2\omega_m)t}
e^{i(\omega_k+2\omega_m)\tau}
+
b^2 b^{\dagger 2} \rho\,
e^{-2i\omega_k t}
e^{i(\omega_k-2\omega_m)\tau}
\\
&+
b^2\left(b^\dagger b + b b^\dagger\right) \rho\,
e^{-2i(\omega_k+\omega_m)t}
e^{i\omega_k\tau}
+
b^{\dagger 2}b^2 \rho\,
e^{-2i\omega_k t}
e^{i(\omega_k+2\omega_m)\tau}
\\
&+
b^{\dagger 2}b^{\dagger 2} \rho\,
e^{-2i(\omega_k-2\omega_m)t}
e^{i(\omega_k-2\omega_m)\tau}
+
b^{\dagger 2}\left(b^\dagger b + b b^\dagger\right) \rho\,
e^{-2i(\omega_k-\omega_m)t}
e^{i\omega_k\tau}
\\
&+
\left(b^\dagger b + b b^\dagger\right)b^2 \rho\,
e^{-2i(\omega_k+\omega_m)t}
e^{i(\omega_k+2\omega_m)\tau}
+
\left(b^\dagger b + b b^\dagger\right)b^{\dagger 2} \rho\,
e^{-2i(\omega_k-\omega_m)t}
e^{i(\omega_k-2\omega_m)\tau}
\\
&+
\left(b^\dagger b + b b^\dagger\right)^2\rho\,
e^{-2i\omega_k t}
e^{i\omega_k\tau}
-
b^2\rho b^2\,
e^{-2i(\omega_k+2\omega_m)t}
e^{i(\omega_k+2\omega_m)\tau}
\\&-
b^2\rho b^{\dagger 2}\,
e^{-2i\omega_k t}
e^{i(\omega_k+2\omega_m)\tau}
-
b^2\rho\left(b^\dagger b + b b^\dagger\right)\,
e^{-2i(\omega_k+\omega_m)t}
e^{i(\omega_k+2\omega_m)\tau}
\\&-
b^{\dagger 2}\rho b^2\,
e^{-2i\omega_k t}
e^{i(\omega_k-2\omega_m)\tau}
-
b^{\dagger 2}\rho b^{\dagger 2}\,
e^{-2i(\omega_k-2\omega_m)t}
e^{i(\omega_k-2\omega_m)\tau}
\\&-
b^{\dagger 2}\rho\left(b^\dagger b + b b^\dagger\right)\,
e^{-2i(\omega_k-\omega_m)t}
e^{i(\omega_k-2\omega_m)\tau}
-
\left(b^\dagger b + b b^\dagger\right)\rho b^2\,
e^{-2i(\omega_k+\omega_m)t}
e^{i\omega_k\tau}
\\
&-
\left(b^\dagger b + b b^\dagger\right)\rho b^{\dagger 2}\,
e^{-2i(\omega_k-\omega_m)t}
e^{i\omega_k\tau}
-
\left(b^\dagger b + b b^\dagger\right)
\rho
\left(b^\dagger b + b b^\dagger\right)\,
e^{-2i\omega_k t}
e^{i\omega_k\tau}
\Big].
\end{align*}

Multiplying the first commutator by the \(M_{\mathbf{k},\lambda}^*\)-part gives
\begin{align*}
&
M_{\mathbf{k},\lambda}^*
e^{2i\omega_k t}e^{-i\omega_k\tau}
\left[
H_{II}^{(m)}(t),
H_{II}^{(m)}(t-\tau)\rho
\right]
\\
=&\;
M_{\mathbf{k},\lambda}^*
\Big[
b^2 b^2 \rho \,
e^{2i(\omega_k-2\omega_m)t}
e^{-i(\omega_k-2\omega_m)\tau}
+
b^2 b^{\dagger 2} \rho\,
e^{2i\omega_k t}
e^{-i(\omega_k+2\omega_m)\tau}
\\
&+
b^2\left(b^\dagger b + b b^\dagger\right) \rho\,
e^{2i(\omega_k-\omega_m)t}
e^{-i\omega_k\tau}
+
b^{\dagger 2}b^2 \rho\,
e^{2i\omega_k t}
e^{-i(\omega_k-2\omega_m)\tau}
\\
&+
b^{\dagger 2}b^{\dagger 2} \rho\,
e^{2i(\omega_k+2\omega_m)t}
e^{-i(\omega_k+2\omega_m)\tau}
+
b^{\dagger 2}\left(b^\dagger b + b b^\dagger\right) \rho\,
e^{2i(\omega_k+\omega_m)t}
e^{-i\omega_k\tau}
\\
&+
\left(b^\dagger b + b b^\dagger\right)b^2 \rho\,
e^{2i(\omega_k-\omega_m)t}
e^{-i(\omega_k-2\omega_m)\tau}
+
\left(b^\dagger b + b b^\dagger\right)b^{\dagger 2} \rho\,
e^{2i(\omega_k+\omega_m)t}
e^{-i(\omega_k+2\omega_m)\tau}
\\
&+
\left(b^\dagger b + b b^\dagger\right)^2\rho\,
e^{2i\omega_k t}
e^{-i\omega_k\tau}
-
b^2\rho b^2\,
e^{2i(\omega_k-2\omega_m)t}
e^{-i(\omega_k-2\omega_m)\tau}
\\
&-
b^2\rho b^{\dagger 2}\,
e^{2i\omega_k t}
e^{-i(\omega_k-2\omega_m)\tau}
-
b^2\rho\left(b^\dagger b + b b^\dagger\right)\,
e^{2i(\omega_k-\omega_m)t}
e^{-i(\omega_k-2\omega_m)\tau}
\\
&-
b^{\dagger 2}\rho b^2\,
e^{2i\omega_k t}
e^{-i(\omega_k+2\omega_m)\tau}
-
b^{\dagger 2}\rho b^{\dagger 2}\,
e^{2i(\omega_k+2\omega_m)t}
e^{-i(\omega_k+2\omega_m)\tau}
\\
&-
b^{\dagger 2}\rho\left(b^\dagger b + b b^\dagger\right)\,
e^{2i(\omega_k+\omega_m)t}
e^{-i(\omega_k+2\omega_m)\tau}
-
\left(b^\dagger b + b b^\dagger\right)\rho b^2\,
e^{2i(\omega_k-\omega_m)t}
e^{-i\omega_k\tau}
\\
&-
\left(b^\dagger b + b b^\dagger\right)\rho b^{\dagger 2}\,
e^{2i(\omega_k+\omega_m)t}
e^{-i\omega_k\tau}
-
\left(b^\dagger b + b b^\dagger\right)
\rho
\left(b^\dagger b + b b^\dagger\right)\,
e^{2i\omega_k t}
e^{-i\omega_k\tau}
\Big].
\end{align*}

Multiplying the second commutator by the \(M_{\mathbf{k},\lambda}\)-part gives
\begin{align*}
&
M_{\mathbf{k},\lambda}
e^{-2i\omega_k t}e^{i\omega_k\tau}
\left[
H_{II}^{(m)}(t),
\rho H_{II}^{(m)}(t-\tau)
\right]
\\
=&\;
M_{\mathbf{k},\lambda}
\Big[
b^2\rho b^2\,
e^{-2i(\omega_k+2\omega_m)t}
e^{i(\omega_k+2\omega_m)\tau}
+
b^2\rho b^{\dagger 2}\,
e^{-2i\omega_k t}
e^{i(\omega_k-2\omega_m)\tau}
\\
&+
b^2\rho\left(b^\dagger b + b b^\dagger\right)\,
e^{-2i(\omega_k+\omega_m)t}
e^{i\omega_k\tau}
+
b^{\dagger 2}\rho b^2\,
e^{-2i\omega_k t}
e^{i(\omega_k+2\omega_m)\tau}
\\
&+
b^{\dagger 2}\rho b^{\dagger 2}\,
e^{-2i(\omega_k-2\omega_m)t}
e^{i(\omega_k-2\omega_m)\tau}
+
b^{\dagger 2}\rho\left(b^\dagger b + b b^\dagger\right)\,
e^{-2i(\omega_k-\omega_m)t}
e^{i\omega_k\tau}
\\
&+
\left(b^\dagger b + b b^\dagger\right)\rho b^2\,
e^{-2i(\omega_k+\omega_m)t}
e^{i(\omega_k+2\omega_m)\tau}
+
\left(b^\dagger b + b b^\dagger\right)\rho b^{\dagger 2}\,
e^{-2i(\omega_k-\omega_m)t}
e^{i(\omega_k-2\omega_m)\tau}
\\
&+
\left(b^\dagger b + b b^\dagger\right)
\rho
\left(b^\dagger b + b b^\dagger\right)\,
e^{-2i\omega_k t}
e^{i\omega_k\tau}
-
\rho b^2b^2\,
e^{-2i(\omega_k+2\omega_m)t}
e^{i(\omega_k+2\omega_m)\tau}
\\
&-
\rho b^2b^{\dagger 2}\,
e^{-2i\omega_k t}
e^{i(\omega_k+2\omega_m)\tau}
-
\rho b^2\left(b^\dagger b + b b^\dagger\right)\,
e^{-2i(\omega_k+\omega_m)t}
e^{i(\omega_k+2\omega_m)\tau}
\\
&-
\rho b^{\dagger 2}b^2\,
e^{-2i\omega_k t}
e^{i(\omega_k-2\omega_m)\tau}
-
\rho b^{\dagger 2}b^{\dagger 2}\,
e^{-2i(\omega_k-2\omega_m)t}
e^{i(\omega_k-2\omega_m)\tau}
\\
&-
\rho b^{\dagger 2}
\left(b^\dagger b + b b^\dagger\right)\,
e^{-2i(\omega_k-\omega_m)t}
e^{i(\omega_k-2\omega_m)\tau}
-
\rho\left(b^\dagger b + b b^\dagger\right)b^2\,
e^{-2i(\omega_k+\omega_m)t}
e^{i\omega_k\tau}
\\
&-
\rho\left(b^\dagger b + b b^\dagger\right)b^{\dagger 2}\,
e^{-2i(\omega_k-\omega_m)t}
e^{i\omega_k\tau}
-
\rho\left(b^\dagger b + b b^\dagger\right)^2\,
e^{-2i\omega_k t}
e^{i\omega_k\tau}
\Big].
\end{align*}

Multiplying the second commutator by the \(M_{\mathbf{k},\lambda}^*\)-part gives
\begin{align*}
&
M_{\mathbf{k},\lambda}^*
e^{2i\omega_k t}e^{-i\omega_k\tau}
\left[
H_{II}^{(m)}(t),
\rho H_{II}^{(m)}(t-\tau)
\right]
\\
=&\;
M_{\mathbf{k},\lambda}^*
\Big[
b^2\rho b^2\,
e^{2i(\omega_k-2\omega_m)t}
e^{-i(\omega_k-2\omega_m)\tau}
+
b^2\rho b^{\dagger 2}\,
e^{2i\omega_k t}
e^{-i(\omega_k+2\omega_m)\tau}
\\
&+
b^2\rho\left(b^\dagger b + b b^\dagger\right)\,
e^{2i(\omega_k-\omega_m)t}
e^{-i\omega_k\tau}
+
b^{\dagger 2}\rho b^2\,
e^{2i\omega_k t}
e^{-i(\omega_k-2\omega_m)\tau}
\\
&+
b^{\dagger 2}\rho b^{\dagger 2}\,
e^{2i(\omega_k+2\omega_m)t}
e^{-i(\omega_k+2\omega_m)\tau}
+
b^{\dagger 2}\rho\left(b^\dagger b + b b^\dagger\right)\,
e^{2i(\omega_k+\omega_m)t}
e^{-i\omega_k\tau}
\\
&+
\left(b^\dagger b + b b^\dagger\right)\rho b^2\,
e^{2i(\omega_k-\omega_m)t}
e^{-i(\omega_k-2\omega_m)\tau}
+
\left(b^\dagger b + b b^\dagger\right)\rho b^{\dagger 2}\,
e^{2i(\omega_k+\omega_m)t}
e^{-i(\omega_k+2\omega_m)\tau}
\\
&+
\left(b^\dagger b + b b^\dagger\right)
\rho
\left(b^\dagger b + b b^\dagger\right)\,
e^{2i\omega_k t}
e^{-i\omega_k\tau}
-
\rho b^2b^2\,
e^{2i(\omega_k-2\omega_m)t}
e^{-i(\omega_k-2\omega_m)\tau}
\\
&-
\rho b^2b^{\dagger 2}\,
e^{2i\omega_k t}
e^{-i(\omega_k-2\omega_m)\tau}
-
\rho b^2\left(b^\dagger b + b b^\dagger\right)\,
e^{2i(\omega_k-\omega_m)t}
e^{-i(\omega_k-2\omega_m)\tau}
\\
&-
\rho b^{\dagger 2}b^2\,
e^{2i\omega_k t}
e^{-i(\omega_k+2\omega_m)\tau}
-
\rho b^{\dagger 2}b^{\dagger 2}\,
e^{2i(\omega_k+2\omega_m)t}
e^{-i(\omega_k+2\omega_m)\tau}
\\
&-
\rho b^{\dagger 2}
\left(b^\dagger b + b b^\dagger\right)\,
e^{2i(\omega_k+\omega_m)t}
e^{-i(\omega_k+2\omega_m)\tau}
-
\rho\left(b^\dagger b + b b^\dagger\right)b^2\,
e^{2i(\omega_k-\omega_m)t}
e^{-i\omega_k\tau}
\\
&-
\rho\left(b^\dagger b + b b^\dagger\right)b^{\dagger 2}\,
e^{2i(\omega_k+\omega_m)t}
e^{-i\omega_k\tau}
-
\rho\left(b^\dagger b + b b^\dagger\right)^2\,
e^{2i\omega_k t}
e^{-i\omega_k\tau}
\Big].
\end{align*}
Ignoring principal-value terms, we have
\begin{equation}
\begin{aligned}
    \int_0^\infty d\tau\, e^{\pm i\omega_k\tau}
    &=
    \pi\delta(\omega_k) = 0, \\
    \int_0^\infty d\tau\,
    e^{\pm i(\omega_k+2\omega_m)\tau}
    &=
    \pi\delta(\omega_k+2\omega_m)
    =
    0.
\end{aligned}
\end{equation}
Namely, they give no dissipative contribution, since $\omega_k>0$ and $\omega_m>0$. Hence the only anomalous terms that
survive the $\tau$-integral are those containing $e^{\pm i(\omega_k-2\omega_m)\tau}$,
\begin{equation}
    \int_0^\infty d\tau \, e^{\pm i(\omega_k-2\omega_m)\tau} = \pi \delta(\omega_k - 2\omega_m).
\end{equation}
With hindsight, we safely conclude that the delta function sets $\omega_k=2\omega_m$. Therefore the remaining $t$-dependent phases are
\begin{equation}
    e^{\pm 2i\omega_m t},
    \qquad
    e^{\pm 4i\omega_m t}.
\end{equation}
As before, these are precisely the terms which we drop in the secular approximation.
Therefore, after the \(\tau\)-integral and the secular approximation, the
anomalous squeezed contribution is
\begin{equation}
\begin{aligned}
\dot\rho_{\mathrm{anom}}
\overset{\mathrm{SA}}{=}
-\frac{\pi}{\hbar^2}
\sum_\lambda
\int d^3k\,
\left(G_{\mathbf{k}}^\lambda\right)^2
\delta(\omega_k-2\omega_m)
\Bigg\{
&
M_{\mathbf{k},\lambda}
\left[
b^{\dagger 2}b^{\dagger 2}\rho
-
2b^{\dagger 2}\rho b^{\dagger 2}
+
\rho b^{\dagger 2}b^{\dagger 2}
\right]
\\
&+
M_{\mathbf{k},\lambda}^*
\left[
b^2b^2\rho
-
2b^2\rho b^2
+
\rho b^2b^2
\right]
\Bigg\}.
\end{aligned}
\end{equation}
On the other hand, the normal part is simply $F(\tau)$ instead of $D(\tau)$ in the secular approximation we did before. In summary
\begin{equation}
\begin{aligned}
\dot\rho
\overset{\mathrm{SA}}{=}
\frac{2\pi}{\hbar^2}
\sum_\lambda
\int d^3k\,
\left(G_{\mathbf{k}}^\lambda\right)^2
\delta(\omega_k-2\omega_m)
\Bigg\{
&
\left(1+N_{\mathbf{k},\lambda}\right)
\left[
b^2\rho b^{\dagger 2}
-
\frac12
\left\{
b^{\dagger 2}b^2,\rho
\right\}
\right]
\\
&+
N_{\mathbf{k},\lambda}
\left[
b^{\dagger 2}\rho b^2
-
\frac12
\left\{
b^2b^{\dagger 2},\rho
\right\}
\right]
\\
&+
M_{\mathbf{k},\lambda}
\left[
b^{\dagger 2}\rho b^{\dagger 2}
-
\frac12
\left\{
b^{\dagger 4},\rho
\right\}
\right]
\\
&+
M_{\mathbf{k},\lambda}^*
\left[
b^2\rho b^2
-
\frac12
\left\{
b^4,\rho
\right\}
\right]
\Bigg\}.
\end{aligned}
\end{equation}
Note that
\begin{equation}\label{eq_spectral_density_gravitons_thesumlambda_integralk}
\begin{aligned}
\sum_\lambda
\int d^3k\,
\left(G_{\mathbf{k}}^\lambda\right)^2
\delta(\omega_k-2\omega_m)
&=
\sum_\lambda
\int k^2dk\,d\Omega\,
\frac{G\hbar^3\omega_k^3}
{64\pi^2c^2\omega_m^2}
\left(e_{xx}^\lambda(\mathbf n)\right)^2
\delta(ck-2\omega_m)
\\
&=
\frac{G\hbar^3}
{64\pi^2c^2\omega_m^2}
\frac{32\pi}{15}
\int_0^\infty dk\,
k^2(ck)^3
\delta(ck-2\omega_m)
\\
&=
\frac{G\hbar^3}
{64\pi^2c^2\omega_m^2}
\frac{32\pi}{15}
\frac{(2\omega_m)^5}{c^3}
\\
&=
\frac{16G\hbar^3\omega_m^3}
{15\pi c^5}.
\end{aligned}
\end{equation}
So,
\begin{equation}
\gamma_{\mathcal{V}}
=\frac{2\pi}{\hbar}J(2\omega_m)=
\frac{2\pi}{\hbar^2}
\sum_\lambda
\int d^3k\,
\left(G_{\mathbf{k}}^\lambda\right)^2
\delta(\omega_k-2\omega_m)
=
\frac{32G\hbar\omega_m^3}{15c^5}.
\end{equation}
The latter is the vacuum decoherence rate. 

In open quantum systems, the frequency distribution of environmental modes weighted by their coupling strength is commonly referred to as the spectral density \cite{breuer_petruccione_2002_the_theory_of_open_quantuma}. We therefore define the gravitational spectral density as
\begin{equation}\label{eq_spectral_density_gravitational_first}
    J^{(g)}(\Omega)
\equiv
\sum_\lambda
\int d^3k\,
\left(G_{\mathbf k}^{\lambda}\right)^2
\delta(\omega_k-\Omega).
\end{equation}
Repeating the calculation above for an arbitrary positive frequency $\Omega$ gives
\begin{equation}
    J^{(g)}(\Omega)
=
\frac{G\hbar^3}
{30\pi c^5\omega_m^2} \Omega^5.
\end{equation}
The corresponding vacuum decoherence rate is consequently,
\begin{equation}
    \gamma_\mathcal{V} = \frac{2\pi}{\hbar^2} J^{(g)}(2\omega_m).
\end{equation}
It is perhaps worthwhile to note that the gravitational spectral density scales as $\Omega^5$, while the electromagnetic spectral density associated with electric-dipole coupling scales as $\Omega^3$ \cite{breuer_petruccione_2002_the_theory_of_open_quantuma}. The additional two powers of frequency in the gravitational case can also be thought as reflecting the quadrupole nature of the gravitational interaction.

We can use the above results for the squeezed case, if we assume that the squeezing parameters only depend on the bath frequency, not on direction or polarization,
\begin{equation}
    N_{\mathbf k, \lambda} = N_{k} \equiv N, \quad M_{\mathbf k, \lambda} = M_{k} \equiv M, \quad k \equiv \frac{2\omega_m}{c}
\end{equation}
Then the master equation reads (dropping the SA),
\begin{equation}
\begin{aligned}
\dot\rho
=
\gamma_\mathcal{V}
\Bigg\{
&
\left(1+N\right)
\left[
b^2\rho b^{\dagger 2}
-
\frac12
\left\{
b^{\dagger 2}b^2,\rho
\right\}
\right]
\\
&+
N
\left[
b^{\dagger 2}\rho b^2
-
\frac12
\left\{
b^2b^{\dagger 2},\rho
\right\}
\right]
\\
&+
M
\left[
b^{\dagger 2}\rho b^{\dagger 2}
-
\frac12
\left\{
b^{\dagger 4},\rho
\right\}
\right]
\\
&+
M^*
\left[
b^2\rho b^2
-
\frac12
\left\{
b^4,\rho
\right\}
\right]
\Bigg\},
\end{aligned}
\end{equation}
Now, because the Kossakowski matrix is rank 1, this means that we can find a single jump operator and write down the equation in Lindblad form. That jump operator is
\begin{equation}
    L = \sqrt{1+N} b^2 + \frac{M}{\sqrt{1+N}} b^{\dagger 2}
\end{equation}
Then, reinstating $\rho \equiv \rho^{(m)}(t)$, for the squeezed graviton bath, the master equation, in the interaction picture, in Lindblad form reads
\begin{equation}
    \dot{\rho}_I^{(m)}(t) = \gamma_\mathcal{V} \, \mathcal D[L] \rho^{(m)}_I(t).
\end{equation}
where the dissipator $\mathcal D[L]$ is defined in Eq.~\eqref{eq_dissipator_definition}.

\subsection{Curing the divergences: the Lamb-shift Hamiltonian and renormalization}\label{sec_lamb_shift_Hamiltonian_renormalized}

For a stationary two-point bath correlator, we obtained
\begin{equation}
\begin{aligned}
        H_\text{LS}
&= \hbar \sum_{\Omega \in \{ 0, \pm 2\omega_m\}} S(\Omega) A^\dagger(\Omega) A(\Omega)   \\
&=
\hbar S(2\omega_m)\,
b^{\dagger 2}b^2
+
\hbar S(-2\omega_m)\,
b^2 b^{\dagger 2}
+
\hbar S(0)\,
(2b^\dagger b+1)^2
\end{aligned}
\end{equation}
In the case of vacuum, we have
\begin{equation}
    S(\omega)
=
\frac{1}{\hbar^2}
\mathcal P
\int_0^\infty d\Omega\,
\frac{J(\Omega)}{\omega -\Omega}, \qquad S(\omega) \in \mathbb R.
\end{equation}
We have denoted it in terms of the gravitational spectral density (which we defined in Eq.~\eqref{eq_spectral_density_gravitational_first}),
\begin{equation}
    J^{(g)}(\Omega)
=
\frac{G \hbar^3 }{30 \pi c^5 \omega_m^2} \Omega^5 \equiv \alpha \, \Omega^5.
\end{equation}
In the case when the Lamb-shift Hamiltonian does not diverge and has a finite value, it just corrects the bare system Hamiltonian. When it diverges, there are two ways forward.

One way amounts to introducing a cut-off function $f(\omega,\Lambda)$ in the spectral density,
\begin{equation}
    J(\omega,\Lambda) = J(\omega) f(\omega,\Lambda).
\end{equation}
A hard cut-off would be
\begin{equation}
    f(\omega,\Lambda) = \theta(\Lambda-\omega).
\end{equation}
This is reminiscent of the one originally introduced by Bethe \cite{bethe_1947_the_electromagnetic_shift_of_energy}, which simply changes the upper integration limit to a finite $\Lambda$. In open-systems literature, one usually introduces a smoother cut-off,
\begin{equation}
    f(\omega,\Lambda) = e^{-\omega/\Lambda}
\end{equation}
This procedure amounts to specifying the ultraviolet behavior of the bath phenomenologically. A hard cut-off removes all modes above $\Lambda$, while a smooth cut-off suppresses them continuously. In either case, the otherwise divergent principal-value integral is replaced by a finite cut-off-dependent Lamb shift. Once a value of $\Lambda$ and a cut-off profile are chosen, the Lamb shift becomes a finite prediction of the effective model, in principle contributing to measurable shifts.

Instead, we regularize the divergent principal-value integral by introducing an ultraviolet cutoff $\Lambda$. The cutoff is treated as an auxiliary regulator and is removed at the end of the calculation by taking the limit $\Lambda \to \infty$. Let us begin by rearranging the Lamb-shift Hamiltonian in the following way,
\begin{equation}
    H_\text{LS} = \chi_1 N^2 + \chi_2 N+\chi_3,
    \qquad N \equiv b^\dagger b,
\end{equation}
where we have defined
\begin{equation}
\begin{aligned}
    \chi_1 &\equiv \hbar S(2\omega_m) + \hbar S(-2\omega_m) + 4 \hbar S(0),\\
    \chi_2 &\equiv - \hbar S(2\omega_m) + 3\hbar S(-2\omega_m) + 4 \hbar S(0),\\
    \chi_3 &\equiv 2 \hbar S(-2\omega_m) + \hbar S(0).
\end{aligned}
\end{equation}
In the master equation, the last term commutes with the system density operator, so one can neglect it; we treat it here. The second term induces a frequency shift. The first term induces a Kerr nonlinearity.

We note that we need the integral
\begin{equation}
    I(\omega,\Lambda) = \mathcal P \int_0^\Lambda d\Omega \, \frac{\Omega^5}{\omega - \Omega}, \qquad I(\omega,\Lambda) \in \mathbb R.
\end{equation}
Its solution is
\begin{equation}
    I(\omega,\Lambda) = -\frac{\Lambda ^5}{5}-\frac{\Lambda ^4 \omega }{4}-\frac{\Lambda ^3 \omega ^2}{3}-\frac{\Lambda ^2 \omega ^3} {2} -\Lambda  \omega ^4+\omega ^5 \log \left| \frac{\omega }{\Lambda -\omega }\right|.
\end{equation}
Thus assuming $0< 2\omega_m<\Lambda$, we obtain
\begin{equation}
\begin{aligned}
    I(2\omega_m,\Lambda) &= 
    -\frac{\Lambda ^5}{5}-\frac{\Lambda ^4 \omega _m}{2}-\frac{4}{3} \Lambda ^3 \omega _m^2-4 \Lambda ^2 \omega _m^3-16 \Lambda  \omega _m^4+32 \omega _m^5 \log \left(\frac{2 \omega _m}{\Lambda -2 \omega _m}\right), \\
    I(-2\omega_m,\Lambda) &=
    -\frac{\Lambda ^5}{5}+\frac{\Lambda ^4 \omega _m}{2}-\frac{4}{3} \Lambda ^3 \omega _m^2+4 \Lambda ^2 \omega _m^3-16 \Lambda  \omega _m^4-32 \omega _m^5 \log \left(\frac{2 \omega _m}{\Lambda +2 \omega _m}\right),
    \\
    I(0,\Lambda) &= -\frac{\Lambda ^5}{5}.
\end{aligned}
\end{equation}
Using these results, for the combinations in the regulated cut-off dependent Lamb-shift Hamiltonian, we get
\begin{equation}
\begin{aligned}
\chi_1^\Lambda
=&
\alpha\hbar
\Bigg[
-\frac{6}{5}\Lambda^5
-\frac{8}{3}\Lambda^3\omega_m^2
-32\Lambda\omega_m^4
+
32\omega_m^5
\log\left(
\frac{\Lambda+2\omega_m}{\Lambda-2\omega_m}
\right)
\Bigg],
\\[1ex]
\chi_2^\Lambda
=&
\alpha\hbar
\Bigg[
-\frac{6}{5}\Lambda^5
+
2\Lambda^4\omega_m
-\frac{8}{3}\Lambda^3\omega_m^2
+
16\Lambda^2\omega_m^3
-32\Lambda\omega_m^4
-
32\omega_m^5
\log\left(
\frac{2\omega_m}{\Lambda-2\omega_m}
\right)
-
96\omega_m^5
\log\left(
\frac{2\omega_m}{\Lambda+2\omega_m}
\right)
\Bigg],
\\[1ex]
\chi_3^\Lambda
=&
\alpha\hbar
\Bigg[
-\frac{3}{5}\Lambda^5
+
\Lambda^4\omega_m
-\frac{8}{3}\Lambda^3\omega_m^2
+
8\Lambda^2\omega_m^3
-32\Lambda\omega_m^4
-
64\omega_m^5
\log\left(
\frac{2\omega_m}{\Lambda+2\omega_m}
\right)
\Bigg].
\end{aligned}
\end{equation}
With the cutoff $\Lambda$ written explicitly, the Lamb-shift Hamiltonian becomes
\begin{equation}
    H_\text{LS}^\Lambda = \chi_1^\Lambda N^2 + \chi_2^\Lambda N+\chi_3^\Lambda.
\end{equation}
Introducing the counterterm Hamiltonian $H_\text{CT}$, we fix it such that we can impose the renormalization condition
\begin{equation}
    \lim_{\Lambda\to\infty}
    \left(
        H_{\rm LS}^{\Lambda}
        +
        H_{\rm CT}^{\Lambda}
    \right)
    =
    0 .
\end{equation}
Equivalently, writing
\begin{equation}
    H_{\rm CT}^{\Lambda}
    =
    \delta\chi_1^\Lambda N^2
    +
    \delta\chi_2^\Lambda N
    +
    \delta\chi_3^\Lambda ,
\end{equation}
we choose the counterterms such that
\begin{equation}
    \lim_{\Lambda\to\infty}
    \left(
        \chi_i^\Lambda
        +
        \delta\chi_i^\Lambda
    \right)
    =
    0,
    \qquad
    i=1,2,3 .
\end{equation}
Thus we find the counterterms coefficients to be
\begin{equation}
\begin{aligned}
    \delta\chi_1^\Lambda
&=
\alpha\hbar
\left[
\frac{6}{5}\Lambda^5
+
\frac{8}{3}\Lambda^3\omega_m^2
+
32\Lambda\omega_m^4
\right], \\
\delta\chi_2^\Lambda
&=
\alpha\hbar
\left[
\frac{6}{5}\Lambda^5
-
2\Lambda^4\omega_m
+
\frac{8}{3}\Lambda^3\omega_m^2
-
16\Lambda^2\omega_m^3
+
32\Lambda\omega_m^4
+
128\omega_m^5
\log\left(\frac{2\omega_m}{\Lambda}\right)
\right],\\
\delta\chi_3^\Lambda
&=
\alpha\hbar
\left[
\frac{3}{5}\Lambda^5
-
\Lambda^4\omega_m
+
\frac{8}{3}\Lambda^3\omega_m^2
-
8\Lambda^2\omega_m^3
+
32\Lambda\omega_m^4
+
64\omega_m^5
\log\left(\frac{2\omega_m}{\Lambda}\right)
\right].
\end{aligned}
\end{equation}
This choice fixes both the power-law and logarithmic UV-sensitive pieces and
corresponds to the fact that the matter Hamiltonian
\begin{equation}
    H^{(m)}
    =
    \hbar\omega_m N = \hbar \omega_m b^\dagger b,
\end{equation}
is expressed in terms of the observed mechanical frequency $\omega_m$.

The counterterm Hamiltonian is of the same perturbative order as the
Lamb-shift Hamiltonian, i.e.~second order in the system--bath coupling. Consequently, including $H_{\rm CT}^{\Lambda}$ in the free Hamiltonian used to
define the interaction picture would modify the Bohr frequencies and the
system eigenoperators only at order $g^2$. Since the Born--Markov dissipator is
itself already of order $g^2$, such modifications would affect the dissipator
only at order $g^4$, which is beyond the accuracy of the present master
equation. It is therefore consistent to derive the Born--Markov dissipator using the harmonic matter Hamiltonian $H^{(m)}$, and to include the counterterms only in the Hamiltonian part of the final
master equation, with the renormalization condition $\lim_{\Lambda\to\infty} \left(H^{(m)}+H_{\rm LS}^{\Lambda}+H_{\rm CT}^{\Lambda}\right)=H^{(m)}$. The renormalized part therefore is simply generated by $H^{(m)}$, while the dissipator is unchanged at the order retained.

The same renormalization prescription applies to number, thermal, and squeezed
graviton states (coherent states are vacuum-like at the level of connected
two-point functions).




\section{State of gravitons: Final master equations and physical consequences}\label{sec_final_master_equations}

\subsection{Coherent bath}\label{sec_coherent_bath_consequences}

For the multi-mode coherent graviton bath, we found that the nonvanishing one-point function generates the additional coherent-shift Hamiltonian \eqref{eq_coherent_shift_Hamiltonian_multimode}, while the dissipative contribution remains identical to that of the vacuum. The expectation value of the metric perturbation is a $c$-number field satisfying the classical linearized wave equation and may therefore be identified with the corresponding classical gravitational wave,
\begin{equation}\label{eq_coherent_state_classical_identification}
    h_{ij}^{\mathrm{CL}}(t,\mathbf x)\equiv\langle h_{ij}(t,\mathbf x)\rangle_\alpha.    
\end{equation} 
We show below that this identification is physically consistent: the coherent-shift Hamiltonian becomes precisely the classical tidal interaction. It also provides a direct relation between the coherent-state amplitude profile and the observable gravitational-wave strain.

\secpar{The tidal Hamiltonian of a classical gravitational wave}
The identification can be made directly at the Hamiltonian level. Namely, using the quantum tidal interaction \eqref{eq_interaction_Hamiltonian_derived}, with the identification \eqref{eq_coherent_state_classical_identification}, gives
\begin{equation}
H_{\mathrm{CS}}(t)
=
-\frac{m}{4}\,
\ddot h_{xx}^{\mathrm{CL}}(t,\mathbf 0)\,x^2.
\end{equation}
This is exactly the tidal Hamiltonian of a classical gravitational wave. Remarkably, the standard classical tidal response is therefore contained directly in the one-point function of the quantized metric perturbation. 

\secpar{The geodesic-deviation equation}
To make the connection with the usual classical description explicit, we now derive the corresponding equation of motion. Take the total matter Hamiltonian at face value,
\begin{equation}
    H^{(m)}+H_{\mathrm{CS}}(t)
=
\frac{p^2}{2m}
+\frac12m\omega_m^2x^2
-\frac{m}{4}\ddot h_{xx}^{\mathrm{CL}}(t,\mathbf0)x^2.
\end{equation}
Hamilton's equations give
\begin{equation}
    \dot x=\frac{p}{m},
\qquad
\dot p
=
-m\omega_m^2x
+\frac{m}{2}\ddot h_{xx}^{\mathrm{CL}}(t,\mathbf0)x.
\end{equation}
Thus, we obtain the equation of motion for the relative coordinate
\begin{equation}
\ddot x+\omega_m^2x
=
\frac12\ddot h_{xx}^{\mathrm{CL}}(t,\mathbf0)x.
\end{equation}
But this is exactly the geodesic-deviation equation supplemented by the mechanical restoring force. In the freely falling limit, $\omega_m \to 0$,
\begin{equation}
    \ddot x
=
\frac12\ddot h_{xx}^{\mathrm{CL}}x.
\end{equation}
This is precisely the standard equation used to describe the relative response of freely falling test masses \cite{parikh_wilczek_zahariade_2021_signatures_of_the_quantization_of_gravity}.

\secpar{The coherent-state eigenvalue is related to the gravitational-wave strain}
We wrote the general solution for a classical (CL) gravitational wave as (Section~\ref{app_gravity_hamiltonian}),
\begin{equation}
    h_{ij}^\text{CL}(t,\mathbf x) = \sum_\lambda \int d^3k\,
    A_\klambda e_{ij}^\lambda(\mathbf n) e^{-i(\omega_k t- \mathbf k \cdot \mathbf x)} + \text{c.c.}.
\end{equation}

For a single monochromatic classical gravitational wave, with a wave vector $\mathbf k_0$, polarization $\lambda_0$, we have
\begin{equation}
    h_{ij}^\text{CL}(t,\mathbf x) = 
    A_\klambdazero e_{ij}^{\lambda_0}(n_0) e^{-i(\omega_{k_0} t- \mathbf k_0 \cdot \mathbf x)} + \text{c.c.}
\end{equation}
Parametrizing the complex amplitude as $A_\klambdazero = |A_\klambdazero|e^{i\theta_\klambdazero}\equiv|A|e^{i \theta}$, and writing the $xx$-component in our local detector frame ($\mathbf x = 0$), we get
\begin{equation}
    h_{xx}^\text{CL}(t,0) = 2|A| \,e_{xx}^{\lambda_0}(\mathbf n)\cos(\omega_{k_0}t-\theta),
\end{equation}
In the case of a plus-polarized gravitational wave propagating in the $z$-direction, we have $e_{xx}^+(\mathbf n_0) = 1$ (Section~\ref{sec_polarization_tensor}), we write
\begin{equation}
    h_{xx}^\text{CL}(t,0) = h_+\cos(\omega_{k_0}t-\theta),
\end{equation}
where we have defined the gravitational-wave strain amplitude (in the plus-polarized case)
\begin{equation}
    h_+ \equiv 2 |A|.
\end{equation}
This can be compared to the quantized gravitational wave via the identification \eqref{eq_coherent_state_classical_identification}.
For a multi-mode coherent state sharply peaked around the mode $(\mathbf k_0,\lambda_0)$, we obtain
\begin{equation}
h_+=2\mathcal G_{xx}^{\mathbf k_0,+}|\alpha|
=
\frac{16\omega_m}{\hbar\omega_{k_0}^2}G_{k_0}^+
|\alpha|.
\end{equation}
This is the quantity which we refer to in Fig.~\ref{fig_fidelity_comparison}. Conversely, this identification allows us to express the mean graviton occupation number of the coherent mode directly in terms of the classical gravitational-wave strain via $\bar n
\equiv
\langle g^\dagger g\rangle_\alpha
=
|\alpha|^2$.

\subsection{Number and thermal baths}\label{sec_number_thermal_consequences}

\secpar{Coherences}
Let us, for convenience, rewrite the common-structure master equation ($\rho \equiv \rho^{(m)}(t)$)
\begin{equation}
    \dot{\rho} = - i \omega_m[b^\dagger b,\rho] + \gamma_\downarrow \mathcal{D}[b^2]\rho + \gamma_\uparrow \mathcal{D}[b^{\dagger 2}]\rho
\end{equation}
For notational convenience, we use $n$ as a state-dependent shorthand, defined by
\begin{equation}
\begin{aligned}
    n =
\begin{cases}
n(2\omega_m), & \text{for the number-state case}, \\
n_\beta(2\omega_m), & \text{for the thermal-state case},
\end{cases}
\end{aligned}
\end{equation}
such that the rates are given by,
\begin{equation}
    \gamma_\downarrow = \gamma_\mathcal{V}(1+n), \qquad \gamma_\uparrow = \gamma_\mathcal{V} n
\end{equation}
In the matter number-state basis we define the matrix element of the density operator as
\begin{equation}\label{eq_density_operator_matrix_elements_number}
    \rho_{pq}(t) \equiv \langle p |\rho(t)|q\rangle
\end{equation}
The diagonal elements $\rho_{pp}$ denote populations, while the off-diagonal elements $\rho_{pq}$ denote the coherences.  

The standard harmonic-oscillator ladder relations are
\begin{equation}
\begin{aligned}
     b \ket{q} &= \sqrt q \,\ket{q-1}, \\
     b^\dagger \ket{q} &= \sqrt{q+1} \, \ket{q+1}.
\end{aligned}
\end{equation}
Relevant to our case, we have
\begin{equation}
\begin{aligned}
    b^2|q\rangle
&=
b\left(\sqrt{q}|q-1\rangle\right)
=
\sqrt{q(q-1)}\,|q-2\rangle, \\
b^{\dagger 2}|q\rangle
&=
b^\dagger\left(\sqrt{q+1}|q+1\rangle\right)
=
\sqrt{(q+1)(q+2)}\,|q+2\rangle.
\end{aligned}
\end{equation}
Using these relations, we obtain the master equation for $\rho_{pq}$,
\begin{equation}\label{eq_generic_rhopq}
\begin{aligned}
    \dot\rho_{pq}
    =&
    - i\omega_m(p-q)\rho_{pq}
    \\
    &+
    \gamma_\downarrow
    \Big[
    \sqrt{(p+1)(p+2)(q+1)(q+2)}\,\rho_{p+2,q+2}
    -\frac12\big(p(p-1)+q(q-1)\big)\rho_{pq}
    \Big]
    \\
    &+
    \gamma_\uparrow
    \Big[
    \sqrt{p(p-1)q(q-1)}\,\rho_{p-2,q-2}
    -\frac12\big((p+1)(p+2)+(q+1)(q+2)\big)\rho_{pq}
    \Big].    
\end{aligned}
\end{equation}
Here it is understood that $\rho_{p-2,q-2}=0$ if $p<2$ or $q<2$.
The first line is the free Hamiltonian phase rotation. In the interaction picture with respect to $H^{(m)}$, this line is absent. This equation describes the decay and redistribution of coherences.

For comparison, the population equation for $p_m \equiv \rho_{mm}$ is
\begin{equation}
\begin{aligned}
    \dot p_m
=
\gamma_\downarrow (m+1)(m+2)p_{m+2}
+
\gamma_\uparrow m(m-1)p_{m-2} \\
-
\Big[
\gamma_\downarrow m(m-1)
+
\gamma_\uparrow (m+1)(m+2)
\Big]p_m .
\end{aligned}
\end{equation}
Note that the Hamiltonian term vanishes. This equation describes relaxation and heating between number states.

\secpar{Coherence ratio}
In order to not bother with the Hamiltonian term, a good diagnostic of the loss of a particular coherence is
\begin{equation}
    R_{pq}(t) \equiv \frac{|\rho_{pq}(t)|}{|\rho_{pq}(0)|}.
\end{equation}
This coherence ratio is the normalized survival of the $(p,q)$ coherence, and in our case, is picture-independent. Since this is the quantity we are going to be interested in, we can continue working in the interaction picture for the density operator (we omit writing the free Hamiltonian). Indeed, since the interaction-picture (I) matrix elements are related to the Schrödinger-picture (S) ones via
\begin{equation}
    \rho_{pq,{I}}(t)
=
e^{i\omega_m(p-q)t}\rho_{pq,{S}}(t),
\end{equation}
we have
\begin{equation}
    |\rho_{pq,{I}}(t)|
=
|e^{i\omega_m(p-q)t}\rho_{pq,{S}}(t)| = |\rho_{pq,{S}}(t)|.
\end{equation}
On the other hand, the normalization removes the dependence on the initial coherence magnitude, so that $R_{pq}(t)$ measures the fraction of the initial coherence that survives at time $t$.

\secpar{Coherence ladder}
Due to the fact that both indices shift together in Eq.~\eqref{eq_generic_rhopq}, $(p,q) \rightarrow (p\pm2, q\pm 2)$, and that $p\neq q$ for coherences, we can define the coherence ladder
\begin{equation}
    c_j(t)\equiv \rho_{2j,2j+1}(t).
\end{equation}
The conjugate ladder is
\begin{equation}
    c_j^*(t)\equiv \rho_{2j+1,2j}(t).
\end{equation}
Namely,
\begin{equation}
    c_0 = \rho_{01}, \qquad c_1 = \rho_{23}, \qquad c_2=\rho_{45},\qquad\ldots.
\end{equation}
Then the master equation \eqref{eq_generic_rhopq}, in the interaction picture, becomes
\begin{equation}\label{eq_coherence_ladder_general}
\begin{aligned}
\dot c_j
=&\,
\gamma_\downarrow
\left[
(2j+2)\sqrt{(2j+1)(2j+3)}\,c_{j+1}
-
4j^2c_j
\right]
\\
&+
\gamma_\uparrow
\left[
2j\sqrt{(2j-1)(2j+1)}\,c_{j-1}
-
4(j+1)^2c_j
\right].
\end{aligned}
\end{equation}

\secpar{The encoded qubit state}
Let us now turn our attention to the encoded qubit state, which we rewrite for convenience,
\begin{equation}\label{eq_bare_number_state_qubit}
    \rho_{\vartheta,\varphi}^{(m)}
\equiv
|\psi_{\vartheta,\varphi}^{(m)}\rangle\langle\psi_{\vartheta,\varphi}^{(m)}|,
\qquad
|\psi_{\vartheta,\varphi}^{(m)}\rangle
=
\cos({\vartheta}/{2})\,|0\rangle
+
e^{i\varphi}\sin({\vartheta}/{2})\,|1\rangle.
\end{equation}
Expanded, this gives
\begin{equation}
\begin{aligned}
\rho_{\vartheta,\varphi}^{(m)}
&=
\cos^2\frac{\vartheta}{2}\ket{0}\bra{0}
+
\sin^2\frac{\vartheta}{2}\ket{1}\bra{1}
\\
&\;\;\;\;\;+
\frac{1}{2}\sin\vartheta\,e^{-i\varphi}\ket{0}\bra{1}
+
\frac{1}{2}\sin\vartheta\,e^{i\varphi}\ket{1}\bra{0}.
\end{aligned}
\end{equation}
In matrix form,
\begin{equation}
    \rho_{\rm enc}(0)
=
\begin{pmatrix}
\rho_{00} & \rho_{01} \\
\rho_{10} & \rho_{11}
\end{pmatrix}
=
\begin{pmatrix}
\cos^2\frac{\vartheta}{2}
&
\frac{1}{2}\sin\vartheta\,e^{-i\varphi}
\\
\frac{1}{2}\sin\vartheta\,e^{i\varphi}
&
\sin^2\frac{\vartheta}{2}
\end{pmatrix}.
\end{equation}
Therefore, all the properties of its coherence are given by $\rho_{01}$, noting that $\rho_{10}= \rho_{01}^*$. Using the master equation \eqref{eq_generic_rhopq}, for a generic $\rho_{01}$, we have
\begin{equation}
    \dot\rho_{01}
=
2\sqrt{3}\,\gamma_\downarrow\rho_{23}
-
4\gamma_\uparrow\rho_{01}.
\end{equation}
The next equation governs the evolution of the $\rho_{23}$, which is given by
\begin{equation}
    \dot\rho_{23}
=
\gamma_{\downarrow}
\left(
4\sqrt{15}\,\rho_{45}
-
4\rho_{23}
\right)
+
\gamma_{\uparrow}
\left(
2\sqrt{3}\,\rho_{01}
-
16\rho_{23}
\right).
\end{equation}
The next equation governs the evolution of the $\rho_{45}$, which is given by
\begin{equation}
    \dot\rho_{45}
=
\gamma_{\downarrow}
\left(
6\sqrt{35}\,\rho_{67}
-
16\rho_{45}
\right)
+
\gamma_{\uparrow}
\left(
4\sqrt{15}\,\rho_{23}
-
36\rho_{45}
\right).
\end{equation}
And so on.

\secpar{Reducing to the vacuum result}
Let us turn our attention to the vacuum, in which case the evolution of the coherences should confirm our previous result. The vacuum limit is immediate. Setting $n=0$, i.e.~$\gamma_\downarrow = \gamma_\mathcal{V}$ and $\gamma_\uparrow = 0$, we obtain
\begin{equation}
\begin{aligned}
\dot\rho_{01}
&=
2\sqrt{3}\,\gamma_{\mathcal V}\rho_{23}, \\
    \dot\rho_{23}
&=
\gamma_{\mathcal V}
\left(
4\sqrt{15}\,\rho_{45}
-
4\rho_{23}
\right),\\
\dot\rho_{45}
&=
\gamma_{\mathcal V}
\left(
6\sqrt{35}\,\rho_{67}
-
16\rho_{45}
\right),\\
&\cdots.
\end{aligned}
\end{equation}
But now suppose that we have an initially encoded qubit. That is, we are assuming the initial condition,
\begin{equation}
    \rho_{23}(0)=\rho_{45}(0)=\rho_{67}(0)=\cdots=0.
\end{equation}
Then the ladder of equations simply reduces to one equation,
\begin{equation}
    \dot\rho_{01}
=
0.
\end{equation}
Therefore we obtain the result that $\rho_{01}$ is constant in time,
\begin{equation}
    \rho_{01}(t) = \rho_{01}(0).
\end{equation}
For the arbitrary encoded qubit,
\begin{equation}
    \rho_{01}^{\vartheta,\varphi}(t) = \rho_{01}^{\vartheta,\varphi}(0) 
=
\frac12\sin\vartheta\,e^{-i\varphi}, \qquad \rho_{10}^{\vartheta,\varphi}(t) = \rho_{01}^{\vartheta,\varphi\,*}(t).  
\end{equation}
Equivalently, since $\gamma_\uparrow$ is real and positive,
\begin{equation}
    R_{01}^\mathcal{V}(t)
\equiv
\frac{|\rho_{01}^{\vartheta,\varphi}(t)|}{|\rho^{\vartheta,\varphi}_{01}(0)|}
=
1.
\end{equation}
For completeness, we note that, in the Schrödinger picture, the free Hamiltonian adds the phase. This explains the oscillations in Fig.~\ref{fig_fidelity_comparison}, since for the fidelity in this case we obtain
\begin{equation}
\mathcal{F}(t)
=
1-\sin^2\vartheta\,
\sin^2\left(\frac{\omega_m t}{2}\right).
\end{equation}
For $\vartheta = \pi/2$,
\begin{equation}
\mathcal{F}(t)
=
\cos^2\left(\frac{\omega_m t}{2}\right),
\end{equation}
which is exactly what the plot shows.

\secpar{Approximate exponential}
Now let us discuss how a number or thermal bath modifies the vacuum result. 
To get a feel for the short-time correction, we start once again from the generic equation for the encoded coherence \(\rho_{01}\),
\begin{equation}\label{eq_rho_01_n}
\begin{aligned}
            \dot\rho_{01}
&=
2\sqrt{3}\,\gamma_\downarrow\rho_{23}
-
4\gamma_\uparrow\rho_{01}, \\
&\ldots.
\end{aligned}
\end{equation}
Assuming that the oscillator is initially prepared as an encoded qubit amounts to saying that the higher coherences are initially zero,
\begin{equation}
    \rho_{23}(0)=\rho_{45}(0)=\cdots=0.
\end{equation}
Evaluating the differential equation \eqref{eq_rho_01_n} at \(t=0\) then gives
\begin{equation}\label{eq_rho_01_initial_t_0}
    \left.\frac{d}{dt}\rho_{01}(t)\right|_{t=0}
=
-4\gamma_{\uparrow}\rho_{01}(0).
\end{equation}
We now Taylor expand the time-dependent function \(\rho_{01}(t)\) around \(t=0\),
\begin{equation}
    \rho_{01}(t)
=
\rho_{01}(0)
+
t\left.\frac{d}{dt}\rho_{01}(t)\right|_{t=0}
+
\mathcal O(t^2).
\end{equation}
Substituting the initial derivative \eqref{eq_rho_01_initial_t_0} into this expansion, and rearranging the terms, we obtain
\begin{equation}
    \rho_{01}(t)
=
\rho_{01}(0)
\left[
1-4\gamma_{\uparrow}t+\mathcal O(t^2)
\right].
\end{equation}
Since
\begin{equation}
    e^{-4\gamma_{\uparrow}t}
=
1-4\gamma_{\uparrow}t+\mathcal O(t^2),
\end{equation}
we can represent the leading short-time behavior by the exponential approximation (which we denote with a star)
\begin{equation}
    \rho_{01}^\star(t)
=
\rho_{01}(0)e^{-4\gamma_{\uparrow}t}.
\end{equation}
From this, we obtain the coherence ratio \eqref{eq_coherence_ratio_p_q},
\begin{equation}
    R_{01}^\star(t) = e^{-4\gamma_{\uparrow}t}.
\end{equation}
Of course, this expression should only be understood as a short-time approximation capturing the initial correction to the vacuum result. The full finite-occupation dynamics is not closed on $\rho_{01}$ alone; in general, one must include the higher coherence ladder contributions.

\secpar{Truncated coherence-ladder solution}
The equation for $c_j$, given in Eq.~\eqref{eq_coherence_ladder_general}, is not closed at any finite value of $j$, since each $c_j$ is coupled to its neighboring coherences $c_{j-1}$ and
$c_{j+1}$. Thus, even if the oscillator is initially prepared in the
encoded subspace, so that $c_{j>0}(0)=0$, the finite-temperature dynamics generates higher coherences. The full problem is therefore an infinite coherence ladder.

To solve the system numerically, we truncate the ladder at a finite value
$M$. We keep
\begin{equation}
    c_0,c_1,\ldots,c_{M-1},
\end{equation}
and omit all coherences $c_j$ with $j\geq M$. Equivalently, we replace
the infinite hierarchy by the finite-dimensional system
\begin{equation}
    \frac{d}{dt}\mathbf c^{(M)}(t)
    =A^{(M)} \mathbf c^{(M)}(t),
\end{equation}
where
\begin{equation}
    \mathbf c^{(M)}(t)
    =
    \big(c_0(t),\ldots,c_{M-1}(t)\big)^T .
\end{equation}
The matrix $A^{(M)}$ contains the ladder couplings appearing above, with the upper boundary imposed by dropping the coupling from $c_{M-1}$ to $c_M$. Its first few rows are
\begin{equation}
    A^{(M)}
=
\begin{pmatrix}
-4\gamma_\uparrow
&
2\sqrt{3}\gamma_\downarrow
&
0
&
0
&
\cdots
\\
2\sqrt{3}\gamma_\uparrow
&
-4\gamma_\downarrow-16\gamma_\uparrow
&
4\sqrt{15}\gamma_\downarrow
&
0
&
\cdots
\\
0
&
4\sqrt{15}\gamma_\uparrow
&
-16\gamma_\downarrow-36\gamma_\uparrow
&
6\sqrt{35}\gamma_\downarrow
&
\cdots
\\
0
&
0
&
6\sqrt{35}\gamma_\uparrow
&
-36\gamma_\downarrow-64\gamma_\uparrow
&
\cdots
\\
\vdots
&
\vdots
&
\vdots
&
\vdots
&
\ddots
\end{pmatrix}.
\end{equation}

The finite-ladder solution is then
\begin{equation}
    \mathbf c^{(M)}(t)
    =
    e^{A^{(M)}t}\mathbf c^{(M)}(0).
\end{equation}
For an initially encoded qubit,
\begin{equation}
    \mathbf c^{(M)}(0)
    =
    \big(\rho_{01}(0),0,\ldots,0\big)^T .
\end{equation}
We define the corresponding normalized coherence ratio by
\begin{equation}
    R_{01,{\mathcal N,\mathcal T}}^{(M)}(t)
\equiv
\frac{|\rho_{01}^{(M)}(t)|}{|\rho_{01}(0)|}.
\end{equation}
In the numerical results we use $M=40$, corresponding to the
truncation $j=0,\ldots,39$. Thus the plotted curve in Fig.~\ref{fig_coherence_ratio_comparison}, $ R_{01,{\mathcal N,\mathcal T}}^{(40)}(t)$, is the exact solution of the finite $M=40$ truncated ladder, and an approximation to the full infinite coherence hierarchy.

\subsection{Squeezed bath}\label{sec_squeezed_bath_consequences}

Let us, for convenience, rewrite the master equation ($\rho \equiv \rho^{(m)}(t)$, $r \equiv r(2\omega_m), \phi \equiv\phi(2\omega_m)$),
\begin{equation}
    \dot{\rho} = - i \omega_m[b^\dagger b, \rho] + \gamma_\mathcal{V} \mathcal{D}[L_\mathcal{S}]\rho,
    \qquad
    L_\mathcal{S} = \cosh r \, b^2 - e^{i\phi}\sinh r \, b^{\dagger,2}
\end{equation}
It is useful to go back to the non-diagonal Kossakowski matrix version of the equation, with
\begin{equation}
    N\equiv\sinh^2r,
\qquad
M\equiv-e^{i\phi}\sinh r\cosh r.
\end{equation}
We then get
\begin{equation}
\begin{aligned}
\dot\rho
={}&
-{i}\omega_m [b^\dagger b, \rho]
\\
&+
\gamma_{\mathcal V}\Bigg\{
(N+1)
\left[
b^2\rho b^{\dagger2}
-\frac12
\left\{
b^{\dagger2}b^2,\rho
\right\}
\right]
\\
&\qquad\qquad
+
N
\left[
b^{\dagger2}\rho b^2
-\frac12
\left\{
b^2b^{\dagger2},\rho
\right\}
\right]
\\
&\qquad\qquad
+
M
\left[
b^{\dagger2}\rho b^{\dagger2}
-\frac12
\left\{
b^{\dagger4},\rho
\right\}
\right]
\\
&\qquad\qquad
+
M^\ast
\left[
b^2\rho b^2
-\frac12
\left\{
b^4,\rho
\right\}
\right]
\Bigg\}.
\end{aligned}
\end{equation}
In the matter number-state basis \eqref{eq_density_operator_matrix_elements_number}
\begin{equation}
    \rho_{pq}(t) \equiv \langle p |\rho(t)|q\rangle
\end{equation}
we obtain the master equation for $\rho_{pq}$,
\begin{equation}
\begin{aligned}
\dot\rho_{pq}
={}&
-i\omega_m(p-q)\rho_{pq}
\\
&+
\gamma_{\mathcal V}(N+1)
\sqrt{(p+1)(p+2)(q+1)(q+2)}
\,\rho_{p+2,q+2}
\\
&+
\gamma_{\mathcal V}N
\sqrt{p(p-1)q(q-1)}
\,\rho_{p-2,q-2}
\\
&-
\frac{\gamma_{\mathcal V}}{2}
\Big[
(N+1)\big(p(p-1)+q(q-1)\big)
\\
&\;\;\;\;\;\;\;\;\;\;\;
+
N\big((p+1)(p+2)+(q+1)(q+2)\big)
\Big]\rho_{pq}
\\
&+
\gamma_{\mathcal V}M
\sqrt{p(p-1)(q+1)(q+2)}
\,\rho_{p-2,q+2}
\\
&+
\gamma_{\mathcal V}M^\ast
\sqrt{(p+1)(p+2)q(q-1)}
\,\rho_{p+2,q-2}
\\
&-
\frac{\gamma_{\mathcal V}M}{2}
\sqrt{p(p-1)(p-2)(p-3)}
\,\rho_{p-4,q}
\\
&-
\frac{\gamma_{\mathcal V}M}{2}
\sqrt{(q+1)(q+2)(q+3)(q+4)}
\,\rho_{p,q+4}
\\
&-
\frac{\gamma_{\mathcal V}M^\ast}{2}
\sqrt{(p+1)(p+2)(p+3)(p+4)}
\,\rho_{p+4,q}
\\
&-
\frac{\gamma_{\mathcal V}M^\ast}{2}
\sqrt{q(q-1)(q-2)(q-3)}
\,\rho_{p,q-4}.
\end{aligned}
\end{equation}
As before, it is understood that terms with negative indices vanish. 

\secpar{Approximate exponential}
Now let us discuss how a squeezed bath modifies the vacuum result, similarly to the number/thermal baths above. 
To get a feel for the short-time correction, we start once again from the generic equation for the encoded coherence \(\rho_{01}\), in the interaction picture,
\begin{equation}\label{eq_rho_01_S}
\begin{aligned}
\dot\rho_{01}
&=
i\omega_m\rho_{01}
-4\gamma_{\mathcal V}N\,\rho_{01}
\\
&\;\;\;\;+
2\sqrt{3}\,\gamma_{\mathcal V}(N+1)\rho_{23}
-\sqrt{30}\,\gamma_{\mathcal V}M\,\rho_{05}
-\sqrt{6}\,\gamma_{\mathcal V}M^\ast\rho_{41}.\\
&\ldots.
\end{aligned}
\end{equation}
Assuming that the oscillator is initially prepared as an encoded qubit amounts to saying that the higher coherences are initially zero,
\begin{equation}
    \rho_{23}(0)=\rho_{05}(0)=\rho_{41}(0)=\cdots=0.
\end{equation}
Evaluating the differential equation \eqref{eq_rho_01_S} at \(t=0\) then gives
\begin{equation}\label{eq_rho_01_initial_t_0_S}
    \left.\frac{d}{dt}\rho_{01}(t)\right|_{t=0}
=-4\gamma_{\mathcal V}N
\rho_{01}(0).
\end{equation}
We now Taylor expand the time-dependent function \(\rho_{01}(t)\) around \(t=0\),
\begin{equation}
    \rho_{01}(t)
=
\rho_{01}(0)
+
t\left.\frac{d}{dt}\rho_{01}(t)\right|_{t=0}
+
\mathcal O(t^2).
\end{equation}
Substituting the initial derivative \eqref{eq_rho_01_initial_t_0_S} into this expansion, and rearranging the terms, we obtain
\begin{equation}
\rho_{01}(t)
=
\rho_{01}(0)
\left[
1+
-4\,\gamma_{\mathcal V}N
\,t
+
\mathcal O(t^2)
\right].
\end{equation}
Since
\begin{equation}
    e^{-4\,\gamma_\mathcal{V} N\,t}
=
1+
-4\,\gamma_{\mathcal V}N
\,t
+
\mathcal O(t^2),
\end{equation}
we can represent the leading short-time behavior by the exponential approximation (which we denote with a star). From it, we obtain the coherence ratio \eqref{eq_coherence_ratio_p_q},
\begin{equation}
    R_{01,\mathcal{S}}^\star(t) = e^{-4\gamma_\mathcal{V} \sinh^2 r \,t}.
\end{equation}
where we have reinstated $N = \sinh^2 r$. For $r = 0$, this reduces to the vacuum result. Similarly, this expression should only be understood as a short-time approximation capturing the initial correction to the vacuum result in the squeezed case. Of course, in general, one must include the higher coherence ladder contributions.

\secpar{Dark sector}
A pure state $\ket{D}$ is dark with respect to the dissipator when
\begin{equation}\label{eq_dark_state__equation_squeezed}
    L_\mathcal{S} \ket{D} = 0.
\end{equation}
More generally, if several states obey this, every density operator supported on their span is also dark.

Since $\cosh r>0$, we can divide Eq.~\eqref{eq_dark_state__equation_squeezed} by $\cosh r$ and define
\begin{equation}
    \lambda\equiv e^{i\phi}\tanh r.
\end{equation}
Then the equation becomes
\begin{equation}\label{eq_eigenvalue_eq_squeezed}
    \left(b^2-\lambda b^{\dagger2}\right)\ket D=0.
\end{equation}
Now let us write the most pure general matter state as
\begin{equation}
    \ket D=\sum_{n=0}^{\infty}c_n\ket n.
\end{equation}
Using
\begin{equation}
    b^2\ket n=\sqrt{n(n-1)}\ket{n-2},
    \qquad
    b^{\dagger2}\ket n=\sqrt{(n+1)(n+2)}\ket{n+2},
\end{equation}
we obtain
\begin{align}
    b^2\ket D
    &=
    \sum_{n=0}^{\infty}
    c_{n+2}\sqrt{(n+1)(n+2)}\ket n,
    \\
    b^{\dagger2}\ket D
    &=
    \sum_{n=0}^{\infty}
    c_{n-2}\sqrt{n(n-1)}\ket n,
\end{align}
where coefficients with negative indices are understood to vanish.

Inserting the above in Eq.~\eqref{eq_eigenvalue_eq_squeezed}, and equating the coefficient of every $\ket n$, gives
\begin{equation}
    \sqrt{(n+1)(n+2)}\,c_{n+2}
    =
    \lambda\sqrt{n(n-1)}\,c_{n-2}.
\end{equation}
Equivalently,
\begin{equation}\label{eq_recurrence_relation_c_n}
    c_{n+2}
    =
    \lambda
    \sqrt{\frac{n(n-1)}{(n+1)(n+2)}}\,c_{n-2}.
\end{equation}
This relation shows that the recurrence advances the occupation number by four. First, we find that $c_2$ and $c_3$ vanish, while the $c_0$ and $c_1$ survive. Starting from each of the latter, we observe two independent solutions which belong to the even and odd parity sectors, respectively.

To find the exact even-parity solution we can set $n = 4j+2$ in Eq.~\eqref{eq_recurrence_relation_c_n}. Then
\begin{equation}
    c_{4j+4}
    =
    \lambda
    \sqrt{
    \frac{(4j+1)(4j+2)}{(4j+3)(4j+4)}
    }\,c_{4j}.
\end{equation}
Iteration gives
\begin{equation}
    c_{4j}
    =
    c_0\lambda^j
    \prod_{k=0}^{j-1}
    \sqrt{
    \frac{(4k+1)(4k+2)}{(4k+3)(4k+4)}
    }.
\end{equation}
Therefore the exact even-parity solution is
\begin{equation}
    \ket{D_+}
=
\mathcal N_+
\sum_{j=0}^{\infty}
\lambda^j
\prod_{k=0}^{j-1}
\sqrt{
\frac{(4k+1)(4k+2)}
{(4k+3)(4k+4)}
}
\ket{4j},
\end{equation}
where we have absorbed $c_0$ into the normalization $\mathcal{N_+}$.
Alternatively, the product can be written more compactly using the Pochhammer symbol,
\begin{equation}
    (a)_j = \frac{\Gamma(a+j)}{\Gamma(a)}, \qquad (a)_j=a(a+1)\cdots(a+j-1),
    \qquad
    (a)_0=1.
\end{equation}
where $\Gamma(z)$ is the Euler gamma function. Then the product becomes
\begin{equation}
    \prod_{k=0}^{j-1}
\frac{(4k+1)(4k+2)}
{(4k+3)(4k+4)}
=
\frac{
\left(\frac14\right)_j
\left(\frac12\right)_j
}{
\left(\frac34\right)_j
j!
},
\end{equation}
Thus the state can be written as
\begin{equation}
\ket{D_+}
=
\mathcal N_+
\sum_{j=0}^{\infty}
\lambda^j
\sqrt{
\frac{
\left(\frac14\right)_j
\left(\frac12\right)_j
}{
\left(\frac34\right)_j
j!
}
}
\ket{4j}.
\end{equation}
Equivalently, we can write it using the definition of the confluent hypergeometric function,
\begin{equation}
    {}_0F_1(;a;z)
=
\sum_{j=0}^{\infty}
\frac{z^j}{(a)_j j!}, \qquad
\end{equation}
In particular, for our case we note
\begin{equation}
    {}_0F_1\!\left(;\frac34;\frac{\lambda}{16}b^{\dagger4}\right)|0\rangle
=
\sum_{j=0}^{\infty}
\frac{1}{\left(\frac34\right)_j j!}
\left(\frac{\lambda}{16}b^{\dagger4}\right)^j
|0\rangle .
\end{equation}
Then the state can be written as
\begin{equation}
\ket{D_+}
=
\mathcal N_+\,
{}_0F_1
\left(
;\frac34;
\frac{\lambda}{16}b^{\dagger4}
\right)
\ket0 .
\end{equation}
Note that the square root appears after acting with the operator. For illustration, the first few terms are,
\begin{equation}
    |D_+\rangle
=
\mathcal N_+
\left[
|0\rangle
+
\frac{\lambda}{\sqrt6}|4\rangle
+
\sqrt{\frac{5}{56}}\lambda^2|8\rangle
+\cdots
\right],
\end{equation}
From the condition $\langle{D_+}\ket{D_+} = 1$, we find the normalization to be
\begin{equation}
\mathcal N_+
=
\left[
{}_2F_1
\left(
\frac14,\frac12;
\frac34;
|\lambda|^2
\right)
\right]^{-1/2}.
\end{equation}
In the small-$r$ expansion, we obtain
\begin{equation}
    |D_+\rangle
=
\left(1-\frac{r^2}{12}\right)|0\rangle
+
\frac{e^{i\phi}r}{\sqrt6}|4\rangle
+
\sqrt{\frac{5}{56}}e^{2i\phi}r^2|8\rangle
+
\mathcal O(r^3).
\end{equation}

Similarly, to find the exact odd-parity solution we can set $n = 4j+3$ in Eq.~\eqref{eq_recurrence_relation_c_n}. Then
\begin{equation}
   c_{4j+5}
=
\lambda
\sqrt{
\frac{(4j+3)(4j+2)}
{(4j+4)(4j+5)}
}
c_{4j+1}.
\end{equation}
Following exactly the same procedure as above, from
\begin{equation}
    \ket{D_-}
=
\mathcal N_-
\sum_{j=0}^{\infty}
\lambda^j
\prod_{k=0}^{j-1}
\sqrt{
\frac{(4k+2)(4k+3)}
{(4k+4)(4k+5)}
}
\ket{4j+1},
\end{equation}
where we have absorbed $c_1$ into the normalization $\mathcal{N}_-$, we obtain, in terms of the Pochhammer symbol,
\begin{equation}
\ket{D_-}
=
\mathcal N_-
\sum_{j=0}^{\infty}
\lambda^j
\sqrt{
\frac{
\left(\frac12\right)_j
\left(\frac34\right)_j
}{
\left(\frac54\right)_j j!
}
}
\ket{4j+1}.
\end{equation}
Equivalently, in terms of the confluent hypergeometric function, we obtain
\begin{equation}
\ket{D_-}
=
\mathcal N_-\,
{}_0F_1
\left(
;\frac54;
\frac{\lambda}{16}b^{\dagger4}
\right)
\ket1.
\end{equation}
Note again that the square root appears after acting with the operator. For illustration, the first few terms are,
\begin{equation}
    \ket{D_-}
=
\mathcal N_-
\left[
\ket1
+
\sqrt{\frac{3}{10}}\lambda\ket5
+
\sqrt{\frac{7}{40}}\lambda^2\ket9
+\cdots
\right].
\end{equation}
From the condition $\langle{D_-}\ket{D_-} = 1$, we find the normalization to be
\begin{equation}
\mathcal N_-
=
\left[
{}_2F_1
\left(
\frac12,\frac34;
\frac54;
|\lambda|^2
\right)
\right]^{-1/2}.
\end{equation}
In the small-$r$ expansion, we obtain,
\begin{equation}
    |D_-\rangle
=
\left(1-\frac{3r^2}{20}\right)|1\rangle
+
\sqrt{\frac{3}{10}}e^{i\phi}r|5\rangle
+
\sqrt{\frac{7}{40}}e^{2i\phi}r^2|9\rangle
+
\mathcal O(r^3).
\end{equation}

Thus we conclude that the dissipator has a two-dimensional dark sector spanned by $\ket{D_+}$ and $\ket{D_-}$. Therefore the most general pure dark density operator is
\begin{equation}\label{eq_squeezed_dark_sector_qubit}
    \rho^{(m)}_{D;\vartheta,\varphi}
=
|\psi^{(m)}_{D;\vartheta,\varphi}\rangle
\langle\psi^{(m)}_{D;\vartheta,\varphi}|, \qquad
|\psi^{(m)}_{D;\vartheta,\varphi}\rangle
=
\cos(\vartheta/2)|D_+\rangle
+
e^{i\varphi}\sin(\vartheta/2)|D_-\rangle
\end{equation}
Expanded, it gives
\begin{equation}
\begin{aligned}
    \rho^{(m)}_{D;\vartheta,\varphi}
&=
\cos^2(\vartheta/2)|D_+\rangle\langle D_+|
+
\sin^2(\vartheta/2)|D_-\rangle\langle D_-|\\
&\;\;\;\;\;+
\frac12\sin\vartheta\, e^{-i\varphi}|D_+\rangle\langle D_-|
+
\frac12\sin\vartheta\, e^{i\varphi}|D_-\rangle\langle D_+|.
\end{aligned}
\end{equation}

\secpar{Bare vs dressed}
We motivationally refer to $\rho^{(m)}_{D;\vartheta,\varphi}$ in Eq.~\eqref{eq_squeezed_dark_sector_qubit} as the dressed (squeezed) dark-sector qubit and $\rho_{\vartheta,\varphi}^{(m)}$ in Eq.~\eqref{eq_bare_number_state_qubit} as the bare (vacuum) dark-sector qubit. For the bare dark qubit we had
\begin{equation}
    \langle0|\rho_{\vartheta,\varphi}^{(m)}|1\rangle
=
\frac12\sin\vartheta\,e^{-i\varphi}.
\end{equation}
While for the dressed dark qubit we obtain,
\begin{equation}
    \langle0|\rho^{(m)}_{D;\vartheta,\varphi}|1\rangle
=
\frac12
\mathcal N_+\mathcal N_-
\sin\vartheta\,e^{-i\varphi}.
\end{equation}
For the dressed dark qubit, we get the same amplitude in the dressed basis, namely,
\begin{equation}
    \langle D_+|\rho^{(m)}_{D;\vartheta,\varphi}|D_-\rangle
=
\frac12\sin\vartheta\,e^{-i\varphi}.
\end{equation}
From the above, we obtain the relation
\begin{equation}
    \langle0|\rho^{(m)}_{D;\vartheta,\varphi}|1\rangle
=
\mathcal N_+\mathcal N_-
\langle D_+|\rho^{(m)}_{D;\vartheta,\varphi}|D_-\rangle .
\end{equation}
To quantify the bare $\ket{0}\bra{1}$ content of the dressed qubit, we define the static ratio
\begin{equation}
    P_{01,\mathcal S}^{D}
\equiv
\frac{
\left|\langle0|\rho^{(m)}_{D;\vartheta,\varphi}(0)|1\rangle\right|
}{
\left|\langle0|\rho_{\vartheta,\varphi}^{(m)}(0)|1\rangle\right|
}
=
\mathcal N_+\mathcal N_-.
\end{equation}
In summary, while the squeezed bath removes the vacuum protection of the bare coherence $\ket{0}\bra{1}$, it restores exact dissipative protection in the dressed basis. The protected coherence is not $\rho_{01}$, but rather the dressed matrix element $\rho_{D_+D_-}$. Namely,
\begin{equation}
    R_{D_+D_-,\mathcal S}(t)
=
\frac{
|\langle D_+|\rho^{(m)}_{\mathcal S}(t)|D_-\rangle|
}{
|\langle D_+|\rho^{(m)}_{\mathcal S}(0)|D_-\rangle|
}=1.
\end{equation}

\secpar{Truncated coherence-block solution}
In the number/thermal case, when we start from $\rho_{01}$, the dynamics stays on the nearest-neighbor ladder
\begin{equation}
    \rho_{01}, \rho_{23}, \rho_{45},\ldots,
\end{equation}
so a single index $c_j = \rho_{2j,2j+1}$ was enough.

In the squeezed case, however, starting from $\rho_{01}$ we not only get $\rho_{23}$ but also coherences such as
\begin{equation}
    \rho_{41}, \rho_{05},\ldots.
\end{equation}
So the natural object is not a single-index ladder but the full even-odd block
\begin{equation}
    C_{jk}(t)\equiv \rho_{2j,2k+1}(t),
\qquad j,k=0,1,2,\ldots .
\end{equation}
Then the bare coherence is
\begin{equation}
    \rho_{01}(t)=C_{00}(t).
\end{equation}
Then we truncate the block at finite $M$. We keep
\begin{equation}
    C_{jk}(t),
\qquad
j,k=0,\ldots,M-1,
\end{equation}
and omit all coherences with $j\geq M$ or $k\geq M$. Thus we replace the infinite hierarchy by the finite-dimensional system
\begin{equation}
    \frac{d}{dt}\mathbf C^{(M)}(t)
=
B^{(M)}\mathbf C^{(M)}(t),
\end{equation}
where
\begin{equation}
    \mathbf C^{(M)}(t)
=
\big(
C_{00},C_{01},\ldots,C_{0,M-1},
C_{10},\ldots,C_{M-1,M-1}
\big)^T .
\end{equation}
The finite-block solution is then
\begin{equation}
    \mathbf C^{(M)}(t)
=
e^{B^{(M)}t}\mathbf C^{(M)}(0).
\end{equation}
For an initially encoded bare qubit,
\begin{equation}
    C_{00}(0)=\rho_{01}(0),
\qquad
C_{jk}(0)=0
\quad
\text{for }(j,k)\neq(0,0).
\end{equation}
Therefore,
\begin{equation}
    \mathbf C^{(M)}(0)
=
\big(\rho_{01}(0),0,\ldots,0\big)^T .
\end{equation}
The corresponding normalized coherence ratio is
\begin{equation}
    R_{01,\mathcal S}^{(M)}(t)
\equiv
\frac{|\rho_{01}^{(M)}(t)|}{|\rho_{01}(0)|}.
\end{equation}
In practice, the squeezed hierarchy is solved in the connected coherence sector generated by the initial coherence \(\rho_{01}\). We write the retained matrix elements as \(\rho_{pq}\), with
\begin{equation}
    p=2a,\qquad q=2b+1,\qquad q-p=1 \mod 4 .
\end{equation}
This sector contains the nearest-neighbor ladder
\begin{equation}
    \rho_{01},\rho_{23},\rho_{45},\ldots,
\end{equation}
but also the off-ladder coherences generated by the anomalous squeezed correlations, such as
\begin{equation}
    \rho_{41},\qquad \rho_{05}.
\end{equation}
We truncate this connected sector by taking \(a,b=0,\ldots,M-1\) and dropping all couplings to coherences outside the retained set. The resulting sparse finite-dimensional system is solved in dimensionless time
\begin{equation}
    \tau=\gamma_{\mathcal V}t,
\end{equation}
as
\begin{equation}
    \frac{d}{d\tau}\mathbf s^{(M)}(\tau)
    =
    A_{\mathcal S}^{(M)}\mathbf s^{(M)}(\tau).
\end{equation}
To repeat, for an initially encoded bare qubit, the initial condition is
\begin{equation}
    \mathbf s^{(M)}(0)
    =
    \big(\rho_{01}(0),0,\ldots,0\big)^T.
\end{equation}
We then define the finite-sector bare-coherence ratio
\begin{equation}
    R_{01,\mathcal S}^{(M)}(t)
    \equiv
    \frac{
    |\rho_{01,\mathcal S}^{(M)}(t)|
    }{
    |\rho_{01}(0)|
    }.
\end{equation}
In the numerical results we use $M=40$. Thus the plotted curve $R_{01,\mathcal S}^{(40)}(t)$ in Fig.~\ref{fig_coherence_ratio_comparison}, is the exact solution of this finite sparse hierarchy and an approximation to the full infinite squeezed coherence network.

\end{document}